\documentclass[twocolumn,usenatbib]{aastex63}

\usepackage{graphicx}	% Including figure files
\usepackage{amsmath}	% Advanced maths commands
\usepackage{amssymb}	% Extra maths symbols

\received{\today}
\revised{\today}
\accepted{\today}
\submitjournal{ApJ}

\shorttitle{Compressive relativistic plasma turbulence}
\shortauthors{Zhdankin}
\begin{document}

\newcommand{\red}{\textcolor{red}}
\newcommand{\blue}{\textcolor{blue}}
\newcommand{\green}{\textcolor{green}}

\title{Particle energization in relativistic plasma turbulence: solenoidal versus compressive driving}

\correspondingauthor{Vladimir Zhdankin}
\email{zhdankin@princeton.edu}

\author{Vladimir Zhdankin}
\thanks{NASA Einstein fellow}
\affiliation{Department of Astrophysical Sciences, Princeton University, 4 Ivy Lane, Princeton, NJ 08544, USA}

%\nocollaboration



\begin{abstract}
Many high-energy astrophysical systems contain magnetized collisionless plasmas with relativistic particles, in which turbulence can be driven by an arbitrary mixture of solenoidal and compressive motions. For example, turbulence in hot accretion flows may be driven solenoidally by the magnetorotational instability or compressively by spiral shock waves. It is important to understand the role of the driving mechanism on kinetic turbulence and the associated particle energization. In this work, we compare particle-in-cell simulations of solenoidally driven turbulence with similar simulations of compressively driven turbulence. We focus on plasma that has an initial beta of unity, relativistically hot electrons, and varying ion temperature. Apart from strong large-scale density fluctuations in the compressive case, the turbulence statistics are similar for both drives, and the bulk plasma is described reasonably well by an isothermal equation of state. We find that nonthermal particle acceleration is more efficient when turbulence is driven compressively. In the case of relativistically hot ions, both driving mechanisms ultimately lead to similar power-law particle energy distributions, but over a different duration. In the case of non-relativistic ions, there is significant nonthermal particle acceleration only for compressive driving. Additionally, we find that the electron-to-ion heating ratio is less than unity for both drives, but takes a smaller value for compressive driving. We demonstrate that this additional ion energization is associated with the collisionless damping of large-scale compressive modes via perpendicular electric fields.
\end{abstract}

%\keywords{acceleration of particles, plasmas, relativistic processes, turbulence} % , pulsars: individual (Crab) magnetohydrodynamics (MHD), 


\section{Introduction} \label{sec:intro}

Turbulence has long been recognized as a candidate process for generating nonthermal populations of high-energy particles in magnetized collisionless plasmas throughout the Universe \citep[e.g.,][]{fermi_1949, tsytovich_1966, kulsrud_ferrari_1971}. Despite this extensive history, a rigorous theoretical understanding of turbulent particle acceleration currently remains elusive. Understanding the quantitative properties of turbulent particle acceleration is essential for interpreting the broadband radiation emission from high-energy astrophysical systems such as pulsar wind nebulae \citep{gaensler_slane_2006}, black-hole accretion flows \citep{yuan_narayan_2014}, blazar jets \citep{bottcher_2007}, and gamma-ray bursts \citep{beloborodov_meszaros_2017}. This task is also critical for constraining possible sources of high-energy cosmic rays \citep{blandford_etal_2014}.

High-energy astrophysical systems typically have substantial populations of relativistic particles, such that the electron temperature $T_e$ or ion temperature $T_i$ exceed the corresponding rest mass energies $m_s c^2$ ($s \in \{ e, i \}$), so that $\theta_s \equiv T_s/m_s c^2 \gtrsim 1$. Furthermore, these systems are also often relativistic ($\sigma > 1$) or trans-relativistic ($\sigma \sim 1$) when characterized by the magnetization parameter $\sigma$, defined as the ratio of magnetic enthalpy to plasma enthalpy, which sets the Alfv\'{e}n velocity (and thus bulk flow velocites) relative to the speed of light, $v_A/c = [\sigma/(1+\sigma)]^{1/2}$. Turbulence in the relativistic regime has been studied significantly less than in the non-relativistic regime ($\theta_e \ll 1$, $\theta_i \ll 1$, $\sigma \ll 1$) relevant to plasmas in the heliosphere and in laboratory experiments.

Recent particle-in-cell (PIC) simulations provide evidence that kinetic plasma turbulence with $\sigma \gtrsim 1$ can accelerate particles nonthermally to high energies \citep{zhdankin_etal_2017, zhdankin_etal_2018b, comisso_sironi_2018, comisso_sironi_2019, nattila_etal_2021}. The energization mechanism in these simulations is broadly consistent with diffusive particle acceleration by gyroresonant-like interactions with turbulent fluctuations \citep{wong_etal_2020}, as predicted by quasilinear analytical theories \citep[e.g.,][and references therein]{schlickeiser_1989, chandran_2000, demidem_etal_2020}. Other processes such as intermittent magnetic reconnection may play a role in heating and injecting particles into the nonthermal population \citep{comisso_sironi_2018, comisso_sironi_2019}. Ions are preferentially energized over electrons across a broad parameter space, consistent with their larger gyroradii enabling stronger interactions with inertial-range fluctuations \citep{zhdankin_etal_2019}.

These previous numerical studies provide an important foundation to our theoretical understanding of turbulent particle energization in relativistic collisionless plasmas, but there are a number of very basic questions that remain unanswered. One aspect that has not yet received careful scrutiny in previous PIC studies is the influence of the driving mechanism (or initial conditions, for the decaying case) on kinetic turbulence and the associated particle energization. In this work, we focus on one of the most basic parameters used to characterize driving mechanisms: the compressibility. 

Turbulence is typically driven at scales much larger than the characteristic kinetic plasma scales (i.e., particle gyroradii and skin depths); the large-scale turbulence is then believed to be accurately described in the framework of magnetohydrodynamics (MHD), assuming that the collisionless plasma is magnetized sufficiently \citep[see, e.g.][]{schekochihin_etal_2009}. In strong MHD turbulence, the cascade can be divided into two channels: an incompressible Alfv\'{e}nic cascade, which is mediated by Alfv\'{e}n waves propagating along the background magnetic field, and a compressive cascade, which is mediated by fast magnetosonic waves. In the simplest standard model, solenoidal driving triggers an Alfv\'{e}nic cascade that exhibits a classical Kolmogorov energy spectrum ($-5/3$ power law) with a scale-dependent anisotropy described by critical balance, such that fluctuations become increasingly elongated along the guide field at smaller scales \citep{goldreich_sridhar_1995, cho_vishniac_2000}. Slow magnetosonic modes are passively mixed with the Alfv\'{e}nic cascade, leading to weak density fluctuations with a similar spectrum \citep{lithwick_goldreich_2001}. Compressive driving, on the other hand, causes a fast-mode cascade that can freely cross magnetic field lines and therefore acquires an isotropic spectrum, as shown by numerical simulations \citep{cho_lazarian_2002, cho_lazarian_2003}. Alfv\'{e}nic and fast-mode cascades are not believed to interact strongly with each other \citep{lithwick_goldreich_2001,schekochihin_etal_2009}, so they are often treated as decoupled channels, with the relative power in each cascade determined by the large-scale driving mechanism. We caution, however, that some degree of mode conversion may occur \citep{makwana_yan_2020}. In particular, recent studies indicated that when $\sigma \gtrsim 1$, the two cascades may freely exchange energy with each other \citep{takamoto_lazarian_2016, takamoto_lazarian_2017}, despite obeying similar phenomenology with regards to critical balance \citep{thompson_blaes_1998, cho_2005}.

Apart from the spectrum of turbulence, the compressibility of driving may influence several other aspects of the resulting dynamics. For example, in MHD models, compressive driving leads to a much broader distribution of density fluctuations \citep{federrath_etal_2008} and different types of coherent structures, such as shocks \citep{yang_etal_2017}. MHD simulations also demonstrated that magnetic field amplification is reduced for compressively driven turbulence when compared to the solenoidal case \citep{federrath_etal_2011, federrath_2016, yang_etal_2016}.

In the collisionless regime, numerous theoretical studies suggested that compressive fluctuations will yield more efficient nonthermal particle acceleration (through diffusive second-order Fermi mechanisms) than solenoidal fluctuations \citep[e.g.,][]{schlickeiser_miller_1998, yan_lazarian_2002, yan_lazarian_2004, lazarian_etal_2012}. The anisotropy associated with critical balance may cause the Alfv\'{e}nic cascade to be inefficient at scattering particles, since the resonance condition becomes difficult to satisfy for most particles \citep{chandran_2000}; the isotropic fast-mode cascade does not suffer from this issue \citep{yan_lazarian_2002}. More recent analytical work found that resonance broadening associated with the finite lifetime of fluctuations can increase the scattering efficiency of the Alfv\'{e}nic cascade, making it a viable accelerator despite anisotropy \citep{demidem_etal_2020}. Supporting this scenario, simulations of test particles in weakly compressible MHD turbulence indicated that resonance broadening may substantially increase the efficiency of particle acceleration for solenoidal driving \citep{lynn_etal_2014}. We note that the presence of non-resonant acceleration mechanisms, intermittency, and magnetic reconnection all complicate the story from idealized analytical scenarios \citep[e.g.,][]{vlahos_etal_2004, brunetti_lazarian_2007, lazarian_etal_2012, isliker_etal_2017, xu_zhang_2017, lemoine_2021}. Overall, there remains significant uncertainty in whether solenoidal and/or compressive cascades can efficiently energize particles in various parameter regimes.

The above considerations hold under the assumption of an MHD-like cascade; however, there are a couple of additional potential complications to nonthermal particle acceleration by compressive kinetic turbulence in collisionless plasmas. The first comes from the fact that fast modes are Landau damped even at large scales \citep{barnes_1966}. If compressive modes are rapidly damped rather than cascaded to smaller scales, this may inhibit diffusive particle acceleration due to the limited spectrum of modes that particles can interact gyroresonantly with. The second complication is that it is {\it a priori} unknown how dissipated energy is partitioned between electrons and ions; this may influence the ability of electrons and/or ions to be injected to energies where they can gain energy from MHD-scale fluctuations. Phenomenological models for electron and ion heating rates from Landau damping of Alfv\'{e}nic turbulence were developed in the non-relativistic regime \citep{quataert_1998, gruzinov_1998, quataert_gruzinov_1999, howes_2010}, and were broadly supported by recent hybrid gyrokinetic simulations \citep{kawazura_etal_2019}. Analytical theory also predicted that fast modes should preferentially heat ions in low $\beta$ plasmas \citep{schekochihin_etal_2019}; this was subsequently confirmed by gyrokinetic simulations of compressively driven turbulence \citep{kawazura_etal_2020}. It is unclear how these results translate to the relativistic regime, where diffusive particle acceleration can absorb a significant fraction of the cascaded energy. These uncertainties about nonthermal particle acceleration in a fully collisionless, relativistic system motivate our present work.

The question that we answer in this work is: how is turbulent particle energization affected by the compressibility of the external driving? To this end, we report the effect of solenoidal and compressive driving on electron and ion energization in PIC simulations of collisionless plasma turbulence. For numerical tractability and applicability to certain high-energy astrophysical systems (e.g., radiatively inefficient accretion flows around supermassive black holes), we focus on the regime where electrons are relativistically hot ($\theta_e \gg 1$), while ions may be relativistic or sub-relativistic. In the relativistic case ($\theta_i \gg 1$, $\sigma \sim 1$), electron and ion dynamics are effectively symmetric, and both particle species are efficiently accelerated for both driving mechanisms; however, particles are accelerated more rapidly in the compressive case. In the sub-relativistic case ($\theta_i \ll 1$, $\sigma \ll 1$), ions are preferentially energized over electrons; nonthermal acceleration occurs for compressive driving but not for solenoidal driving. This indicates that a fast mode cascade may be essential for accelerating particles in many systems that are not strongly relativistic. We apply diagnostics to demonstrate that the extra ion energization in the compressive case is associated with the damping of compressive fluctuations at large scales via perpendicular electric fields.

This concludes Section~\ref{sec:intro}. In Section~\ref{sec:methods}, we describe the parameter space and numerical setup. In Section~\ref{sec:results1}, we set the stage by providing a synopsis of the turbulence properties in our PIC simulations. In Section~\ref{sec:results2}, we reveal the principal results on the electron and ion energization (including overall electron-ion energy partition and nonthermal particle acceleration). Finally, we conclude in Section~\ref{sec:conclusions} by summarizing our primary results, stating implications of our study, and pointing out potential future directions.

\section{Methods} \label{sec:methods}

\subsection{Parameter space}

In this subsection, we describe the parameter space explored by the numerical simulations in our study. In particular, we define the {\it relativistic} and {\it semirelativistic} regimes, which are two distinct plasma physical regimes that will be compared in the results and are relevant to different classes of high-energy astrophysical systems.

In this work, we consider collisionless plasmas with $\beta \equiv 8 \pi (n_{i,0} T_i + n_{e,0} T_e)/B_{\rm rms}^2 \sim 1$, ultra-relativistically hot electrons ($\theta_e \equiv T_e/m_e c^2 \gg 1$), and ions with a temperature that may be either sub-relativistic ($\theta_i \equiv T_i/m_i c^2 \ll 1$) or relativistic ($\theta_i \gtrsim 1$). In these definitions, $B_{\rm rms}$ is the characteristic (root-mean-square) magnetic field, $n_{s,0} = n_0/2$ is the average particle number density per species, and $m_s$ is the particle rest mass for species $s \in \{ e, i \}$; we focus on an electron-proton composition so that $m_i/m_e = 1836$. We always consider an initial ion-to-electron temperature ratio of unity, $T_{i.0}/T_{e,0} = 1$, although this is free to evolve as the particles are heated by the turbulence. Throughout the paper, initial values of the parameters are denoted with a subscript $0$.

The plasma is characterized by several relevant kinetic scales. The characteristic Larmor radii are given by $\rho_{s} = (\overline{\gamma}_{s}^2 - 1)^{1/2} m_s c^2/e B_{\rm rms}$, where $\overline{\gamma}_{s} = 1 + \overline{E}_s/m_s c^2$ are the mean particle Lorentz factors and $\overline{E}_s$ are the mean particle kinetic energies (for species $s$). Note that for a thermal plasma, $\overline{\gamma}_s \sim 3 \theta_s$ for $\theta_s \gg 1$ and $\overline{\gamma}_s \sim 1 + (3/2) \theta_s$ for $\theta_s \ll 1$. The skin depths are given by $d_s = (\overline{\gamma}_s m_s c^2/4\pi n_{s,0} e^2)^{1/2}$.

We call $\theta_i \gg 1$ the {\it relativistic} regime, because both electrons and ions are relativistically hot in this case. In this situation, the particle rest masses $m_s$ become negligible compared to their relativistic mass $\gamma_s m_s$. The plasma then acts like a pair (electron-positron) plasma, as long as radiative cooling effects are neglected. Thus, there is no electron-ion kinetic scale separation, i.e., $\rho_i/\rho_e = 1$ and $d_i/d_e = 1$. The dynamics are identical (in a statistical sense) for electrons and ions.

We call the regime $m_e/m_i \ll \theta_i \ll 1$ the {\it semirelativistic\rm} regime, a term that was previously coined in \cite{werner_etal_2018}. In the semirelativistic regime, electrons are relativistically hot ($\theta_e \gg 1$) while ions are subrelativistic.  The separation between the electron and ion Larmor radii is given by $\rho_{e}/\rho_{i} \sim \theta_{i}^{1/2} T_e/T_i$, while the separation between the skin depths is given by $d_e/d_i \sim \theta_i^{1/2} (T_e/T_i)^{1/2} $.  Due to the computational benefits of having relativistic electrons (which effectively reduces the ion-to-electron mass ratio) and also because of its relevance for systems such as hot accretion flows, this semirelativistic regime was extensively studied by PIC simulations of magnetic reconnection \citep[e.g.,][]{rowan_sironi_narayan_2017, ball_etal_2018, werner_etal_2018}. We also recently studied electron and ion energization in PIC simulations of electromagnetically driven turbulence in the semirelativistic regime in \cite{zhdankin_etal_2019}, which was a precursor to our present work. 

The characteristic magnetization $\sigma \equiv B_{\rm rms}^2/4\pi \overline{h}$ is a function of the other physical parameters stated above, and thus cannot be considered as independent. The average relativistic plasma enthalpy density is given by $\overline{h} = \overline{h}_i + \overline{h}_e$ where $\overline{h}_s  = n_{s,0} \overline{\gamma}_s m_s c^2 + \overline{P}_s$ and $\overline{P}_s$ is the average pressure for species $s$. Note that $\overline{h}_s$ is a nontrivial function of temperature in the trans-relativistic temperature regime ($\theta_s \sim 1$). However, for relativistic particles ($\theta_s \gg 1$) it has the simple expression $\overline{h}_s \sim (4/3) n_{s,0} \overline{\gamma}_s m_s c^2$ while for non-relativistic particles ($\theta_s \ll 1$) it is $\overline{h}_s \sim n_{s,0} m_s c^2$.

\subsection{Numerical setup}

In this subsection, we describe the numerical setup for our PIC simulations, including common parameters and a description of the driving mechanisms.

We perform a series of 3D simulations of externally driven turbulence using the PIC code {\sc Zeltron} \citep{cerutti_etal_2013}. The domain is a periodic cubic box of volume $L^3$ with mean magnetic field $\boldsymbol{B}_0 = B_0 \hat{\boldsymbol{z}}$. Particles are initialized from a uniform Maxwell-J\"{u}ttner distribution with particle density per species $n_0/2$ and equal electron and ion temperatures, $T_{e0} = T_{i0}$, with $T_{i0}$ specified by the dimensionless temperature parameter $\theta_{i0} = T_{i0}/m_i c^2$ (which will be varied between simulations). All simulations have an initial plasma beta (based on the guide field) of $\beta_0 = 1$. To speed up the formation of turbulence, we initialize all simulations with a weak magnetic field perturbation (with amplitude $\delta B \ll B_0$) at the largest scale.

We apply an external body force on particles to drive bulk motions at large scales; this is in contrast to our previous studies \citep[e.g.,][]{zhdankin_etal_2019}, where we used an external current density to drive the turbulence electromagnetically. The use of an external body force is necessary for controlling the compressibility of the driving. In our simulations, we apply a perpendicular driving force $\boldsymbol{F}_{\rm ext} = F_{{\rm ext},x} \hat{\boldsymbol{x}} + F_{{\rm ext},y} \hat{\boldsymbol{y}}$ for all modes with perpendicular wavenumbers satisfying $k_\perp \equiv (k_x^2 + k_y^2)^{1/2} \le 6\pi/L$ and parallel wavenumbers at the largest scale, $k_z = \pm 2\pi/L$ (necessary to break symmetry along the mean field direction). The force at each value of $\boldsymbol{k}$ has a random phase that is evolved independently using the Langevin equation in \cite{tenbarge_etal_2014}. The reason that we constrain $\boldsymbol{F}_{\rm ext}$ to be perpendicular to $\boldsymbol{B}_0$ is to avoid generation of large-scale flows parallel to $\boldsymbol{B}_0$, which complicate the analysis. Each mode has an amplitude of $B_0^2 k/(8\pi n_0 N_{\rm dr}^{1/2})$, frequency $\omega_{\rm dr} = 0.5 v_{A0} k$, and decorrelation rate $\gamma_{\rm dr}  = 0.4 v_{A0} k$, where $N_{\rm dr} = 56$ is the number of modes. These driving amplitudes make the fluctuating magnetic field comparable to mean magnetic field, $\delta B_{\rm rms} \sim B_0$, so that turbulence is strong. As a result, the turbulent flow velocities are Alfv\'{e}nic, ${\mathcal V}_{\rm rms}/v_A \sim 1$. Since $\beta_0 = 1$, the characteristic turbulent Mach number is order unity: $M \equiv {\mathcal V}_{\rm rms}/c_s \sim \beta^{-1} {\mathcal V}_{\rm rms}/v_A \sim 1$, where $c_s$ is the sound speed. In practice, the turbulence is subsonic.

In general, $\boldsymbol{F}_{\rm ext}$ can drive an arbitrary mixture of solenoidal modes and compressive modes, depending on the orientation of the force and the wavevector for each mode. In this work, we focus on two limiting cases: solenoidal driving (with $\nabla_\perp \cdot \boldsymbol{F}_{\rm ext} = 0$) and compressive driving (with $\nabla_\perp \times \boldsymbol{F}_{\rm ext} = 0$), where $\nabla_\perp$ is the gradient perpendicular to $\boldsymbol{B}_0$. For the solenoidal (incompressible) cases, we thus choose the direction of $\boldsymbol{F}_{\rm ext}$ at each wavevector to be perpendicular to $\boldsymbol{k}_\perp$, which enforces $\nabla_\perp \cdot \boldsymbol{F}_{\rm ext} = 0$. For the compressive cases, we instead choose $\boldsymbol{F}_{\rm ext}$ to be parallel to $\boldsymbol{k}_\perp$, thus enforcing $\nabla_\perp \times \boldsymbol{F}_{\rm ext} = 0$. Note that since $\boldsymbol{F}_{\rm ext}$ does not have a component in $\hat{\boldsymbol{z}}$, we cannot enforce $\partial _z F_{{\rm ext},x} = 0$ and $\partial_z F_{{\rm ext},y} = 0$, so the total curl is nonzero, $\nabla \times \boldsymbol{F}_{\rm ext} \ne 0$, for the compressive case; there is thus a small solenoidal component and this case is not ``purely'' compressive. We have also performed simulations with isotropic driving by including $F_{{\rm ext},z}$ components, which enforce exact compressibility, with qualitatively similar results. Finally, we also performed simulations with fewer driven modes, which yielded similar results to the ones described in this paper (although with stronger statistical variability).

\subsection{Numerical simulations}

\begin{table}
\centering \caption{List of simulations and parameters. \newline
Common parameters: $\beta_0 = 1$, $32$ particles per cell,~$dx=\rho_{e0}/1.5$, and $dt=3^{-1/2} dx/c$.} \label{table:sims}
\begin{tabular}{|c|c|c|c|c|} 
	\hline
\hspace{0.5 mm} Case \hspace{0.5 mm}  & \hspace{1 mm} $N_x^3$ \hspace{1 mm}   & $L/2\pi\rho_i$  &   \hspace{1 mm} $\theta_{i0}$ \hspace{1 mm} &  Sol. or Comp.?   \\
	\hline
rL10s & $768^3$ & 91.0 & $10$ & Sol. \\
rL10c & $768^3$ & 91.0 & $10$ & Comp. \\
rL1d256s & $768^3$ & 9.8 & $1/256$ & Sol. \\
rL1d256c & $768^3$ & 9.8 & $1/256$ & Comp. \\
rS10s & $384^3$ & 45.5 & $10$ & Sol.  \\
rS10c & $384^3$ & 45.5 & $10$ & Comp.  \\
rS1s & $384^3$ & 42.5 & $1$ & Sol.  \\
rS1c & $384^3$ & 42.5 & $1$ & Comp.  \\
rS1d4s & $384^3$ & 31.8 & $1/4$ & Sol.  \\
rS1d4c & $384^3$ & 31.8 & $1/4$ & Comp.  \\
rS1d16s & $384^3$ & 18.6 & $1/16$ & Sol.  \\
rS1d16c & $384^3$ & 18.6 & $1/16$ & Comp.  \\
rS1d64s & $384^3$ & 9.7 & $1/64$ & Sol.  \\
rS1d64c & $384^3$ & 9.7 & $1/64$ & Comp.  \\
rS1d256s & $384^3$ & 4.9 & $1/256$ & Sol.  \\
rS1d256c & $384^3$ & 4.9 & $1/256$ & Comp.  \\
	\hline
\end{tabular}
\centering
\label{table-sims}
\end{table}

The simulations described in this paper, along with their lattice sizes ($N_x^3$), dimensionless physical parameters, driving type are listed in Table~\ref{table:sims}. There are four large fiducial cases (with $768^3$ cells) and twelve small cases (with $384^3$ cells), which involving parallel scans in $1/256 \le \theta_{i0} \le 10$ with compressive and solenoidal driving.

For the majority of the paper, we focus on the set of four fiducial simulations: compressive and solenoidal cases at $\theta_{i0} = 10$ (the relativistic regime), as well as compressive and solenoidal cases at $\theta_{i0} = 1/256$ (the semirelativistic regime). Since $\beta_0 = 1$, these two regimes also have different initial magnetizations: $\sigma_0 \approx 0.5$ for the $\theta_{i0} = 10$ cases while $\sigma_0 \approx 0.02$ for the $\theta_{i0} = 1/256$ cases. These four fiducial cases thus allow us to compare the effect of the driving mechanism both in the relativistically hot, $\sigma \sim 1$ regime (which is effectively a pair plasma) and in the semirelativistic, $\sigma \ll 1$ regime (which has an electron-ion kinetic scale separation of $\rho_{i0}/\rho_{e0} = 9.3$). Since the resolution is fixed with respect to electron kinetic scales, the $\theta_{i0} = 10$ cases have a relatively long MHD inertial range ($L/2\pi\rho_{i0} = 91.0$), while the $\theta_{i0} = 1/256$ cases have a limited MHD inertial range ($L/2\pi\rho_{i0} = 9.8$).

In addition to the fiducial simulations, we conduct a parameter scan in $\theta_{i0} \in \{1/256, 1/64, 1/16, 1/4, 1, 10 \}$ for both types of driving at twice smaller system sizes, which is used in Section~\ref{eiheating}. Apart from the simulations listed in Table~\ref{table:sims}, we conducted numerous additional simulations to confirm the numerical accuracy of the simulations, by varying resolution, particles per cell, driving parameters, etc.

All simulations have $32$ particles per cell per species, cell size $dx = \rho_{e0}/1.5$, and time step $dt = 3^{-1/2} dx/c$. All cases run for a duration of at least $6 L/v_{A0}$; the fiducial $\theta_{i0} = 10$ cases run for more than $14 L/v_{A0}$, while the $\theta_{i0} = 1/256$ cases run for a little over $6 L/v_{A0}$.

\section{Turbulence properties} \label{sec:results1}

\subsection{Evolution} \label{subsec:evo}

 \begin{figure}
\includegraphics[width=0.95\columnwidth]{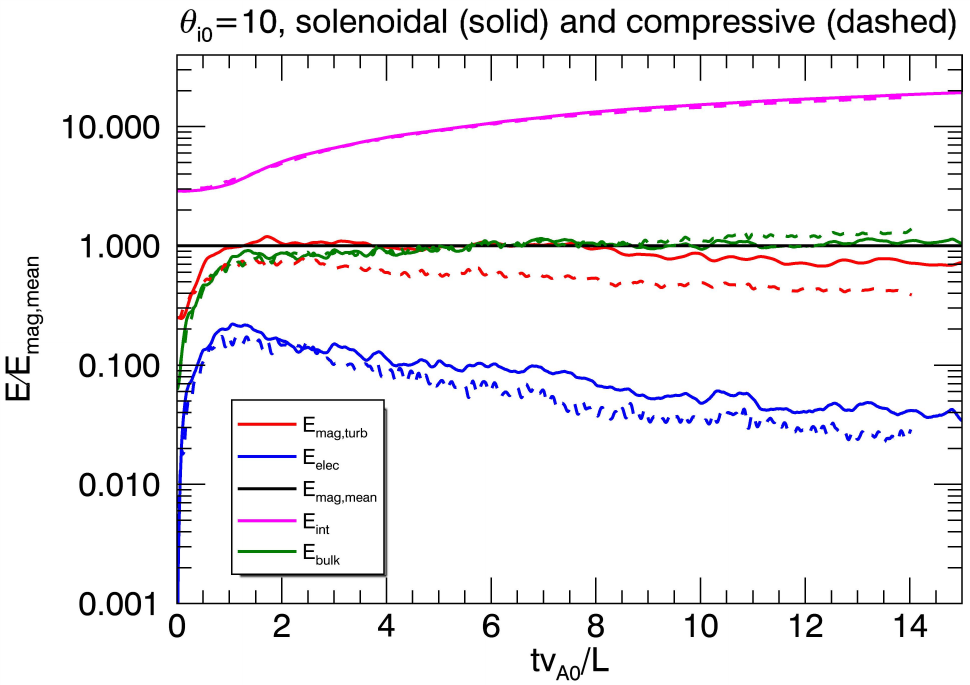}
\includegraphics[width=0.95\columnwidth]{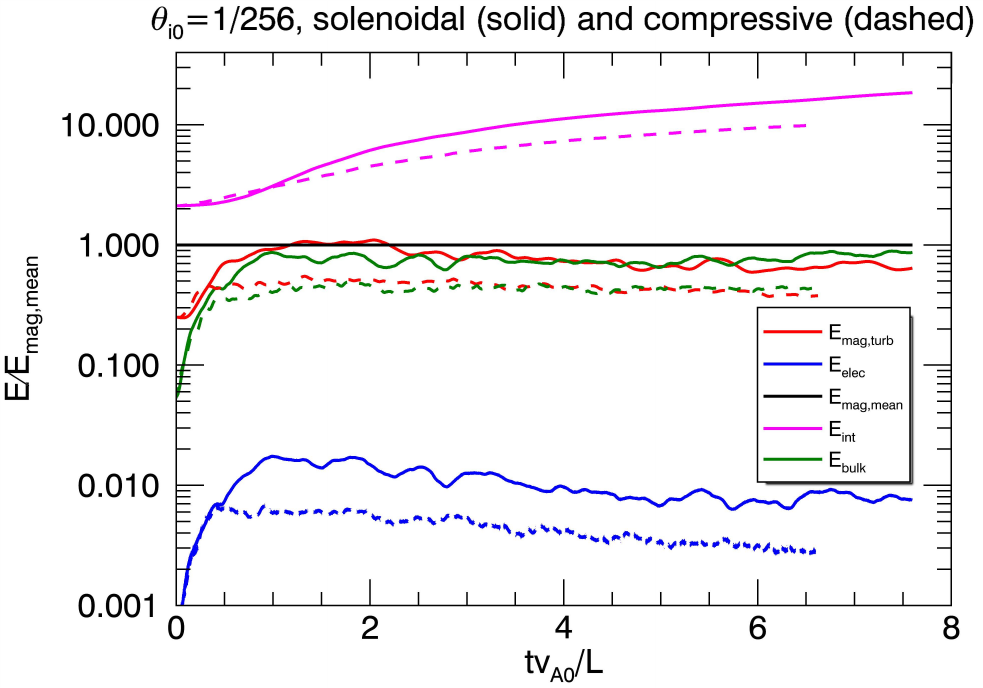}
   \centering
  \caption{\label{fig:energy} Energy evolution in the $\theta_{i0}=10$ simulations (top panel) and $\theta_{i0} = 1/256$ simulations (bottom panel) with solenoidal driving (solid lines) and compressive driving (dashed lines). The quantities are turbulent magnetic energy (red), electric energy (blue), mean field magnetic energy (black), internal energy (magenta), and turbulent bulk kinetic energy (green).}
 \end{figure}
 
 In this subsection, we commence the presentation of numerical results by describing the overall evolution of the turbulence energies. In the following, we use relativistic definitions for the internal and bulk fluid energies introduced in \cite{zhdankin_etal_2018a}, with appropriate subtractions of the rest mass energy. Specifically, the internal and bulk energies for species $s$ are defined respectively by
 \begin{align}
 E_{{\rm int},s} &= \int d^3x \left[ \left( {\mathcal E}_{f,s}^2 - |\boldsymbol{\mathcal P}_{f,s}|^2 c^2 \right)^{1/2} - n_s m_s c^2 \right] \, , \nonumber \\
 E_{{\rm bulk},s} &= \int d^3x \left({\mathcal E}_{f,s} - n_s m_s c^2 \right) - E_{{\rm int},s} \, ,
 \end{align}
 where~${\mathcal E}_{f,s}(\boldsymbol{x},t) = \int d^3p m_s^2 c^4 + p^2 c^2 )^{1/2} f_s(\boldsymbol{p},\boldsymbol{x},t)$ is the local species energy density,~$\boldsymbol{\mathcal P}_{f,s}(\boldsymbol{x},t) = \int d^3p \boldsymbol{p} f_s(\boldsymbol{p},\boldsymbol{x},t)$ is the species momentum density,~$n_s(\boldsymbol{x},t)=\int d^3p f_s(\boldsymbol{p},\boldsymbol{x},t)$ is the species number density, and~$f_s(\boldsymbol{p},\boldsymbol{x},t)$ is the species distribution function. The total internal and bulk energies are obtained by combining the separate species contributions: $E_{\rm int} = E_{{\rm int},i} + E_{{\rm int},e}$ and $E_{\rm bulk} = E_{{\rm bulk},i} + E_{{\rm bulk},e}$.
 
 Turbulence develops from the initial state after a transient phase that takes roughly one large-scale Alfv\'{e}n crossing time, $L/v_{A0}$. The subsequent evolution of the overall plasma energy partition is shown in Fig.~\ref{fig:energy} for the four fiducial simulations; the top panel shows the $\theta_{i0} = 10$ cases while the bottom panel shows the $\theta_{i0} = 1/256$ cases. In all cases, once turbulence is fully developed, the energy in the magnetic field fluctuations ($E_{\rm mag, turb} = \int d^3x \delta B^2/8\pi$, where $\delta \boldsymbol{B} = \boldsymbol{B}-\boldsymbol{B}_0$ is the fluctuating magnetic field) and in the bulk turbulent motions $E_{\rm bulk}$ are both comparable (within a factor of two) to the energy in the mean magnetic field, $E_{\rm mag,mean} = \int d^3x B_0^2/8\pi$, as governed by the driving amplitude. The internal energy $E_{\rm int}$ increases in time due to net plasma heating; we note that since $E_{\rm int} \gg E_{\rm bulk}$, it follows that the turbulent Mach number is significantly less than unity. The electric energy $E_{\rm elec} = \int d^3x \boldsymbol{E}^2/8\pi$ is subdominant in all cases; this is because the electric field is mainly from advective motions, $\boldsymbol{E}_{\rm ideal} = - (\boldsymbol{\mathcal V}_f/c) \times \boldsymbol{B}$, where $\boldsymbol{\mathcal V}_f$ is the flow velocity, which gives the ratio $E_{\rm ideal}^2/B^2 \sim ({\mathcal V}_f/c)^2 \sim (v_A/c)^2 = \sigma/(1+\sigma)$, which is significantly below unity in our simulations.
 
We now hone in on the differences between the various cases in Fig.~\ref{fig:energy}. For the solenoidal cases, $E_{\rm mag, turb}$ and $E_{\rm bulk}$ are both very similar, as expected for Alfv\'{e}nic fluctuations. In contrast, for the compressive case with $\theta_{i0} = 10$, $E_{\rm mag, turb}$ is smaller than $E_{\rm bulk}$ and $E_{\rm mag, mean}$ by roughly a factor of 2. This indicates that turbulence is non-Alfv\'{e}nic at large scales; one may read this result as indicating that turbulent magnetic field amplification is weaker for compressive driving than solenoidal driving, as previously observed in the MHD case \citep[e.g.,][]{federrath_etal_2011}. For the compressive $\theta_{i0} = 1/256$ case, $E_{\rm mag, turb}$ and $E_{\rm bulk}$ are both below $E_{\rm mag, mean}$ by a factor of 2, indicating a reduced capacity for sustaining turbulent fluctuations. For the $\theta_{i0} = 10$ case, $E_{\rm elec}$ decreases in time since advective motions slow down as the plasma heats up and the relativistic mass increases (decreasing $v_A/c$). For $\theta_{i0} = 1/256$, $E_{\rm elec}$ stays essentially constant because $v_A/c$ is determined mainly by the rest mass, which is fixed in time.

The external driving is designed such that $E_{\rm int}$ has a similar evolution for both the solenoidal case and compressive case at $\theta_{i0} = 10$. For the $\theta_{i0} = 1/256$ case, on the other hand, the compressive case has a slower increase of $E_{\rm int}$ than the solenoidal case, indicating that the driving injects less energy into the system. The differences in the energy injection rate between the relativistic and semirelativistic cases are not surprising because the external force couples differently with the plasma in the two regimes. For a relativistic plasma, the external force acts symmetrically on both species, and thus drives a directed flow in the plasma. For a semirelativistic plasma, however, the external force accelerates electrons more rapidly due to their lower effective mass; this causes an electric current to be driven in addition to the plasma flow.

   \begin{figure}
\includegraphics[width=0.95\columnwidth]{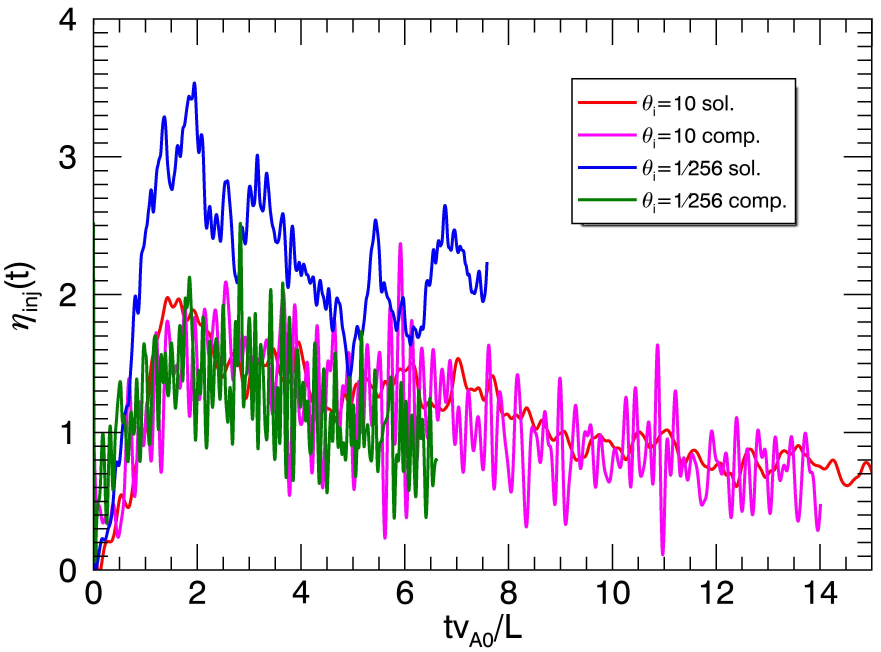}
\includegraphics[width=0.95\columnwidth]{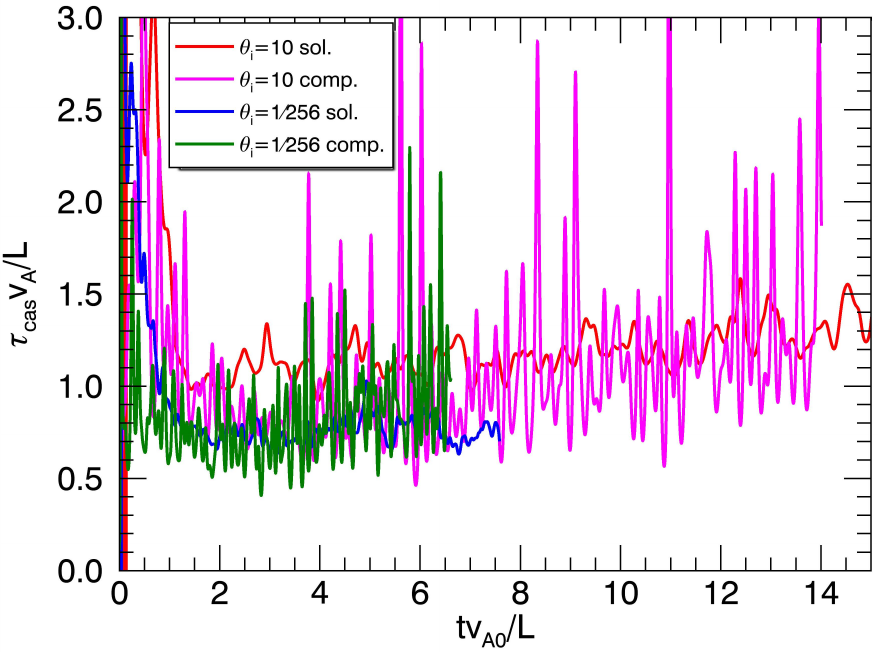}
   \centering
  \caption{\label{fig:etainj} Top panel: The evolution of the injection efficiency $\eta_{\rm inj}$ for the fiducial simulations (indicated in legend). Bottom panel: The cascade timescale $\tau_{\rm cas}$ relative to the Alfv\'{e}n crossing time $L/v_A$.}
 \end{figure}
 
 The internal energy evolution can be characterized quantitatively by measuring the dimensionless {\it injection efficiency}, $\eta_{\rm inj} \equiv (\dot{E}_{\rm int} / E_{\rm mean,mag}) v_{A0}/L$, where $\dot{E}_{\rm int}$ is the time derivative of the internal energy. We expect $\eta_{\rm inj} \sim 1$ for Alfv\'{e}nic turbulence with $\delta B_{\rm rms} \approx B_0$. We show $\eta_{\rm inj}$ for the fiducial cases in the top panel of Fig.~\ref{fig:etainj}. We find that $\eta_{\rm inj} \sim 1$ for all cases except for the $\theta_{i0} = 1/256$ solenoidal case, which is a factor of $\sim 2$ larger. 
 
 The differences in $\eta_{\rm inj}$ from case to case may either due to the cascade timescale changing, or the amplitude of the turbulence changing. To isolate these effects, we estimate the cascade timescale by taking the ratio of the turbulent energy to the heating rate,
 \begin{align}
 \tau_{\rm cas} = \frac{E_{\rm mag, turb} + E_{\rm bulk}}{\dot{E}_{\rm int}} \, .
 \end{align}
 We show the evolution of $\tau_{\rm cas}(t) v_A(t)/L$ in the bottom panel of Fig.~\ref{fig:etainj}. Although there are moments of rapid variability, on average $\tau_{\rm cas} v_{A}/L \sim 1$ for all cases. This indicates that after accounting for the varying amplitude of turbulent fluctuations, the compressive and solenoidal cases both cascade at a similar rate --- namely, the Alfv\'{e}nic rate.

 \subsection{Morphology}
 
In this subsection, we describe qualitative features of the turbulence in the four fiducial simulations by showing visuals of the PIC simulations.

 \begin{figure}
  \includegraphics[width=0.95\columnwidth]{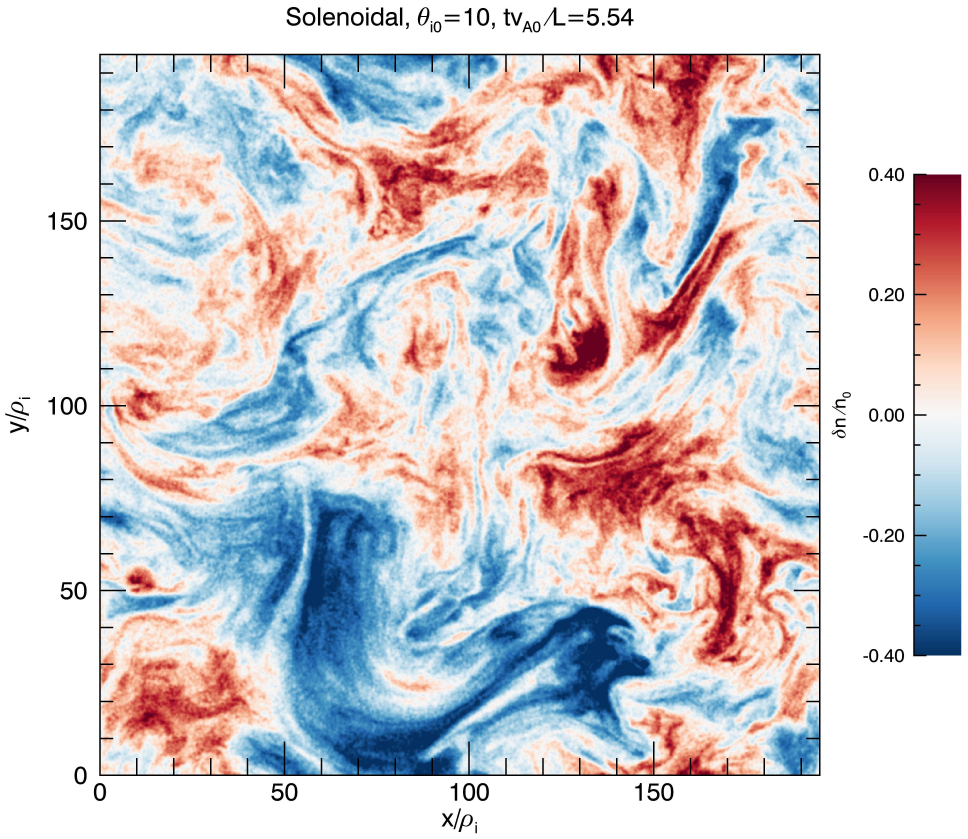}
\includegraphics[width=0.95\columnwidth]{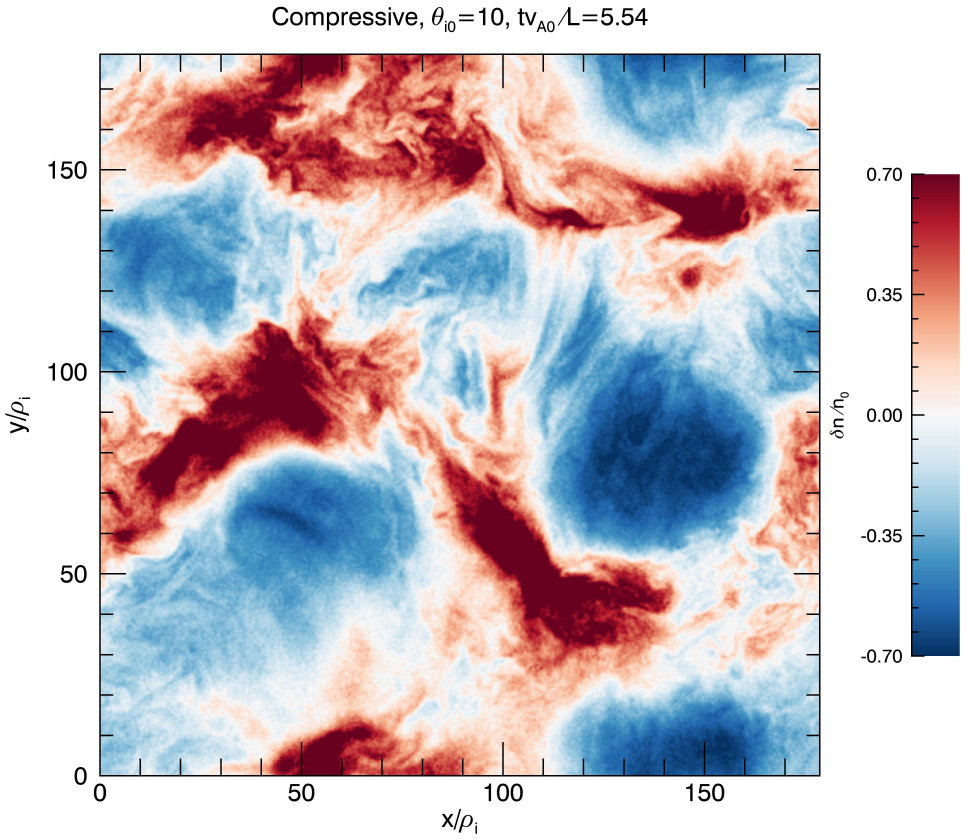}
   \centering
  \caption{\label{fig:visual_rel} Image of the fluctuations in particle number density $\delta n$ for the $\theta_{i0}=10$ simulations with solenoidal driving (top panel) and compressive driving (bottom panel) in an arbitrary $xy$ slice of the domain, at $t = 5.5 L/v_{A0}$. Note that the colorbar is rescaled between the two cases.}
 \end{figure}
 
  \begin{figure}
  \includegraphics[width=0.95\columnwidth]{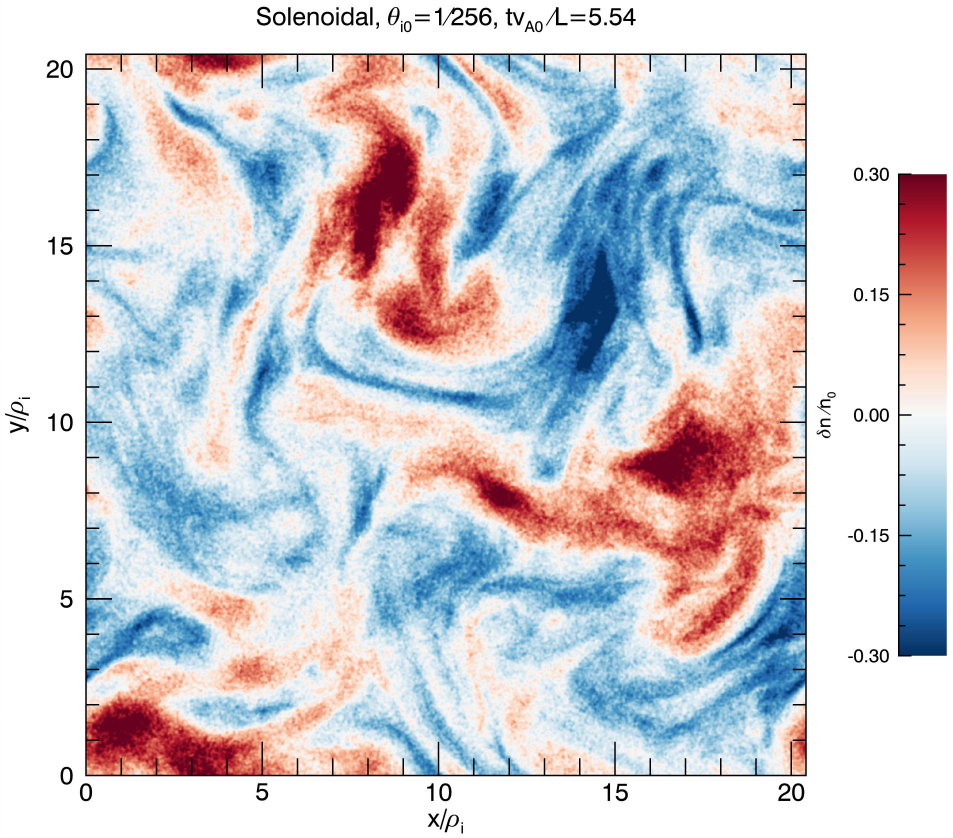}
\includegraphics[width=0.95\columnwidth]{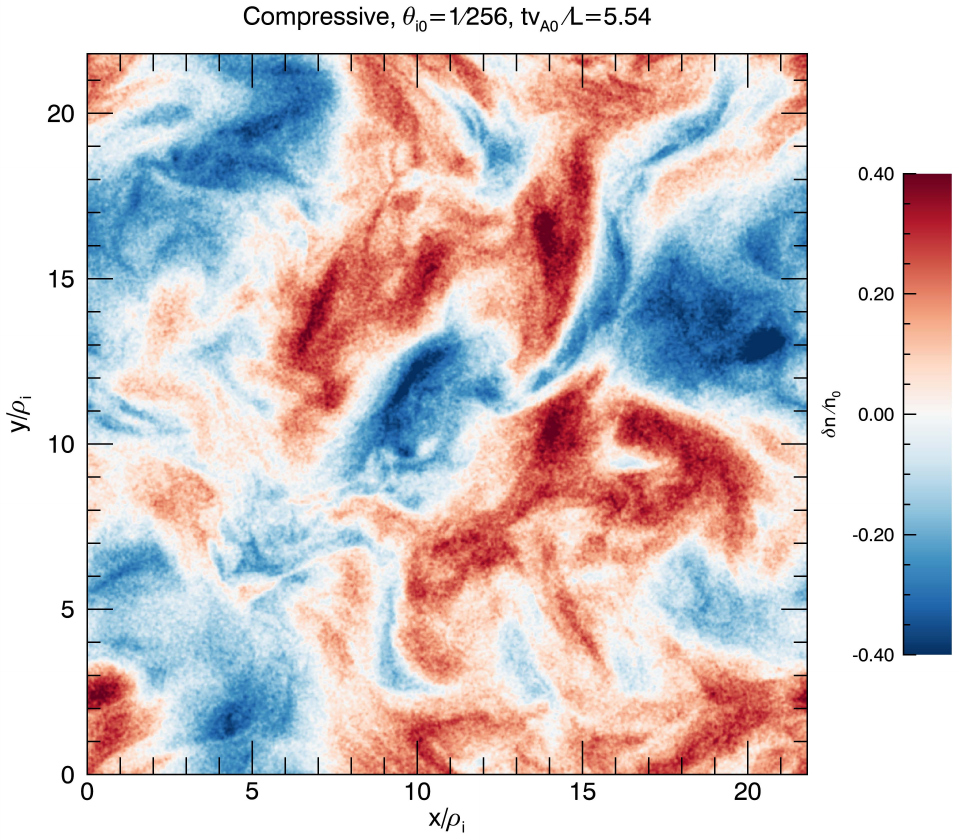}
   \centering
  \caption{\label{fig:visual} Same as Fig.~\ref{fig:visual_rel} but for the $\theta_{i0}=1/256$ cases.}
 \end{figure}
 
We focus on the particle number density, $n(\boldsymbol{x},t) \equiv n_e(\boldsymbol{x},t) + n_i(\boldsymbol{x},t)$. We construct several images of the density fluctuations, $\delta n = n - n_0$, in arbitrary $xy$ slices of the simulation domain at a select time after turbulence has fully developed, $t = 5.5 L/v_{A0}$. We first show $\delta n$ for the relativistic ($\theta_{i0} = 10$) cases, which are expected to have the largest inertial range, in Fig.~\ref{fig:visual_rel}. The solenoidal case (top panel) exhibits density structure with complex morphology across a broad range of scales, as expected from density fluctuations being passively mixed by the turbulence. The compressible case (bottom panel), on the other hand, shows large-scale density clumps and voids. We also show $\delta n$ for the semirelativistic ($\theta_{i0} = 1/256$) cases in Fig.~\ref{fig:visual}. These cases exhibit smoother structure than the $\theta_{i0} = 10$ cases with the corresponding driving, evidently due to damping of the density fluctuations by the ion-scale kinetic physics. Morphological differences between the compressive and solenoidal cases are less conspicuous for $\theta_{i0} = 1/256$, likely due to the limited macroscopic scale separation ($L/\rho_i$).

  \begin{figure}
  \includegraphics[width=0.95\columnwidth]{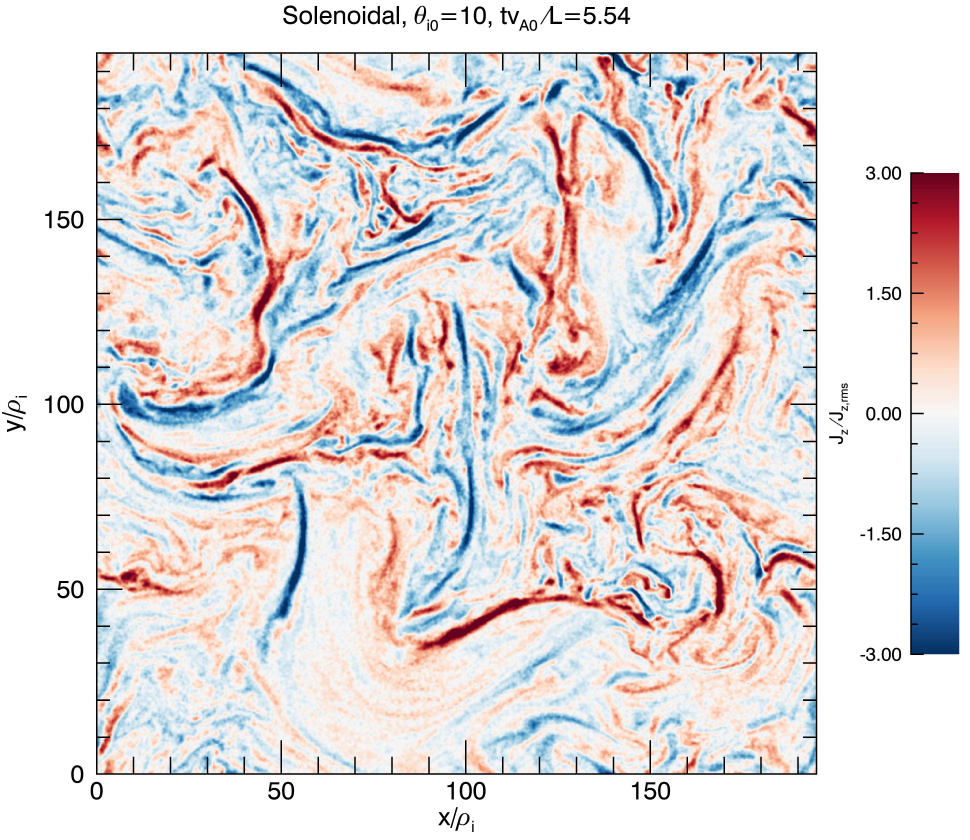}
\includegraphics[width=0.95\columnwidth]{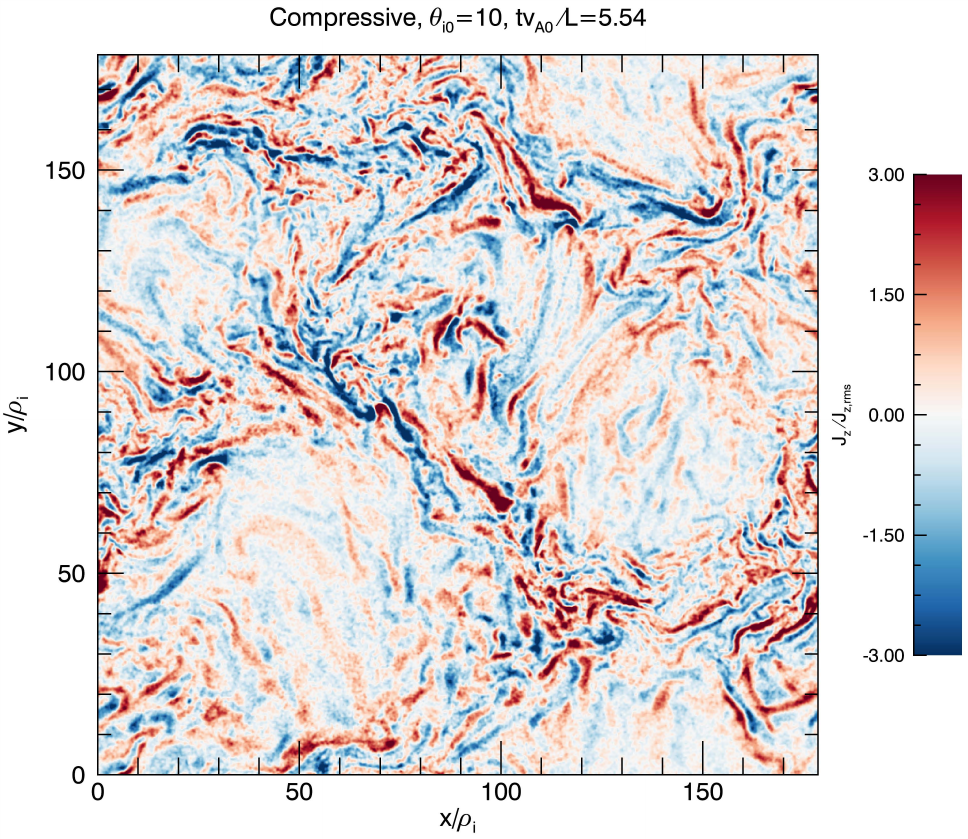}
   \centering
  \caption{\label{fig:visual_jz} Image of the out-of-plane electric current density $J_z$ for the $\theta_{i0}=10$ simulation with solenoidal driving (top panel) and compressive driving (bottom panel) in the same $xy$ slice as Fig.~\ref{fig:visual_rel}.}
 \end{figure}
 
The compressive driving influences the morphology of other quantities, in addition to density. As an example, we show the $J_z$ --- the current density component along $\boldsymbol{B}_0$ --- for the $\theta_{i0} = 10$ cases in Fig.~\ref{fig:visual_jz}. The solenoidally driven case exhibits intermittent current sheets that have thicknesses comparable to the kinetic scales, while maintaining widths and lengths at MHD scales; these structures are thus fairly coherent and may serve as localized sites of magnetic reconnection.  The $J_z$ structures in the compressively driven case, while also intermittent, have irregular shapes and are often broken up into thin filaments. These filaments tend to be clustered in regions of high particle density. This indicates that compressive fluctuations can disrupt current sheets. For the solenoidal case, it is natural to expect the maximum aspect ratio of these current sheets to increase in proportion with the inertial range of the system \citep{zhdankin_etal_2014}; they may then become unstable to the tearing instability at sufficiently large system sizes, as observed in MHD and PIC simulations of 2D turbulence \citep[e.g.,][]{dong_etal_2018, walker_etal_2018, comisso_sironi_2018}. The role of the tearing instability on the turbulence statistics is under active theoretical study for both MHD turbulence \citep{loureiro_boldyrev_2017, mallet_etal_2017} and kinetic turbulence \citep{loureiro_boldyrev_2017b, mallet_etal_2017b}. We speculate that the disruption of current sheets by compressive fluctuations may prevent tearing instability from occurring in strongly compressive turbulence. Such a disruption of current sheets by compressive fluctuations may affect the properties of magnetic reconnection in the system and thus the particle energization and/or injection mechanisms. We return to this issue in Section~\ref{sec:entrans}.

 \subsection{Density fluctuations}
 
  \begin{figure}
  \includegraphics[width=0.95\columnwidth]{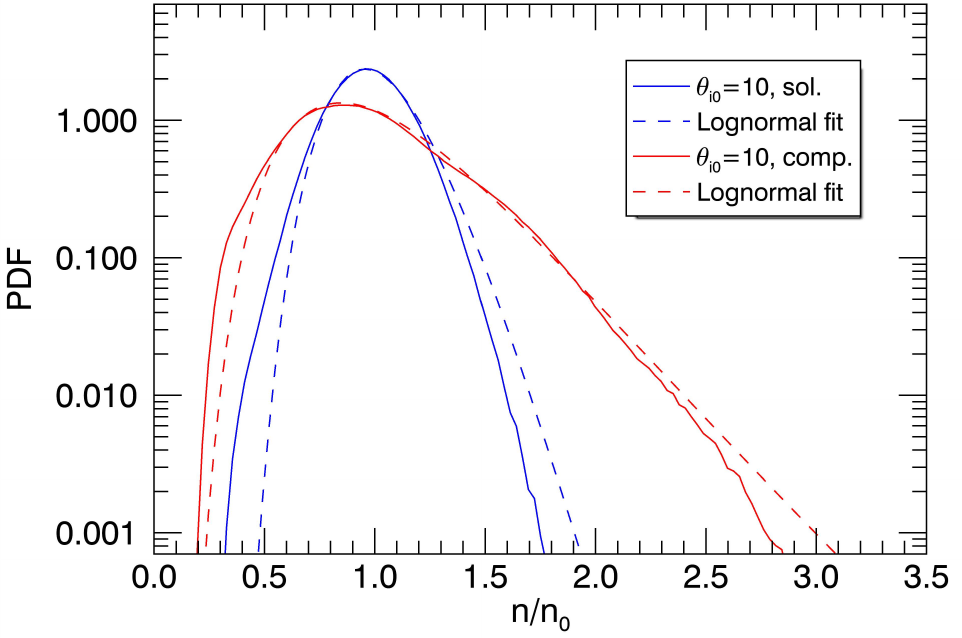}
\includegraphics[width=0.95\columnwidth]{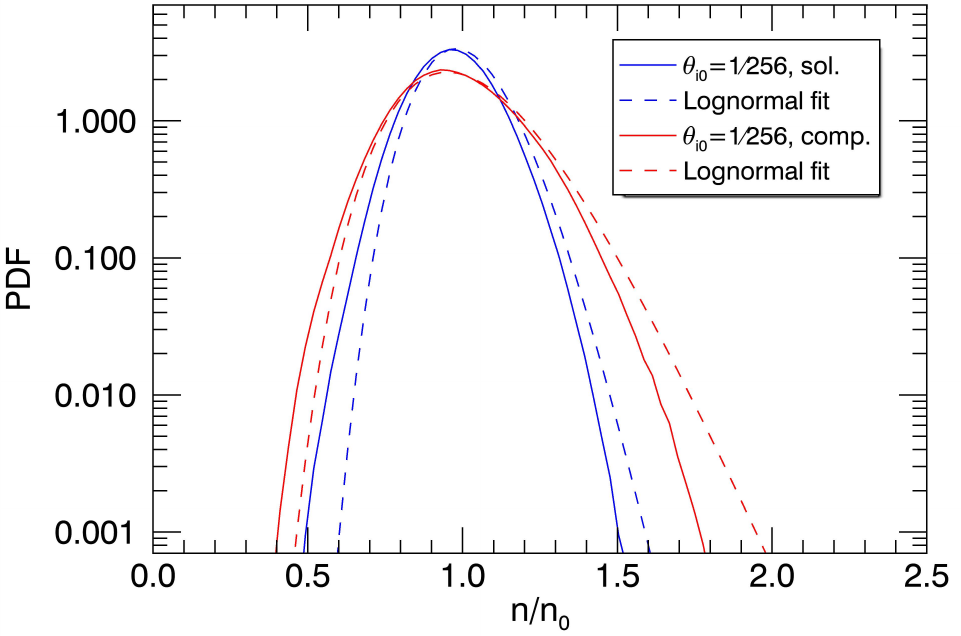}
   \centering
  \caption{\label{fig:npdf} The PDF for particle number density $n$. Solenoidal (blue) and compressive (red) cases are shown for $\theta_{i0} = 10$ (top panel) and $\theta_{i0} = 1/256$ (bottom panel); log-normal fits are also shown for each case (dashed lines).}
 \end{figure}
 
Since strong density fluctuations are the primary manifestation of compressive driving, in this subsection, we describe the particle density distributions in our simulations.
 
 We first compare the probability density function (PDF) for $n/n_0$ from the four fiducial cases in Fig.~\ref{fig:npdf}. These PDFs are averaged from $t v_{A0}/L = 4.7$ to $t v_{A0}/L = 6.5$. We find that the PDFs can be approximately fit with a log-normal distribution for all cases, which is typically expected for a stochastically mixed isothermal gas \citep[e.g.,][]{passot_etal_1998, hopkins_2013}. The compressive cases have a broader distribution than the solenoidal cases at a given value of $\theta_{i0}$. In addition, the $\theta_{i0} = 10$ cases (top panel) have a broader distribution than $\theta_{i0} = 1/256$ (bottom panel) with the same driving; this is consistent with compressive fluctuations being more strongly coupled to Alfv\'{e}nic fluctuations in the relativistic regime \citep{takamoto_lazarian_2016, takamoto_lazarian_2017}. Specifically, we find that the time-averaged rms fluctutations $\delta n_{\rm rms}/n_0$ are $0.18$ for the $\theta_{i0} = 10$ solenoidal case, $0.34$ for the $\theta_{i0} = 10$ compressive case, $0.12$ for the $\theta_{i0} = 1/256$ solenoidal case, and $0.18$ for the $\theta_{i0} = 1/256$ compressive case.
 
  \subsection{Equation of State}
 
In this subsection, we briefly comment on the equation of state observed in our PIC simulations. This is motivated by the fact that theoretical and numerical studies of compressible MHD (as well as hybrid kinetic models) require the user to specify an equation of state as an input. A typical closure is to assume that the plasma acts as an ideal gas with isotropic pressure, such that the pressure scales as a power law with density, $P \propto n^{\kappa}$ where $\kappa$ is the adiabatic index. Special cases are $\kappa = 1$ for an isothermal gas, $\kappa = 4/3$ for a relativistic monatomic gas, and $\kappa = 5/3$ for a non-relativistic monatomic gas. More generally, weakly collisional plasmas may be modeled using an anisotropic pressure tensor with respect to the magnetic field \citep{cgl_1956, gedalin_1991}.
 
For a collisionless plasma, it is not {\it a priori} guaranteed that the fluid equations can be rigorously closed by an equation of state. In PIC simulations, the thermodynamics of the plasma is described self-consistently by the Vlasov-Maxwell equations. An important question is: can the plasma be described empirically by a simple equation of state? To address this question, we directly measure the equation of state in our compressively driven simulations (solenoidal cases exhibit similar scalings, but over a narrower range of $n$). 
 
     \begin{figure}
  \includegraphics[width=0.95\columnwidth]{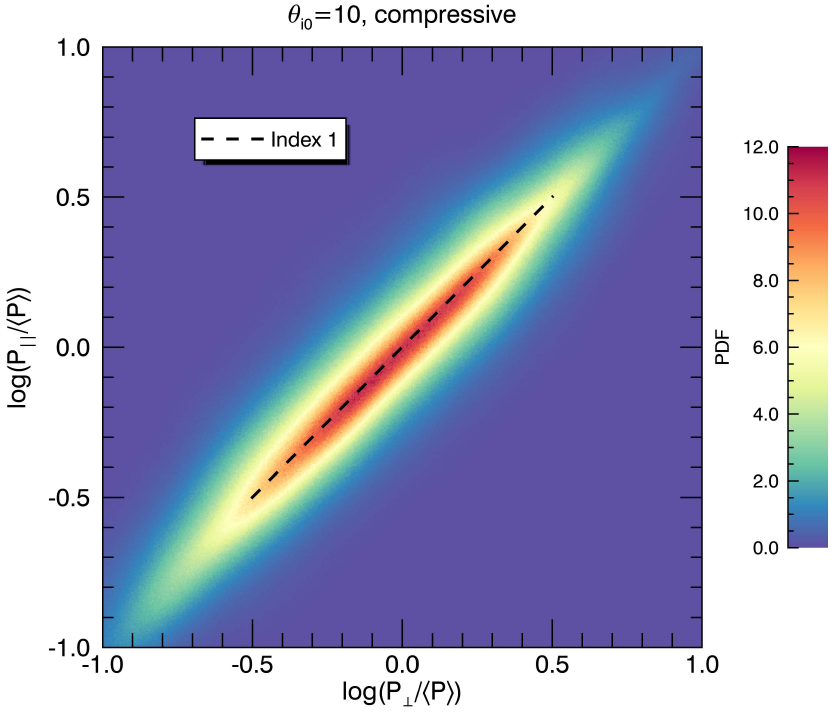}
\includegraphics[width=0.95\columnwidth]{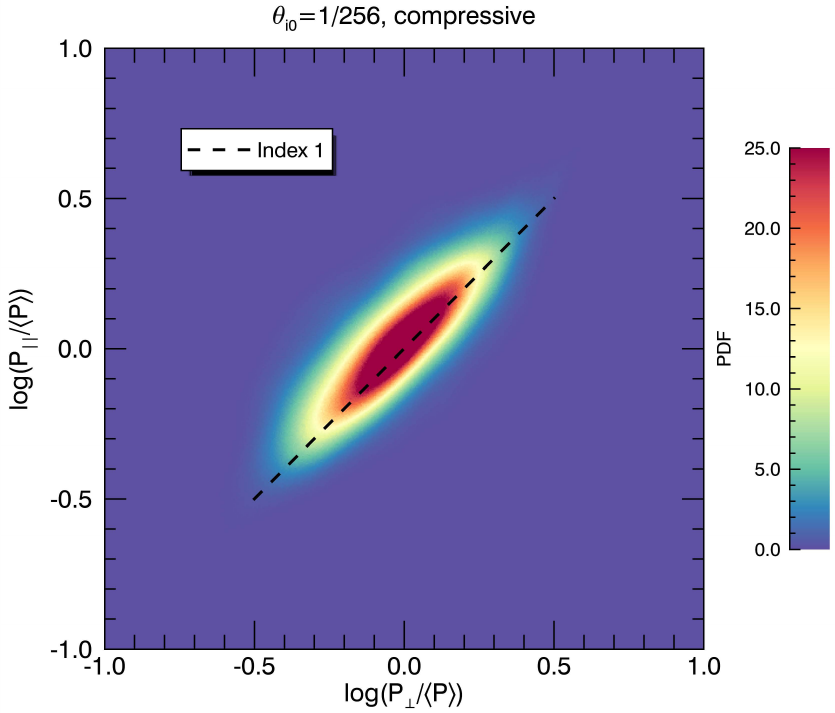}
   \centering
  \caption{\label{fig:pressure} 2D PDF of the pressure component parallel to the magnetic field, $P_{\parallel}$, versus perpendicular component, $P_\perp$, for the compressively driven simulation with $\theta_{i0} = 10$ (top panel) and $\theta_{i0} = 1/256$ (bottom panel).}
 \end{figure}
 
 We first consider the pressure anisotropy by measuring components parallel and perpendicular to the local magnetic field $\boldsymbol{B}(\boldsymbol{x},t)$, defined by
\begin{align}
P_{{\parallel},s} &= \hat{\boldsymbol{B}} \hat{\boldsymbol{B}} : {\bf P}_s \, , \nonumber \\
P_{{\perp},s} &= \frac{1}{2} \left( {\bf I} - \hat{\boldsymbol{B}} \hat{\boldsymbol{B}} \right) : {\bf P}_s \, ,
\end{align}
where ${\bf P}_s(\boldsymbol{x},t) = \int d^3p f_s(\boldsymbol{x},\boldsymbol{p},t) \boldsymbol{p} \boldsymbol{p} c/\sqrt{m_s^2 c^2 + p^2}$ is the pressure tensor for species $s$ (with ram pressure terms from bulk flows, which are sub-dominant, included) and $\hat{\boldsymbol{B}} = \boldsymbol{B}/B$ is the magnetic field direction. We show $P_{\parallel}$ versus $P_{\perp}$ (combined for both species), sampled throughout the simulation domain at a given time of $t v_{A0}/L = 5.5$, for the compressive $\theta_{i0}=10$ and $\theta_{i0}=1/256$ cases in Fig.~\ref{fig:pressure}. On average, $P_{\parallel} \sim P_{\perp}$, indicating that the pressure is approximately isotropic. The $\theta_{i0} = 1/256$ has a moderately broader statistical spread, suggesting stronger deviations from isotropy at any given location.

\begin{figure}
 \includegraphics[width=0.95\columnwidth]{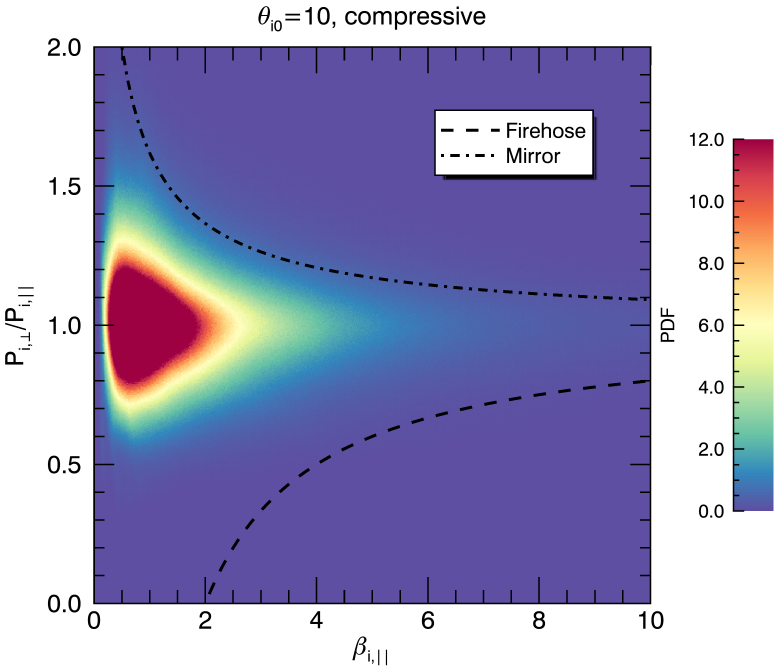}
   \centering
  \caption{\label{fig:pressureanisotropy1} 2D PDF of the local ion pressure anisotropy $P_{i,\perp}/P_{i,\parallel}$, versus parallel plasma beta, $\beta_{i,\parallel}$, for the compressively driven simulation with $\theta_{i0} = 10$. The non-relativistic firehose (dashed) and mirror (dash-dotted) instability thresholds are also shown.}
 \end{figure}
 
 \begin{figure}
 \includegraphics[width=0.95\columnwidth]{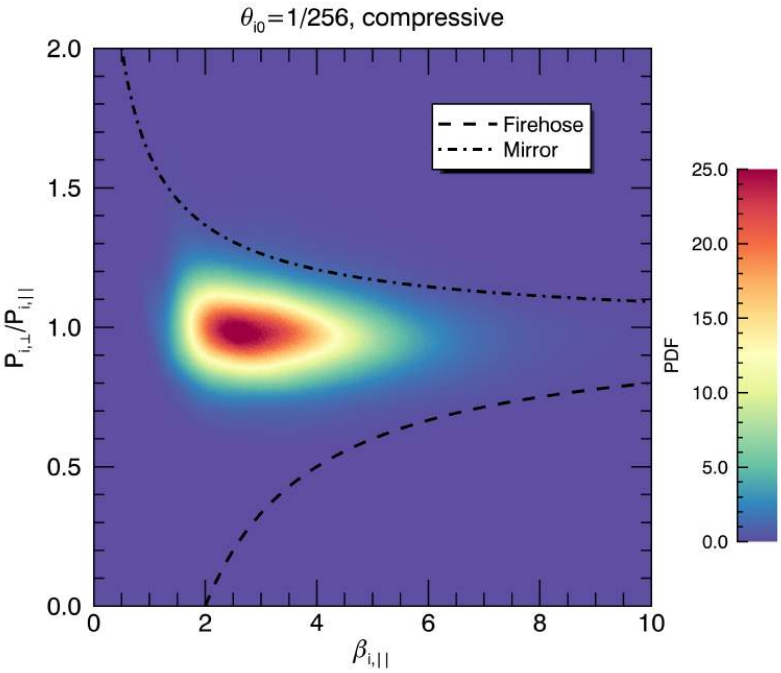}
  \includegraphics[width=0.95\columnwidth]{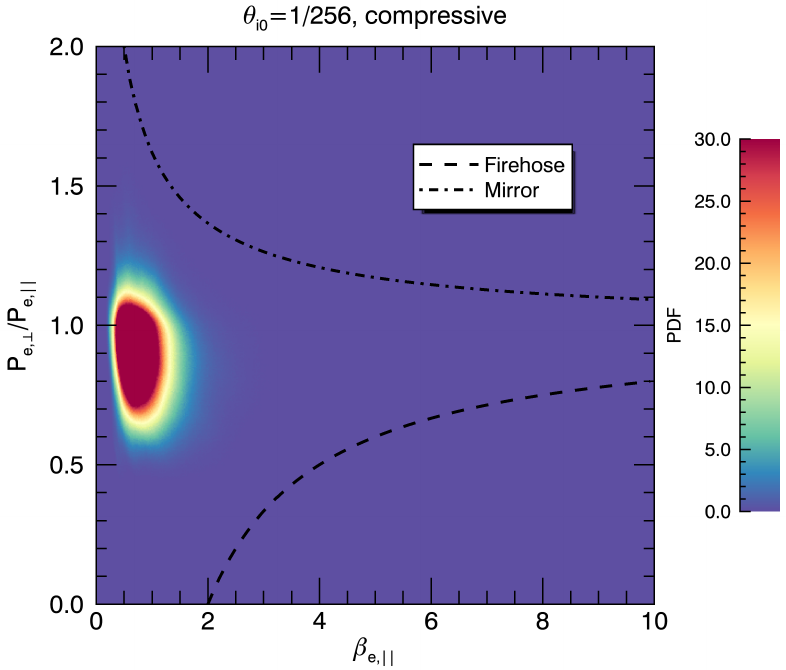}
   \centering
  \caption{\label{fig:pressureanisotropy2} 2D PDF of the local pressure anisotropy $P_{s,\perp}/P_{s,\parallel}$, versus parallel plasma beta, $\beta_{s,\parallel}$, for ions (top panel) and electrons (bottom panel) in the compressively driven simulation with $\theta_{i0} = 1/256$. The non-relativistic firehose (dashed) and mirror (dash-dotted) instability thresholds are also shown.}
 \end{figure}

Pressure anisotropy may influence the system dynamics and energetics if it becomes sufficiently strong to trigger the kinetic firehose, mirror, or ion-cyclotron instabilities \citep[see, e.g.,][]{kunz_etal_2014}. For example, the ion-cyclotron instability has been previously suggested as a mechanism of nonthermal particle acceleration \citep{ley_etal_2019}. To judge whether or not these instabilities occur in our simulations, we consider the local pressure anisotropy $P_{s,\perp}/P_{s,\parallel}$ versus plasma beta calculated with parallel pressure component, $\beta_{s,\parallel} = 8\pi P_{s,\parallel}/B^2$ for each particle species. In a non-relativistic plasma, the firehose instability occurs when,
\begin{align}
\frac{P_\perp}{P_{\parallel}} \lesssim \frac{P_\perp}{P_{\parallel}} \Big|_{\rm Firehose} \equiv 1 - \frac{2}{\beta_{\parallel}} \, , \label{eq:firehose}
\end{align}
while the mirror instability occurs when
\begin{align}
\frac{P_\perp}{P_{\parallel}} \gtrsim \frac{P_\perp}{P_{\parallel}} \Big|_{\rm Mirror} &\equiv \frac{1}{2} \left( 1 + \sqrt{1 + \frac{4}{\beta_{\parallel}}} \right) \, . \label{eq:mirror}
\end{align}
For a discussion of relativistic corrections, see \cite{chou_hau_2004}. We show a 2D PDF of $P_{i,\perp}/P_{i,\parallel}$ versus $\beta_{i,\parallel}$ for the $\theta_{i0} = 10$ case in Fig.~\ref{fig:pressureanisotropy1}; the result is similar for electrons in this simulation. We find that the bulk of the plasma remains far from the instability thresholds given by Eq.~\ref{eq:firehose} and Eq.~\ref{eq:mirror}, indicating that the instabilities should play a minimal role at these plasma parameters.

Similarly, we show $P_{s,\perp}/P_{s,\parallel}$ versus $\beta_{s,\parallel}$ in the $\theta_{i0} = 1/256$ case for ions and electrons pressures separately in the $\theta_{i0} = 1/256$ case in Fig.~\ref{fig:pressureanisotropy2}. The ions lie closer to the instability thresholds than the electrons, due to the fact that ions absorb more of the turbulent energy than electrons (as later discussed in Sec.~\ref{eiheating}), which causes $\beta_{i,\parallel}$ to reach large values. However, the bulk of the plasma still remains far from the thresholds.
 
     \begin{figure}
       \includegraphics[width=0.95\columnwidth]{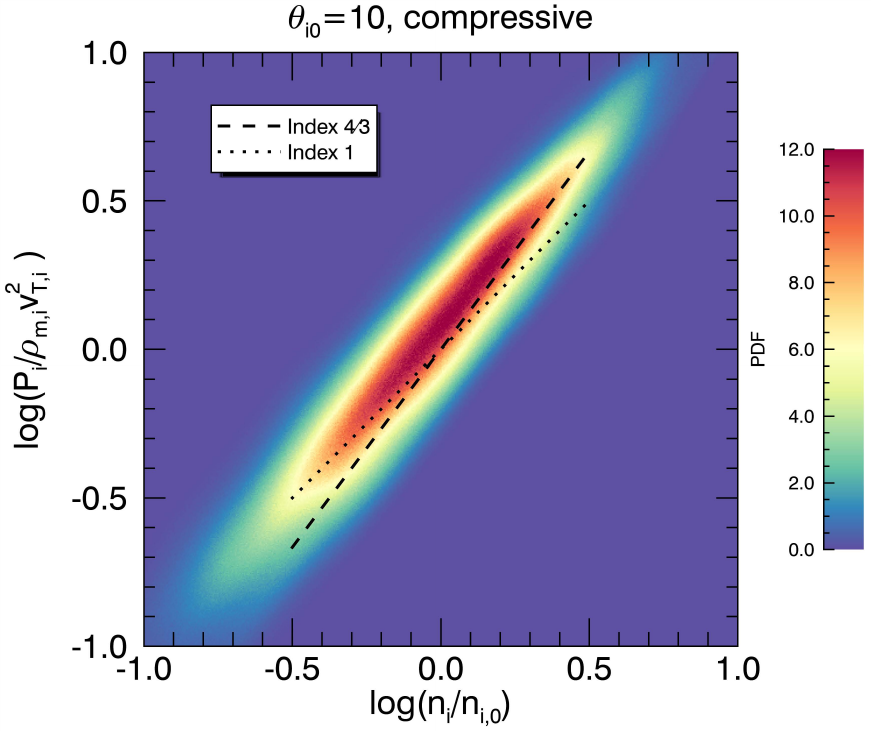}
   \centering
  \caption{\label{fig:eos_thi10} 2D PDF of the local ion pressure $P_i$ versus density $n_i$ for the $\theta_{i0}=10$ simulation with compressive driving. Power-law scalings with an index of $1$ (dotted) and $4/3$ (dashed) are also shown. Pressure is normalized with respect to the relativistic mass density times ion sound speed squared, $\rho_{m,i} v_{T,i}^2$.}
 \end{figure}

Having established that pressure is approximately isotropic, we next measure the equation of state for each species, $P_s(n_s)$, using the isotropic part of the pressure ($P_s = {\rm Tr }{\bf P}_s/3$). We first show a 2D PDF of $P_i$ versus $n_i$ for the $\theta_{i0} = 10$ simulation, in Fig.~\ref{fig:eos_thi10}; the result is nearly identical for electrons (not shown) due to relativistic mass symmetry ($\theta_{i0} \gg 1$). Intriguingly, we find that the scaling is wedged between the isothermal case (linear scaling) and relativistic ideal gas case ($4/3$ power law).

Using the isothermal fit, we could infer the thermal velocity from the expression $P_s \sim v_{T,s}^2 \rho_{m,s} n/n_{s,0}$, where $\rho_{m,s} = m_s n_{s,0} + E_{{\rm int},s}/L^3$ is the average relativistic mass density and $v_{T,s}$ is the thermal speed for species $s$. We thus only account for the spatial variation in mass density due to particle number, not internal energy (using the latter gives similar results). To test whether the results are consistent with the classical ultra-relativistic sound speed, $v_{T,s} = c/3^{1/2}$, in Fig.~\ref{fig:eos_thi10} we normalized pressure to $v_{T,s}^2 \rho_{m,s}$ with $v_{T,s} = c/3^{1/2}$ and density to $n_{i,0}$.  The fact that the data then goes through the origin (in logarithmic coordinates) indicates that the simulations are described well by an isothermal model with the classical sound speed.
 
    \begin{figure}
  \includegraphics[width=0.95\columnwidth]{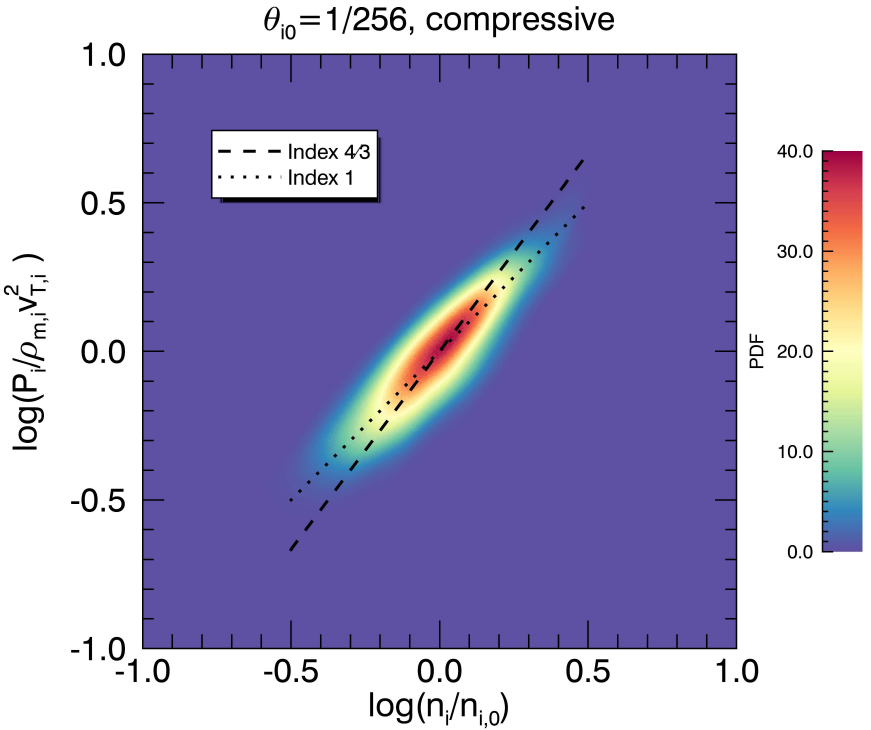}
\includegraphics[width=0.95\columnwidth]{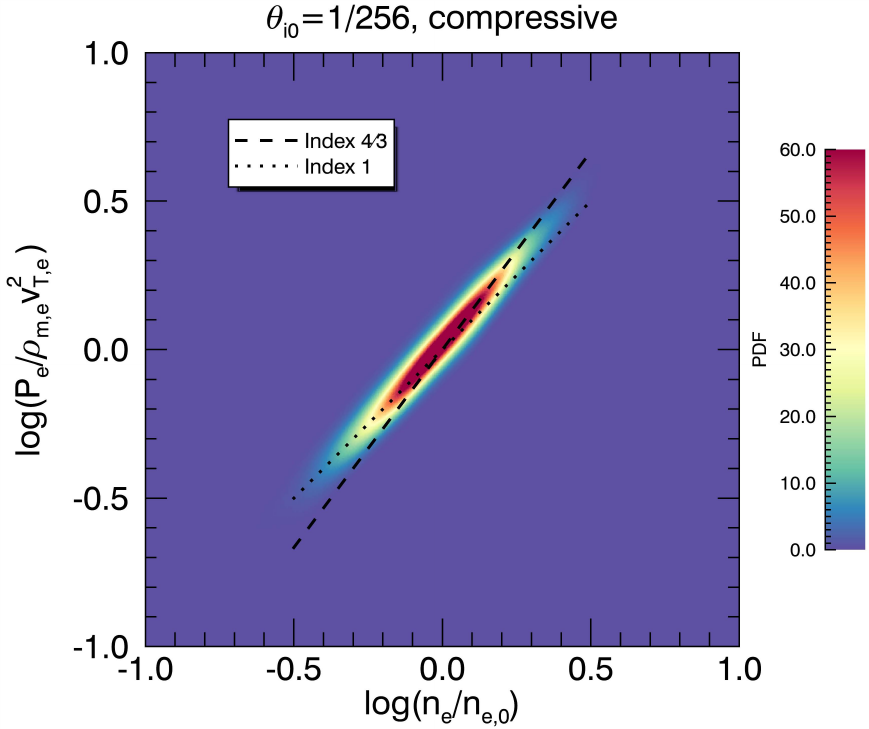}
   \centering
  \caption{\label{fig:eos} 2D PDF of the local pressure $P_s$ versus density $n_s$ for ions (top panel) and electrons (bottom panel) in the $\theta_{i0}=1/256$ simulation with compressive driving. Power-law scalings with an index of $1$ (dotted) and $4/3$ (dashed) are also shown. Pressure is normalized with respect to the corresponding relativistic mass density times sound speed squared, $\rho_{m,s} v_{T,s}^2$, where $v_{T,e} = c/3^{1/2}$ and $v_{T,i} = (T_{{\rm eff},i}/m_i)^{1/2}$.}
 \end{figure}
 
For the semirelativistic case ($\theta_{i0} = 1/256$), $P_s(n_s)$ is similarly between the isothermal and relativistic gas scalings, for both electrons and ions separately, as shown in Fig.~\ref{fig:eos}. This result is contrary to the naive expectation that ions would be described by a non-relativistic ideal gas (with power-law index $\kappa = 5/3$). The scaling coefficients are consistent with unity when normalizing pressure in Fig.~\ref{fig:eos} with respect to $\rho_m v_{T,s}^2$ using $v_{T,e} = c/3^{1/2}$ and $v_{T,i} = (T_{{\rm eff},i}/m_i)^{1/2}$ where $T_{{\rm eff},i} = (2/3) E_{{\rm int},i}/L^3 n_{i,0}$ is the effective non-relativistic temperature based on ion internal energy. This confirms that the semirelativistic equations of state are consistent with an ultra-relativistic electron thermal velocity and a non-relativistic ion thermal velocity.
 
 The equation of state may be skewed toward an isothermal one due to efficient collisionless heat transfer throughout the plasma, on the timescale of turbulent fluctuations. Characterizing non-ideal corrections to the equation of state (including effects of pressure anisotropy and nonthermal particles) is left to future work. In summary, the existence of simple near-isothermal equations of state validate the use of MHD phenomenology for modeling the fluid dynamics of our system.

 \subsection{Turbulence spectra}

We next consider the power spectra of turbulent fluctuations. When describing these spectra, we focus on the scaling with respect to the perpendicular wavenumber $k_\perp = (k_x^2 + k_y^2)^{1/2}$, to account for the fact that the turbulence may be anisotropic with respect to $\boldsymbol{B}_0$. The spectra are thus reduced by integrating over $k_z$ and over directions of $\boldsymbol{k}$ in the perpendicular plane. 
 
    \begin{figure}
  \includegraphics[width=0.95\columnwidth]{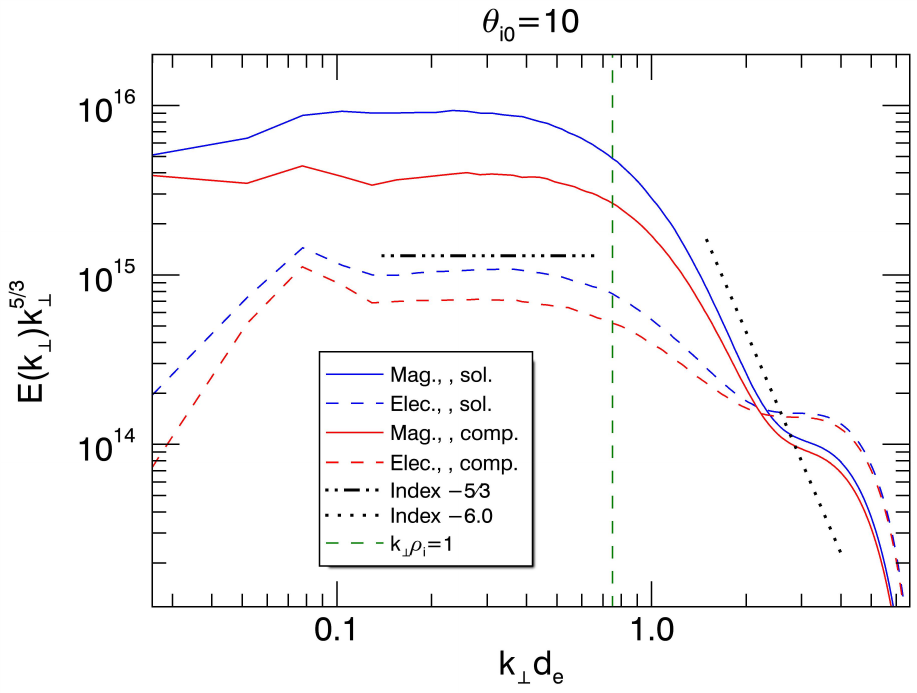}
\includegraphics[width=0.95\columnwidth]{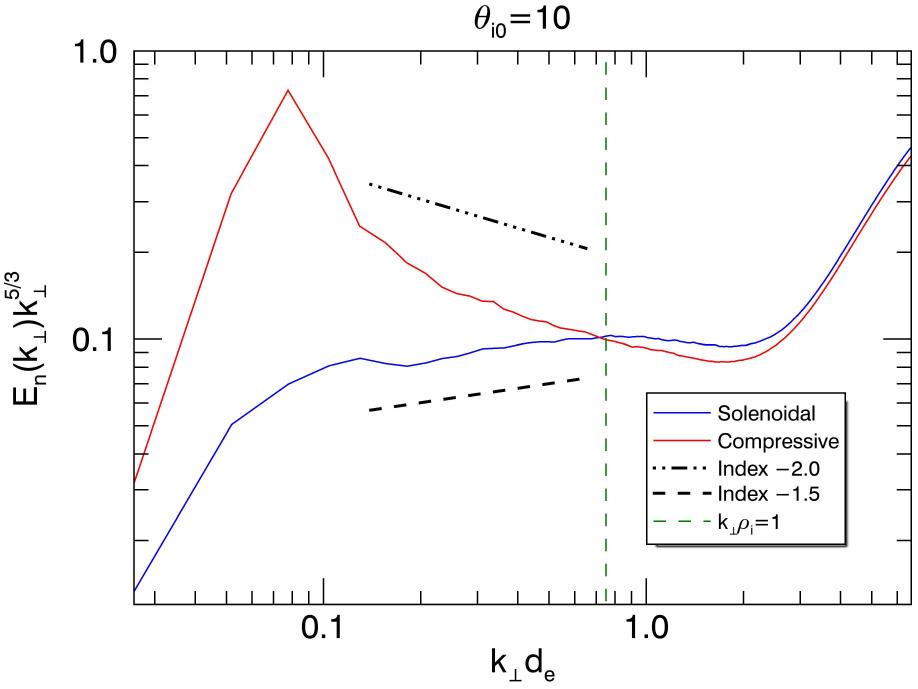}
   \centering
  \caption{\label{fig:spec10} Top panel: Compensated magnetic (solid line) and electric (dashed line) energy spectra for solenoidal (blue) and compressive (red) simulations in the relativistic regime, $\theta_{i0} = 10$. Power laws with pre-compensated indices of $-5/3$ (dash-dotted) and $-6$ (dotted) are shown; the characteristic electron and ion gyroradii are indicated (green vertical line). Bottom panel: Similar compensated power spectrum but for fluctuations in the particle number density $n$, with power laws of index $-2$ (dash-dotted) and $-1.5$ (dashed) indicated.}
 \end{figure}

We first describe spectra for the relativistic ($\theta_{i0}=10$) cases, which have the largest $L/\rho_{e0}$ and thus the longest MHD inertial range. The top panel of Fig.~\ref{fig:spec10} shows the magnetic and electric energy spectra for both solenoidal and compressive driving, averaged from $t v_{A0}/L = 4.8$ to $t v_{A0}/L = 6.7$. For clarity, we have compensated by $k_\perp^{5/3}$, since the classical MHD turbulence theories predict a $k_\perp^{-5/3}$ scaling in the inertial range \citep{goldreich_sridhar_1995}. Although the plasma is relativistically hot, we expect non-relativistic phenomenology for the turbulence to be applicable since $\sigma \lesssim 1$, so bulk motions are essentially sub-relativistic. We find that the magnetic and electric energy spectra are both consistent with a $k_\perp^{-5/3}$ power law at scales $k_\perp d_e \approx 0.1$ to $k_\perp d_e \approx 0.4$. At scales $k_\perp d_e \gtrsim 1$ the spectra steepen; there is not a clear power law in this range, but we show a $k_\perp^{-6}$ power law for comparison. Overall, the spectra are remarkably similar for both solenoidal and compressive driving. The primary difference is that the solenoidal case has a higher amplitude at all wavenumbers, consistent with the larger overall magnetic energy described in Sec.~\ref{subsec:evo}. Another difference is that the magnetic energy spectrum is somewhat steeper for the compressive case in the driving range, $k_\perp d_e \lesssim 0.1$, indicating possible damping near the driving scale.

In contrast, the spectrum of particle density fluctuations is drastically different between the compressive and solenoidal cases, as shown in the bottom panel of Fig.~\ref{fig:spec10}. The compressive case exhibits a strong spike in power near the the driving scale, $k_\perp d_e \sim 0.1$, confirming that density fluctuations are robustly generated by the external force. The density spectrum declines steeply in the inertial range, possibly approaching a $k_\perp^{-2}$ scaling. By contrast, the solenoidal case has a scaling close to $k_\perp^{-5/3}$ throughout the inertial range, consistent with MHD phenomenology based on passive mixing of slow modes and entropy modes with Alfv\'{e}nic turbulence \citep{lithwick_goldreich_2001}. Note, however, that $k_\perp^{-1.5}$ provides a slightly better fit, which is incidentally the spectrum expected from a fast-mode cascade \citep{cho_lazarian_2002, cho_lazarian_2003}; this could suggest that an undamped fast-mode cascade is intertwined with the Alfv\'{e}nic cascade due to relativistic mode conversion \citep{takamoto_lazarian_2016, takamoto_lazarian_2017}. The compensated density spectra also show a spike at $k_\perp d_e \gtrsim 2$, but this is due to the noise floor (from finite number of particles per cell). 

   \begin{figure}
  \includegraphics[width=0.95\columnwidth]{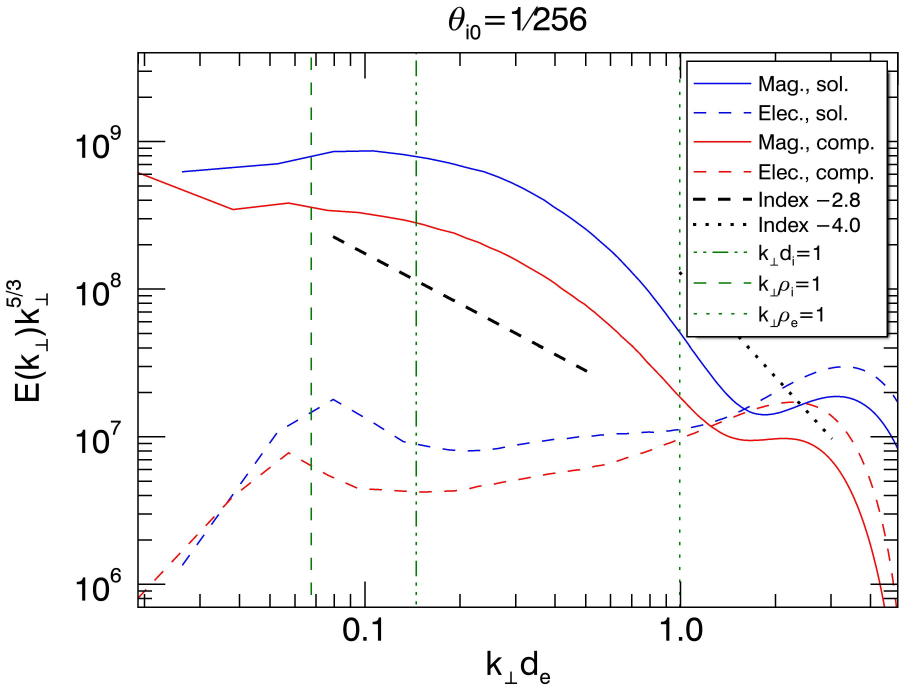}
\includegraphics[width=0.95\columnwidth]{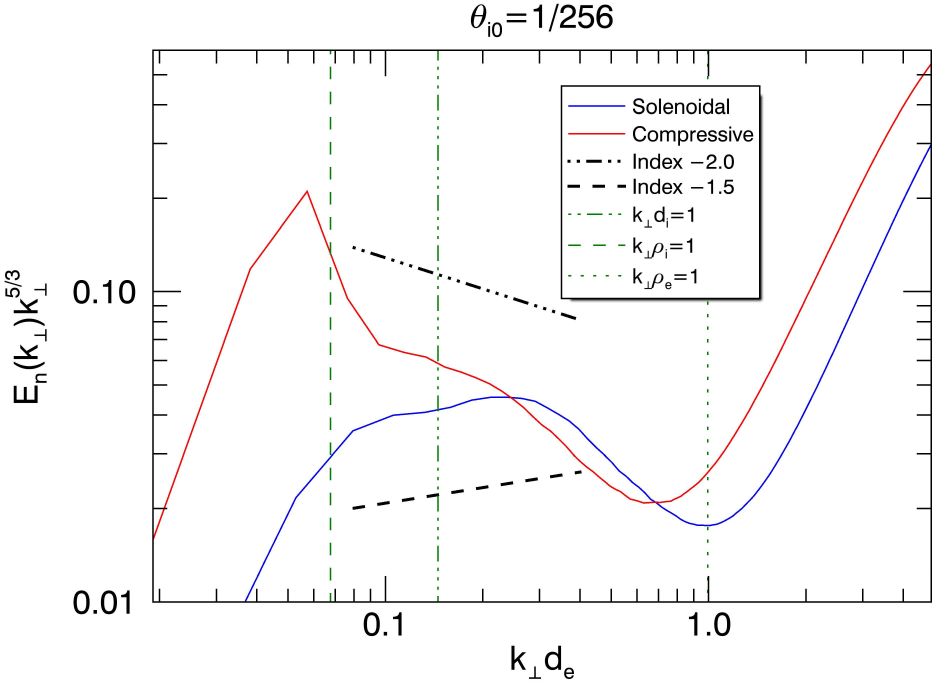}
   \centering
  \caption{\label{fig:spec1d256} Similar as Fig.~\ref{fig:spec10} except for the semirelativistic regime, $\theta_{i0} = 1/256$. In this case, the top panel shows power laws with index $-2.8$ (dashed) and $-4.0$ (dotted).}
 \end{figure}

We next describe spectra for the semirelativistic ($\theta_{i0} = 1/256$) cases, for which $L/\rho_{e0}$ is so small that the MHD inertial range is practically absent; instead, the dynamical range of the simulation is used primarily to resolve the kinetic range of scales between $\rho_e$ and $\rho_i$. We show the magnetic and electric energy spectra for the fiducial $\theta_{i0} = 1/256$ cases, averaged from $t v_{A0}/L = 4.7$ to $t v_{A0}/L = 6.5$, in the top panel of Fig.~\ref{fig:spec1d256}. The magnetic energy spectra drop off more quickly than $k_\perp^{-5/3}$ at scales $k_\perp \rho_i > 1$, while the electric energy spectra stay close to $k_\perp^{-5/3}$. Although this kinetic range of turbulence is not understood well theoretically even in the non-relativistic regime, one typically expects a cascade of kinetic Alfv\'{e}n waves or whistler waves; solar wind measurements indicate a corresponding power-law index of $-2.8$ or so \citep[e.g.,][]{alexandrova_etal_2009, sahraoui_etal_2009}, which we overlay for reference. In our simulations, the spectrum is not a power law in this range, which we attribute to the limited scale separation. Whereas $\rho_i/\rho_e \sim (m_i/m_e)^2 \approx 43$ in a non-relativistic electron-proton plasma with $T_i/T_e = 1$, it is barely a factor of 10 in our case due to the relativistic effects increasing $\rho_e$. Furthermore, it is likely that the MHD and kinetic ranges both need to be simulatenously fully resolved to produce an asymptotic power-law spectrum in the kinetic range. The magnetic energy spectrum in the $\theta_{i0} = 1/256$ cases is qualitatively similar for both driving mechanisms, although it does appear to be slightly steeper for the compressive case, indicating stronger damping. Overall, we conclude that the kinetic cascade at $k \rho_i \gtrsim 1$ is mostly insensitive to compressive fluctuations driven at larger scales. This conclusion is consistent with previous works that used hybrid kinetic simulations to study the effect of varying driving mechanisms on two-dimensional turbulence at $k \rho_i \gtrsim 1$ in the non-relativistic regime \citep{cerri_etal_2017b}.

As shown in the bottom panel of Fig.~\ref{fig:spec1d256}, the density spectra for the $\theta_{i0} = 1/256$ cases are qualitatively similar to the $\theta_{i0} = 10$ cases at large scales ($k_\perp \rho_i \sim 1$). In particular, the density spectrum for the $\theta_{i0} = 1/256$ compressive case shows a strong spike at driving scales, and declines very quickly (more strongly than a power law) at $k_\perp \rho_i \gtrsim 1$. The density spectrum for the solenoidal case does not show any spike, and is initially close to a $k_\perp^{-5/3}$ power law (once again, better fit by $k_\perp^{-1.5}$). Curiously, the density spectra appear to have a spectral break at $k_\perp d_e \sim 0.3$ for both cases, indicating a possible regime transition at intermediate scales in the kinetic range. Inferring the asymptotic scaling of the spectra would require larger simulations.

In summary, the spectra of turbulent fluctuations indicate that while strong density fluctuations are produced at large scales by the compressive driving, this has a minimal affect on the magnetic and electric energy spectra at smaller scales (throughout the inertial and kinetic range). This suggests that compressive fluctuations are localized at large (driving) scales, while the cascade is a predominantly incompressible one.

\subsection{Structure functions}

To conclude our discussion of turbulence statistics, we briefly describe the two-point structure functions of the magnetic field, as a means to characterize the scale-dependent anisotropy of the turbulence at large scales. Note that multi-point structure functions are necessary to accurately measure the scalings of spectra steeper than $k^{-3}$ \citep[e.g.,][]{cerri_etal_2019}; thus, we focus on the relativistic ($\theta_{i0} = 10$) cases, where the inertial-range magnetic energy spectrum is much shallower than $k_\perp^{-3}$.  In this work, we limit our analysis to the second-order structure function, defined as
 \begin{align}
 S_2(\delta\boldsymbol{x},t) = \langle |\boldsymbol{B}(\boldsymbol{x}+\delta\boldsymbol{x},t)-\boldsymbol{B}(\boldsymbol{x},t)|^2 \rangle_{\boldsymbol{x}} \, ,
 \end{align}
 where the average is performed over all points $\boldsymbol{x}$ in the domain. The structure function can be expressed in a coordinate system relative to the local magnetic field, by defining 
  \begin{align}
 \delta x_\parallel &= \boldsymbol{\delta x} \cdot \hat{\boldsymbol{B}}_{\rm loc} \, , \nonumber \\
 \delta x_\perp &= [(\delta x)^2 - (\delta x_\parallel)^2]^{1/2} \, ,
 \end{align}
 where $\boldsymbol{B}_{\rm loc}(\boldsymbol{x},\delta \boldsymbol{x},t) = [\boldsymbol{B}(\boldsymbol{x}+\delta\boldsymbol{x},t) + \boldsymbol{B}(\boldsymbol{x},t)]/2$ is the local magnetic field \citep[see][]{cho_vishniac_2000}. 

   \begin{figure}
  \includegraphics[width=0.95\columnwidth]{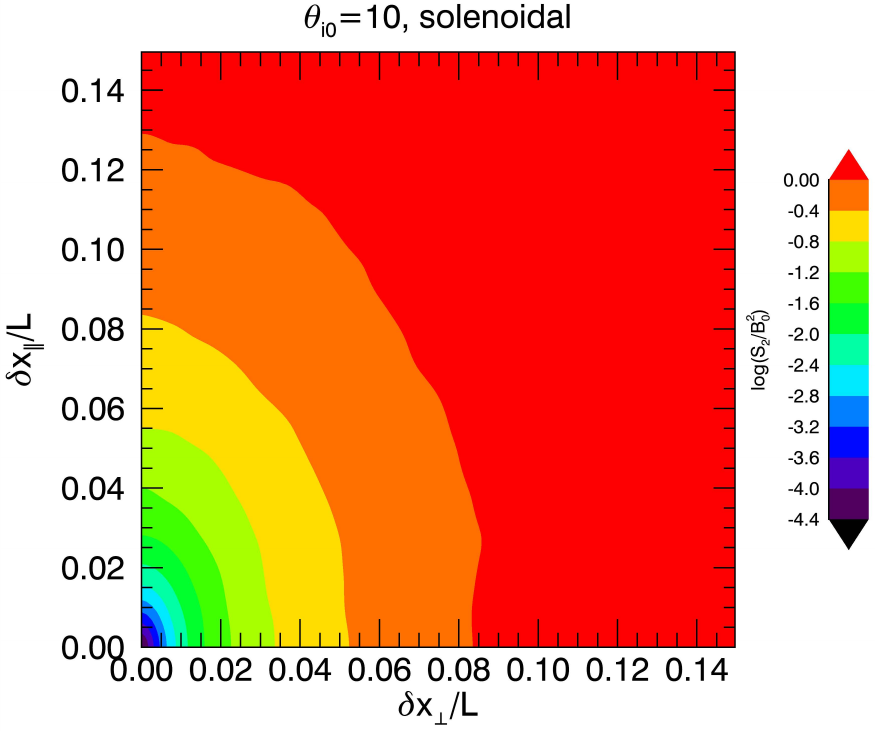}
\includegraphics[width=0.95\columnwidth]{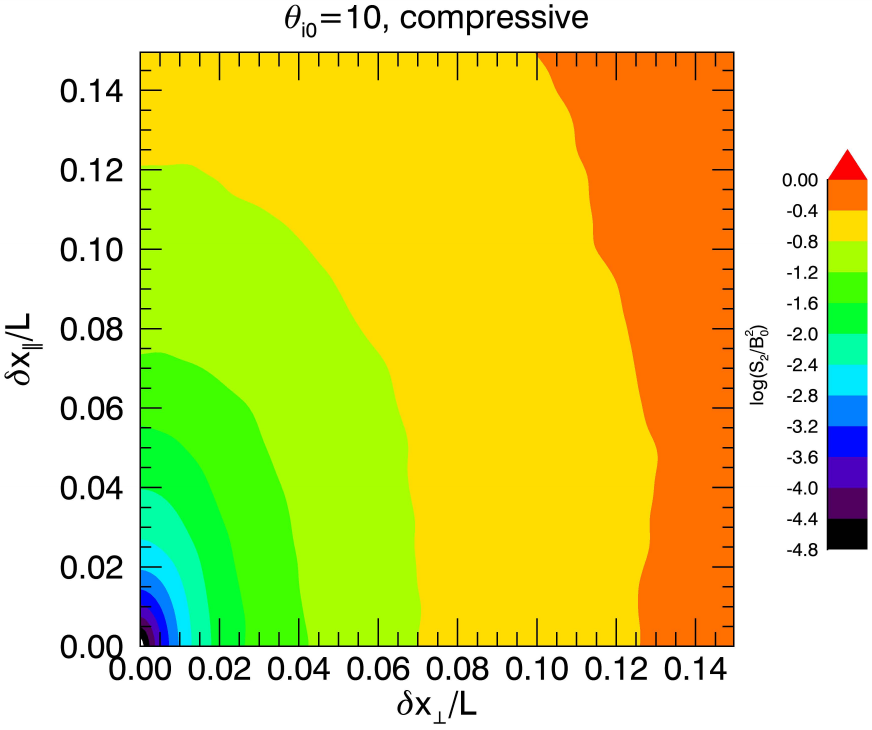}
   \centering
  \caption{\label{fig:sfs_contour} Contours of the second-order structure function for the magnetic field, $S_2(\delta x_\perp, \delta x_{\parallel})$. The $\theta_{i0}=10$ case with solenoidal driving (top panel) and with compressive driving (bottom panel) are shown.}
 \end{figure}

We show the contours of $S_2(\delta x_\perp, \delta x_\parallel)$ for the fiducial $\theta_{i0} = 10$ cases in Fig.~\ref{fig:sfs_contour} (averaged from $t v_{A0}/L = 4.8$ to $t v_{A0}/L = 6.7$). We find that the contours are extended in the direction of $\delta x_\parallel$, indicating that turbulent structures are elongated along the local background magnetic field. This anisotropy is qualitatively similar for both types of driving.

   \begin{figure}
  \includegraphics[width=0.95\columnwidth]{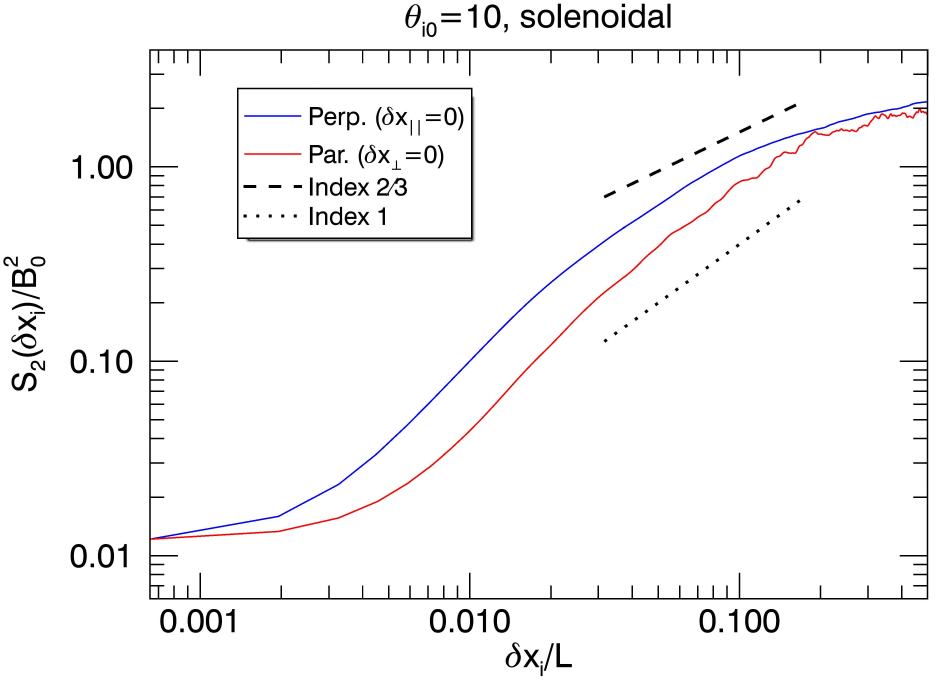}
\includegraphics[width=0.95\columnwidth]{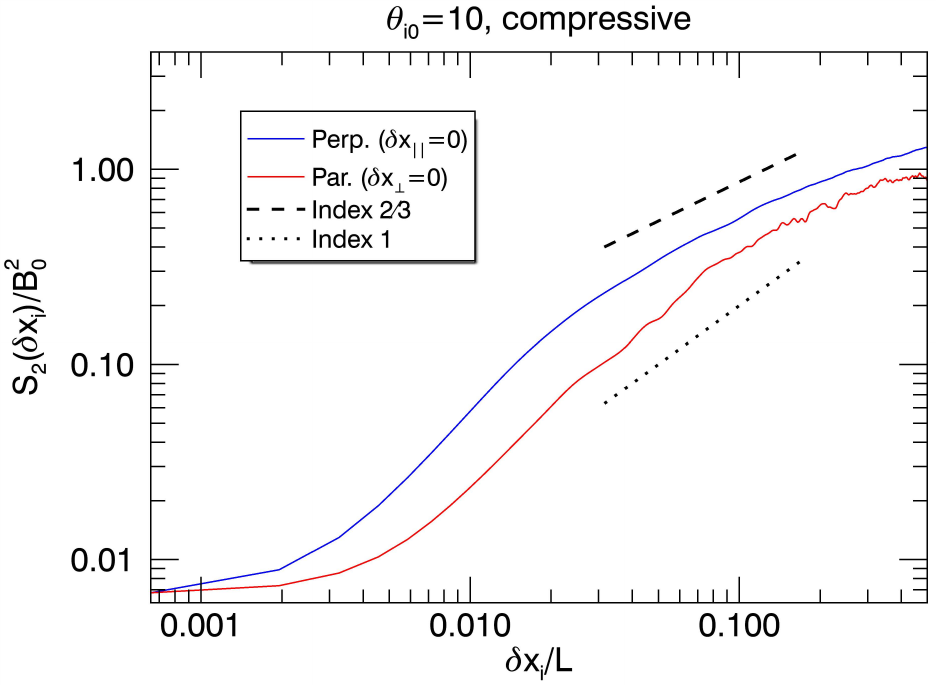}
   \centering
  \caption{\label{fig:sfs} Second-order structure function for the magnetic field, $S_2(\delta x_\perp, \delta x_{\parallel})$, along the principal axes ($\delta x_{\parallel} = 0$ in blue and $\delta x_{\perp} = 0$ in red), for the $\theta_{i0}=10$ cases with solenoidal driving (top panel) and with compressive driving (bottom panel). For reference, the critical balance predictions of $S_2(\delta x_\perp, 0) \sim \delta x_\perp^{2/3}$ (dashed) and $S_2(0, \delta x_{\parallel}) \sim \delta x_{\parallel}$ (dotted) are also shown.}
 \end{figure}
 
 To characterize the anisotropy quantitatively, we measure $S_2(\delta x_\perp,\delta x_\parallel)$ along the principal axes and show the resulting scalings in Fig.~\ref{fig:sfs}. The theory of critical balance in MHD turbulence predicts $S_2(\delta x_\perp,0) \sim \delta x_\perp^{2/3}$ and $S_2(0,\delta x_\parallel) \sim \delta x_\parallel$, corresponding to scalings of $k_\perp^{-5/3}$ and $k_\parallel^{-2}$ in the magnetic energy spectrum \citep{goldreich_sridhar_1995}. While the scaling range, we find that our PIC simulations are consistent with the Goldreich-Sridhar theory in the inertial range, for both types of driving. 
 
Previous works in the literature found that fast-mode cascades are isotropic in MHD turbulence \citep{cho_lazarian_2002, cho_lazarian_2003}. Since we do not see a clear difference in the structure functions between the solenoidally driven case and the compressively driven case, with both being consistent with critical balance predictions, this suggests that the fast mode cascade is subdominant. The cascade in our case may be dominated by Alfv\'{e}n and slow modes either due to mode conversion or due to collisionless damping of fast modes at large scales. In principle, further progress on understanding the nature of turbulent fluctuations can be made by decomposing them in terms of ideal MHD mode polarizations \citep{cho_lazarian_2002, cho_lazarian_2003, makwana_yan_2020}; since the focus of this work is on particle energization, we defer such an analysis to future works.

 \section{Analysis of energy dissipation} \label{sec:results2}

\subsection{Electron and ion energy partition} \label{eiheating}

We now turn to the properties of energy dissipation (heating, nonthermal particle acceleration) in our PIC simulations. In this subsection, we begin by describing the partitioning of the dissipated energy between electrons and ions.

\begin{figure}
\includegraphics[width=0.95\columnwidth]{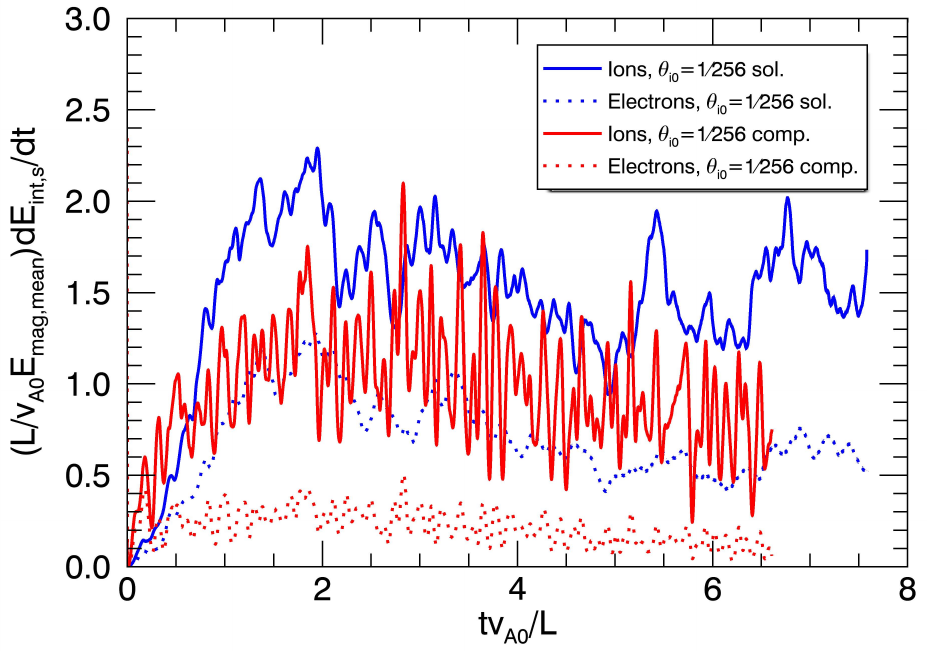}
   \centering
  \caption{\label{fig:eint_evo} Evolution of the overall heating rate $\dot{E}_{{\rm int},s}$ for ions (solid) and electrons (dotted) for the solenoidal (blue) and compressive (red) cases with sub-relativistic ions ($\theta_{i0} = 1/256$).}
 \end{figure}
 
We first provide an example of the evolution of the electron and ion heating rates. In Fig.~\ref{fig:eint_evo}, we show the evolution of the heating rate $\dot{E}_{{\rm int},s}$ (normalized to $E_{\rm mag, mean} v_{A0}/L$) for ions and electrons in the fiducial $\theta_{i0} = 1/256$ simulations. We find that the heating rates enter a quasi-steady state after a time of $\sim 2 L/v_{A0}$. During developed turbulence, $\dot{E}_{{\rm int},s}$ randomly fluctuates by roughly a factor of two in time, with faster variability in the compressive case.

We define $\Delta E_{{\rm int},s}$ as the change in the overall internal energy in particles of species $s$ between time $t v_{A0}/L = 2$ and $t v_{A0}/L = 6$. We then refer to $\Delta E_{{\rm int},e}/\Delta E_{{\rm int},i}$ as the electron-to-ion heating ratio. As shown later in this paper (Section~\ref{sec:ntpa}), a significant fraction of the dissipated energy in fact goes into nonthermal particle acceleration, rather than thermal heating. We also note that the following results on electron-ion heating are almost identical when using total particle kinetic energy ($\Delta E_s$) in place of internal energy. The former does not distinguish between energy going into irreversible dissipation or into bulk motions; but the contribution to $\Delta E_s$ from bulk motions is negligible, because the bulk flow kinetic energy is statistically constant in time (once turbulence has fully developed) and also because it is subdominant to the internal energy, as shown previously in Fig.~\ref{fig:energy}.

 \begin{figure}
\includegraphics[width=0.95\columnwidth]{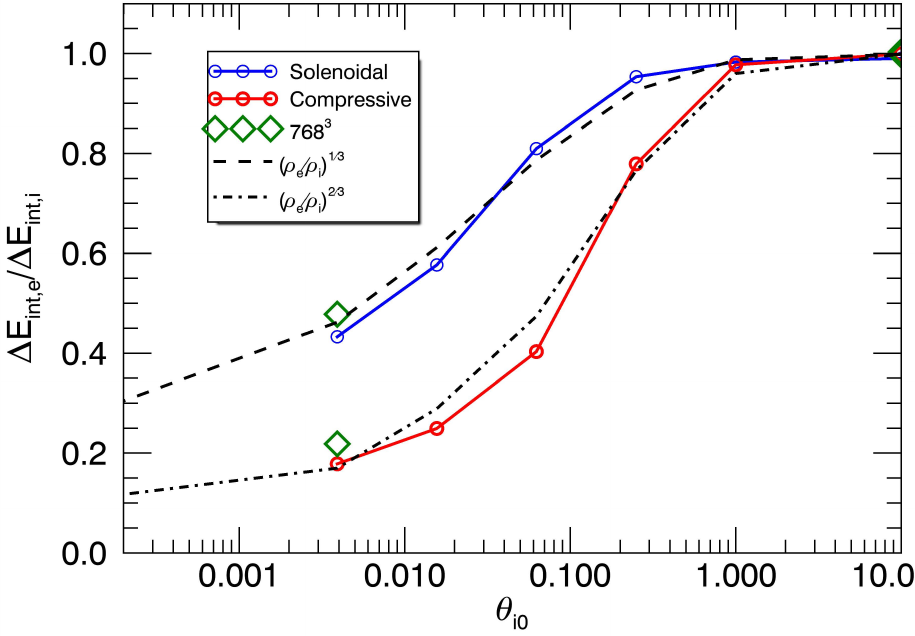}
   \centering
  \caption{\label{fig:partition} Electron-to-ion heating ratio $\Delta E_{{\rm int},e}/\Delta E_{{\rm int},i}$ for simulations with varying initial ion temperature $\theta_{i0}$ for solenoidal (blue) and compressive (red) driving. The fiducial simulations with large system size are indicated by green diamonds. For reference, scalings of $(\rho_e/\rho_i)^\alpha$ with $\alpha = 1/3$ (dashed) and $\alpha = 2/3$ (dash-dotted) are also shown.}
 \end{figure}
 
We show the electron-to-ion heating ratio~$\Delta E_{{\rm int},e}/\Delta E_{{\rm int},i}$ for solenoidal and compressive simulations with varying $\theta_{i0}$ in Fig.~\ref{fig:partition}. Here, we utilize the parameter scan with the $384^3$-cell simulations to better compare to previous results in \cite{zhdankin_etal_2019}; the four large ($768^3$-cell) fiducial simulations are overlaid (green diamond markers). We find that $\Delta E_{{\rm int},e}/\Delta E_{{\rm int},i}$ varies from a minimum of $\sim 0.2$ in the compressive case (and $\sim 0.45$ for the solenoidal case) at our lowest ion temperature of $\theta_{i0} = 1/256$ to a maximum of $1$ at $\theta_{i0} = 10$; there is a monotonic increase of $\Delta E_{{\rm int},e}/\Delta E_{{\rm int},i}$ with $\theta_{i0}$ between these two extremes. Thus, ions are preferentially heated throughout the semirelativistic regime. The asymptotic value of $\Delta E_{{\rm int},e}/\Delta E_{{\rm int},i} = 1$ in the ultra-relativistic regime is easily explained by the relativistic mass symmetry of the electrons and ions: both particles must gain the same amount of energy since they obey identical dynamical equations in the ultra-relativistic limit. The energy partition in the semirelativistic regime, however, is a nontrivial result. As indicated by the fiducial simulations, there is only a weak sensitivity of the results to system size at $\theta_{i0} = 1/256$, with the larger simulations exhibiting slightly less ion heating; this is likely due to the limited inertial range in our simulations at low $\theta_{i0}$. Larger simulations would be necessary to demonstrate the convergence of the heating ratio in the limit of large $L/\rho_{i0}$ relevant to astrophysical systems. There is not a strong sensitivity of our results to the time period analyzed, as long as turbulence is fully developed.

Due to the lack of an established analytical theory for $\Delta E_{{\rm int},e}/\Delta E_{{\rm int},i}$ in this physical regime with the relevant energization mechanisms, we focus on describing the data with a purely empirical fit (with the hope that this could eventually be easily compared to future analytical theories). It is natural to expect $\Delta E_{{\rm int},e}/\Delta E_{{\rm int},i}$ to decrease with an increasing scale separation between electrons and ions, which motivates considering fits to $\Delta E_{{\rm int},e}/\Delta E_{{\rm int},i}$ using functions parameterized by the ratio of the particle gyroradii, $\rho_e/\rho_i$ (which in turn is a function of $T_i/T_e$ and $\theta_i$, both of which evolve in time). Therefore, we consider empirical fits to the data based on a power law in the scale separation,
\begin{align}
\Delta E_{{\rm int},e}/\Delta E_{{\rm int},i} \sim (\rho_e/\rho_i)^{\alpha} \, ,
\end{align}
where $\rho_s$ are averaged over the same time interval that $\Delta E_{{\rm int},s}$ are measured. As indicated by the black lines in Fig.~\ref{fig:partition}, this scaling is able to represent the available data well, with $\alpha \approx 1/3$ for the solenoidal case and $\alpha \approx 2/3$ for the compressive case. 

We now pause to make a few points about these empirical fits. First, we emphasize that our simulations have $\beta \sim 1$ and $\delta B_{\rm rms}/B_0 \sim 1$, and so the dependence of $\Delta E_{{\rm int},e}/\Delta E_{{\rm int},i}$ on these two additional parameters is not established in our present study. However, in our numerical setup, $\beta \ll 1$ would be unsustainable because the plasma would heat to $\beta \sim 1$ over one turnover time; thus, the $\beta$ dependence may be rather weak \citep[see also the $\beta$ scan in][]{zhdankin_etal_2019}. Second, the empirical scaling with $\alpha \approx 2/3$ was previously found in \cite{zhdankin_etal_2019}, which studied electron-ion heating in turbulence with electromagnetic driving and similar plasma physical parameters as considered here; the implication of our present study is that there was a significant compressive component in those simulations. Third, we note that analytical theories of transit-time damping of Alfv\'{e}n waves from MHD turbulence modes with $k_\perp \rho_i \lesssim 1$ derived a non-relativistic heating ratio $\Delta E_{{\rm int},e}/\Delta E_{{\rm int},i} \sim (m_e T_e/m_i T_i)^{1/2} e^{1/\beta_i}$; in terms of the particle gyroradii, this translates to $\Delta E_{{\rm int},e}/\Delta E_{{\rm int},i} \sim (\rho_e/\rho_i) e^{1/\beta_i}$, where the ion beta dependence becomes negligible when $\beta_i \gg 1$ \citep{quataert_1998, quataert_gruzinov_1999}. The analytical model thus naively predicts a more extreme scaling of $\Delta E_{{\rm int},e}/\Delta E_{{\rm int},i}$ with $\rho_e/\rho_i$ than observed in our PIC simulations. A more sophisticated prescription was proposed by \cite{howes_2010}, which has significant differences near $\beta \sim 1$ and predicts preferential electron heating for $\beta_i \lesssim 1$ and $T_i/T_e = 1$, with preferential ion heating only at higher $\beta_i$. Likewise, the empirical prescription in \cite{kawazura_etal_2020}, based on gyrokinetic simulations, predicts preferential electron heating at these parameters. Note that these analytical models are inherently non-relativistic and therefore do not satisfy the necessary limit of $\Delta E_{{\rm int},e}/\Delta E_{{\rm int},i} = 1$ for $\theta_i \gg 1$ or for $m_i/m_e \to 1$. They also do not account for the significant amount of energy absorbed through diffusive particle acceleration, which may preferentially energize ions (see Section \ref{sec:ntpa}). Fourth, we observe that deep in the semirelativistic regime, to a fair approximation, our results are consistent with $\Delta E_{{\rm int},e}/\Delta E_{{\rm int},i}$ being reduced by a factor of 2 when changing from solenoidal to compressive driving. This may be an indication that roughly half of the injected energy, which is held in compressive fluctuations, is damped into ions, and the rest is converted into a quasi-Alfv\'{e}nic cascade; this picture is consistent with the model of \cite{kawazura_etal_2020}. 
 
   \subsection{Electron and ion energy transfer} \label{sec:entrans}
 
 To gain some physical understanding of the preferential ion heating in our semirelativistic simulations, we next evaluate the energy exchange between electrons, ions, and the electromagnetic fields. The external force directly injects a similar amount of kinetic energy into both electrons and ions, since it couples with the bulk motion of the plasma [we have verified this by measuring energy injection rates for the two species individually, given by~$\dot{E}_{{\rm inj},s}(t) \equiv \int d^3x d^3p \boldsymbol{F}_{\rm ext}(\boldsymbol{x},t) \cdot \boldsymbol{p}c(m_s^2 c^2 + p^2)^{-1/2} f_s(\boldsymbol{x},\boldsymbol{p},t)$]. The only way that this symmetrically injected energy (contained in large-scale flows) can be redistributed from electrons to ions is by intermediary transfer via the electric field. This transfer is quantified by the rate of energy exchange between the electric field and particles, given by $\boldsymbol{E} \cdot \boldsymbol{J}_s$, where $\boldsymbol{J}_s$ is the current density for species $s$.
 
 Before proceeding, we emphasize that for force-driven turbulence in a (rigorous) statistical steady state, the total electromagnetic energy is constant in time, so~$\partial_t (E_{\rm mag} + E_{\rm elec}) = -\int d^3x \boldsymbol{E} \cdot \boldsymbol{J} \approx 0$. This is because electromagnetic fluctuations are produced at the same average rate as they are dissipated. This differs from decaying turbulence, where $\int d^3x \boldsymbol{E} \cdot \boldsymbol{J} > 0$ due to the decay of fields into internal energy, leading to $\boldsymbol{E} \cdot \boldsymbol{J}$ often being interpreted as a ``heating'' term. It also differs from electromagnetically driven turbulence, where the electromagnetic fields have a source term to compensate for $\int d^3x \boldsymbol{E} \cdot \boldsymbol{J} > 0$. In our situation, the steady-state condition implies $\int d^3x \boldsymbol{E} \cdot \boldsymbol{J}_e \sim - \int d^3x \boldsymbol{E} \cdot \boldsymbol{J}_i$. For preferential ion heating (as observed in Section~\ref{eiheating}), there must be a net energy transfer from electrons ($\int d^3x \boldsymbol{E} \cdot \boldsymbol{J}_e < 0$) to ions ($\int d^3x \boldsymbol{E} \cdot \boldsymbol{J}_i > 0$). Note that a net energy transfer from electrons to ions (via electromagnetic fields) does not preclude the overall heating of electrons, because this diagnostic does not include the energy injected into electrons by the external driving, $\int d^3x \boldsymbol{F}_{\rm ext} \cdot \boldsymbol{\mathcal{V}}_e n_e$, where $\boldsymbol{\mathcal{V}}_e = - \boldsymbol{J}_e/(e n_e)$ is the electron fluid velocity. The external energy injection typically exceeds the electromagnetic energy transfer.  Thus, $\boldsymbol{E} \cdot \boldsymbol{J}_s$ is useful for understanding inter-species energy transfer, but an alternative diagnostic would be needed to measure conversion of bulk flow energy into internal energy within a single particle species.
 
   \begin{figure}
\includegraphics[width=0.95\columnwidth]{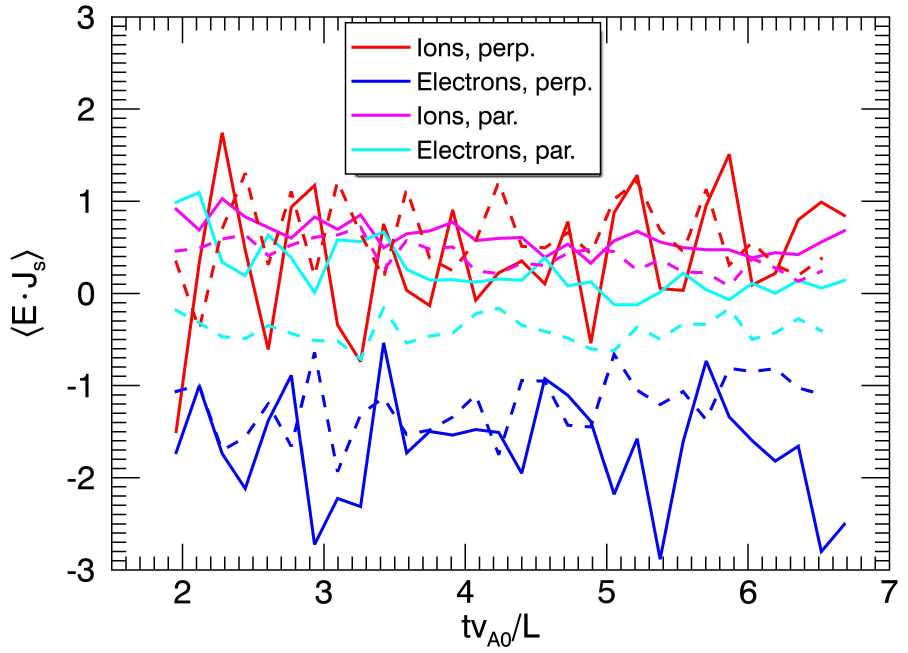}
      \centering
   \caption{\label{fig:edotj_evo} Evolution of the volume-averaged electromagnetic energy exchange term, for solenoidal (solid) and compressive (dashed) simulations, normalized to the time-averaged $\langle \boldsymbol{E} \cdot \boldsymbol{J}_i \rangle$. The different colors indicate perpendicular energization of ions $\langle \boldsymbol{E} \cdot \boldsymbol{J}_{\perp,i} \rangle$ (red), perpendicular energization of electrons $\langle \boldsymbol{E} \cdot \boldsymbol{J}_{\perp,e} \rangle$, parallel energization of ions $\langle \boldsymbol{E} \cdot \boldsymbol{J}_{\parallel,i} \rangle$, and parallel energization of electrons $\langle \boldsymbol{E} \cdot \boldsymbol{J}_{\parallel,e} \rangle$}
 \end{figure}
 
 We first decompose the fields into components that are parallel and perpendicular to $\boldsymbol{B}$, by defining~$\boldsymbol{E}_{\parallel} = \boldsymbol{E} \cdot \hat{\boldsymbol{B}} \hat{\boldsymbol{B}}$,~$\boldsymbol{E}_\perp = \boldsymbol{E} - \boldsymbol{E}_{\parallel}$, and similar for $\boldsymbol{J}_s$. We then decompose $\boldsymbol{E} \cdot \boldsymbol{J}_s = \boldsymbol{E}_\perp \cdot \boldsymbol{J}_{\perp,s} + \boldsymbol{E}_{\parallel} \cdot \boldsymbol{J}_{{\parallel},s}$. We show the evolution of the volume-averaged quantities $\langle \boldsymbol{E}_\perp \cdot \boldsymbol{J}_{\perp,s}  \rangle$ and $\langle \boldsymbol{E}_{\parallel} \cdot \boldsymbol{J}_{\parallel,s} \rangle$ in Fig.~\ref{fig:edotj_evo} for the $\theta_{i0}=1/256$ fiducial simulations with solenoidal driving (solid lines) and compressive driving (dashed lines). We find that ions gain energy from both perpendicular and parallel fields, while electrons lose energy mainly through perpendicular fields. In the solenoidal case, electrons appear to gain a very small amount of energy from parallel fields, while in the compressive case, they lose a small amount of energy through parallel fields. The time-averaged values of $\langle \boldsymbol{E}_\perp \cdot \boldsymbol{J}_{\perp,s}  \rangle$ and $\langle \boldsymbol{E}_{\parallel} \cdot \boldsymbol{J}_{\parallel,s} \rangle$, normalized to the time-averaged $\langle \boldsymbol{E} \cdot \boldsymbol{J}_i \rangle$, are tabulated in Table~\ref{table:transfer}. Note that $\langle \boldsymbol{E} \cdot \boldsymbol{J} \rangle \neq 0$, contrary to the naive expectations, which is likely due to the slight tendency of electrons to be accelerated faster than ions by the external force (due to their smaller inertia).
 
 \begin{table}%[h!b!p!]
\centering \caption{Average energy transfer terms \newline} \label{table:transfer}
\begin{tabular}{|c|c|c|} 
	\hline
\hspace{0.5 mm} Quantity \hspace{0.5 mm}  & Solenoidal (rL1d256s)  & Compressive (rL1d256c)   \\
	\hline
$\langle\boldsymbol{E}_\perp\cdot\boldsymbol{J}_{\perp,i}\rangle$ & $0.38$ & 0.60 \\
$\langle\boldsymbol{E}_{\parallel}\cdot\boldsymbol{J}_{\parallel,i}\rangle$ & $0.62$ & 0.40 \\
$\langle\boldsymbol{E}_\perp\cdot\boldsymbol{J}_{\perp,e}\rangle$ & $-1.68$ & -1.21  \\
$\langle\boldsymbol{E}_{\parallel}\cdot\boldsymbol{J}_{\parallel,e}\rangle$ & $0.25$ & -0.40 \\
	\hline
\end{tabular}
\centering
\end{table} % \hspace{1 mm} 
 
To understand the scale dependence of the energy transfer, we expand the electric field and current density in Fourier modes,~$\boldsymbol{E}(\boldsymbol{x},t) = \int d^3k \tilde{\boldsymbol{E}}(\boldsymbol{k},t) e^{i \boldsymbol{k} \cdot \boldsymbol{x}}/(2\pi)^3$ and~$\boldsymbol{J}_s(\boldsymbol{x},t) = \int d^3k \tilde{\boldsymbol{J}}_s(\boldsymbol{k},t) e^{i \boldsymbol{k} \cdot \boldsymbol{x}}/(2\pi)^3$. The rate of energy transfer from electric and magnetic fields to particles of species $s$ is given by $\partial_t E_s = \int d^3x \boldsymbol{E} \cdot \boldsymbol{J}_s$. We can express this integral in Fourier space as
\begin{align}
\int d^3x \boldsymbol{E}(\boldsymbol{x},t) \cdot \boldsymbol{J}_s(\boldsymbol{x},t) = \int \frac{d^3k}{(2 \pi)^3} \tilde{\boldsymbol{E}}(\boldsymbol{k},t) \cdot \tilde{\boldsymbol{J}}_s^*(\boldsymbol{k},t) \, .
\end{align}
Thus, the integrand $\tilde{\boldsymbol{E}} \cdot \tilde{\boldsymbol{J}}_s^*$ describes the rate of energy transfer from the electric field mode at wavenumber $\boldsymbol{k}$ to the kinetic energy of the particle species $s$. This kinetic energy may involve bulk flows, adiabatic compressions, heating, and nonthermal particle acceleration. The energy transfer rate associated with flows/compressions will fluctuate between positive and negative values, while irreversible energy dissipation (i.e., heating and nonthermal particle acceleration) will have a net positive value. Thus, the signatures of flows and compressions are removed after integrating over directions of $\boldsymbol{k}$ and averaging over sufficiently long times. We are therefore led to define the {\it energy transfer spectrum\rm} by
\begin{align}
{\mathcal D}_s(k,t) = k^2 \int d\Omega \tilde{\boldsymbol{E}}(\boldsymbol{k},t) \cdot \tilde{\boldsymbol{J}}_s^*(\boldsymbol{k},t) \, ,
\end{align}
where $d\Omega$ is the solid angle differential in $\boldsymbol{k}$ space (we do not take into account anisotropy with respect to $\boldsymbol{B}_0$ here).

  \begin{figure}
  \includegraphics[width=0.95\columnwidth]{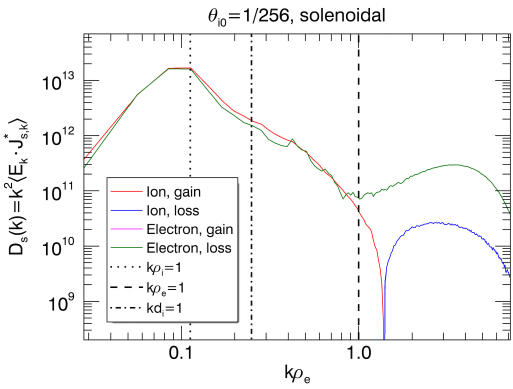}
\includegraphics[width=0.95\columnwidth]{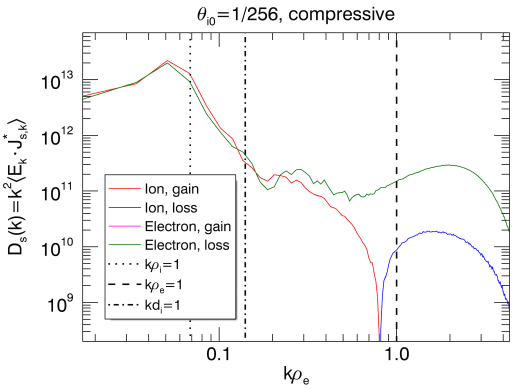}
      \centering
   \caption{\label{fig:edotjspec} Absolute value of the energy transfer spectrum ${\mathcal D}_s(k)$ for the solenoidal case (top panel) and for the compressive case (bottom panel) with $\theta_{i0} = 1/256$. Ion energy gain (${\mathcal D}_i > 0$) is shown in red, ion energy loss (${\mathcal D}_i < 0$) in blue, and electron energy loss (${\mathcal D}_e < 0$) in green. The ion gyroscale (dotted), ion skin depth (dash-dotted), and electron gyroscale (dashed) are shown for reference.}
 \end{figure}

We show the energy transfer spectrum ${\mathcal D}_s(k)$ for the fiducial $\theta_{i0}=1/256$ cases in Fig.~\ref{fig:edotjspec}. We average ${\mathcal D}_s$ over the period $t v_{A0}/L = 3.7$ to $t v_{A0}/L = 6.1$, sufficiently long to eliminate fluctuations from reversible processes; without a time average, the spectrum rapidly fluctuates from positive to negative values at varying $k$. We find that the averaged ${\mathcal D}_s$ is a smooth function for both cases, with a scale-by-scale balance ${\mathcal D}_i \approx - {\mathcal D}_e > 0$ at $k\rho_e < 1$, indicating that energy is transferred from electrons to ions via the electric fields, consistent with the overall partitioning of $\Delta E_e/\Delta E_i < 1$. The peak value of ${\mathcal D}_i$ occurs at $k\rho_i \approx 1$, indicating that most of the energization occurs in the vicinity of the ion gyroscale. In the kinetic range of scales, $\rho_i^{-1} < k < \rho_e^{-1}$, the spectrum falls off more strongly than $k^{-1}$, indicating a diminishing contribution to energy exchange from modes with $k\rho_i \gtrsim 1$. The scaling for the solenoidal and compressive cases are both qualitatively similar in this range, although the compressive case has a steeper drop off. At scales $k\rho_e \gtrsim 1$, the antisymmetry between electrons and ions is broken, with both species undergoing net cooling; this range may be sensitive to numerical resolution \citep[see also][]{zhdankin_etal_2020}.

The most significant difference in ${\mathcal D}_s(k)$ between the solenoidal and compressive cases occurs at scales $k\rho_i \lesssim 1$, the nominal MHD range. Whereas ${\mathcal D}_i(k)$ is very small when $k \rho_i \lesssim 1$ for the solenoidal case, it retains a large value in the compressive case. This suggests that much, if not most, of the electron-to-ion energy exchange occurs at large scales $k \rho_i \lesssim 1$ in the compressive case.

 \begin{figure}
   \includegraphics[width=0.95\columnwidth]{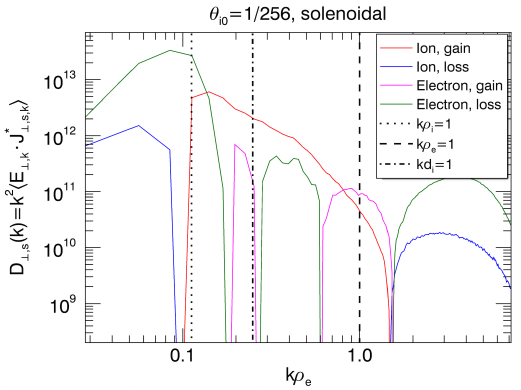}
\includegraphics[width=0.95\columnwidth]{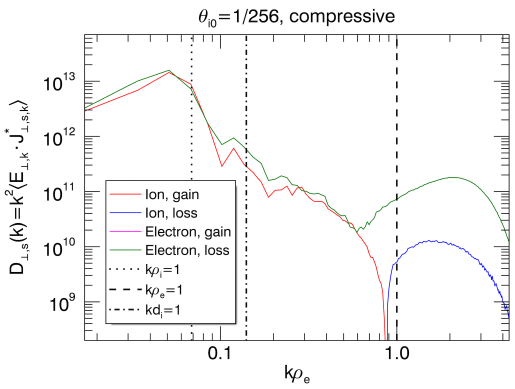}
      \centering
   \caption{\label{fig:edotjspecperp} Absolute value of the perpendicular energy transfer spectrum ${\mathcal D}_{\perp,s}(k)$ for the solenoidal case (top panel) and for the compressive case (bottom panel) with $\theta_{i0} = 1/256$. Ion energy gain (${\mathcal D}_{\perp,i} > 0$) is shown in red, ion energy loss (${\mathcal D}_{\perp,i} < 0$) in blue, electron energy gain (${\mathcal D}_{\perp,e} > 0$) in magenta, and electron energy loss (${\mathcal D}_{\perp,e} < 0$) in green. The ion gyroscale (dotted), ion skin depth (dash-dotted), and electron gyroscale (dashed) are shown for reference.}
 \end{figure}
 
To enhance the features in the energy transfer spectra, we next look at the spectra for perpendicular and parallel fields separately, by defining
\begin{align}
{\mathcal D}_{\perp,s}(k,t) = k^2 \int d\Omega \tilde{\boldsymbol{E}}_\perp(\boldsymbol{k},t) \cdot \tilde{\boldsymbol{J}}_{\perp,s}^*(\boldsymbol{k},t) \, , \nonumber \\
{\mathcal D}_{\parallel,s}(k,t) = k^2 \int d\Omega \tilde{\boldsymbol{E}}_{\parallel}(\boldsymbol{k},t) \cdot \tilde{\boldsymbol{J}}_{\parallel,s}^*(\boldsymbol{k},t)  \, .
\end{align}
 In Fig.~\ref{fig:edotjspecperp}, we show the perpendicular part ${\mathcal D}_{\perp,s}(k)$ for the solenoidal (top panel) and compressive (bottom panel) cases. For the solenoidal case, ions lose energy to perpendicular fields at large scales ($k\rho_i \lesssim 1$), while gaining energy in the kinetic range ($\rho_i^{-1} \lesssim k \lesssim \rho_e^{-1}$). Electrons lose energy throughout most of spectrum, except for pockets of heating in the vicinity of $k d_i = 1$ and $k d_e = 1$. Since magnetic reconnection is typically expected to occur at the skin depth scales \citep[e.g.,][]{shay_etal_2007}, it is tempting to associate the electron heating with intermittent diffusion regions on these scales. For the compressive case, the perpendicular energy transfer spectrum is much less structured than the parallel case, with scale-by-scale energy transfer from electrons to ions throughout the entire range $k\rho_e \le 1$, much like the total energy transfer spectrum that was previously shown in Fig.~\ref{fig:edotjspec}.
 
  \begin{figure}
    \includegraphics[width=0.95\columnwidth]{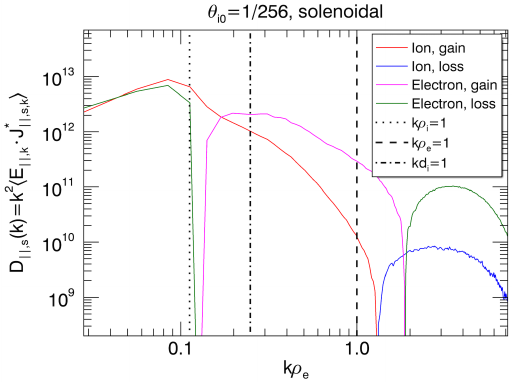}
\includegraphics[width=0.95\columnwidth]{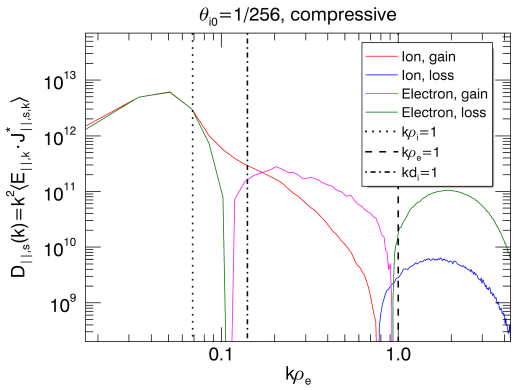}
      \centering
   \caption{\label{fig:edotjspecpar} Absolute value of the parallel energy transfer spectrum ${\mathcal D}_{\parallel,s}(k)$ for the solenoidal case (top panel) and for the compressive case (bottom panel) with $\theta_{i0} = 1/256$. Ion energy gain (${\mathcal D}_{\parallel,i} > 0$) is shown in red, ion energy loss (${\mathcal D}_{\parallel,i} < 0$) in blue, electron energy gain (${\mathcal D}_{\parallel,e} > 0$) in magenta, and electron energy loss (${\mathcal D}_{\parallel,e} < 0$) in green. The ion gyroscale (dotted), ion skin depth (dash-dotted), and electron gyroscale (dashed) are shown for reference.}
 \end{figure}
 
 For completeness, we show the parallel part of the energy transfer spectrum ${\mathcal D}_{\parallel,s}(k)$ in Fig.~\ref{fig:edotjspecpar}. At large scales, $k\rho_i \lesssim 1$, energy is transferred from electrons to ions for both drives. In the kinetic range ($\rho_i^{-1} \lesssim k \lesssim \rho_e^{-1}$), electrons and ions are both energized, with electrons being preferentially energized; the electron heating is stronger and occurs over a broader range of scales in the solenoidal case than in the compressive case. Since parallel electric fields are often associated with heating by magnetic reconnection \citep[e.g.,][]{dahlin_etal_2016}, it is natural to conclude that electrons gain significant energy from magnetic reconnection throughout the kinetic range. The stronger parallel electron energization in the solenoidal case is then consistent with the presence of coherent current sheets visible in Fig.~\ref{fig:visual_jz}
 
 In summary, by studying the spectra of electron and ion energy transfer through perpendicular and parallel electric fields separately (${\mathcal D}_{\perp,s}$ and ${\mathcal D}_{\parallel,s}$), we arrive at the following interpretation. In the solenoidal case, very little energy is transferred from electrons to ions at large scales ($k\rho_i \lesssim 1$); instead, most of the electron-to-ion energy transfer occurs in the kinetic range ($k\rho_i \gtrsim 1$). The signatures highlight magnetic reconnection as a mechanism of energy transfer, due to spikes in electron heating at the skin depth scales and strong heating of both species by parallel electric fields. In the compressive case, on the other hand, there is significant electron-to-ion energy transfer at large scales ($k\rho_i \lesssim 1$) by perpendicular fields, and the heating in the kinetic range is consequently diminished. This large-scale perpendicular ion heating may be associated with efficient diffusive particle acceleration, as will be discussed in Section~\ref{sec:ntpa}.
 
Thus, the fundamental difference in the compressive case (when compared to the solenoidal case) is that ion heating occurs at larger scales and electron heating is reduced in the kinetic range. These dual effects lead to stronger prefential ion heating in the compressive case, as described previously in Section~\ref{eiheating}. To conclusively identify the mechanisms of electron and ion heating, more sophisticated diagnostics will need to be applied. Such diagnostics may be based on characterizing the local particle dynamics \citep{arzamasskiy_etal_2019, cerri_etal_2021} or measuring field-particle correlations using the local distribution function \citep{klein_howes_2016, li_etal_2019, klein_etal_2020}. In addition, a more rigorous, predictive analytical phenomenology is necessary for interpreting the results.

\subsection{Nonthermal particle acceleration} \label{sec:ntpa}

We now turn to analysis of the global particle distributions, as a means of characterizing nonthermal particle acceleration. Previous works already investigated nonthermal particle acceleration in PIC simulations of relativistic plasma turbulence in great detail \citep[e.g.,][]{zhdankin_etal_2017, zhdankin_etal_2018b, comisso_sironi_2018, comisso_sironi_2019, zhdankin_etal_2019, wong_etal_2020}. These studies applied ensembles of tracked particles to demonstrate that the nonthermal acceleration process at high energies is primarily diffusive in momentum space \citep{zhdankin_etal_2018b, comisso_sironi_2019, wong_etal_2020}, which can be attributed to gyroresonant-like interactions between particles and the spectrum of turbulent fluctuations \citep[e.g.,][]{demidem_etal_2020}. While this nonthermal particle acceleration relies on energization by perpendicular electric fields, the particle injection may be influenced by parallel electric fields \citep{comisso_sironi_2019}. We expect this general paradigm to hold in the simulations described in the present work. The new aspect that will be considered in this work is the effect of the solenoidal and compressive driving mechanisms on the overall particle distributions.

 \begin{figure}
 \includegraphics[width=0.95\columnwidth]{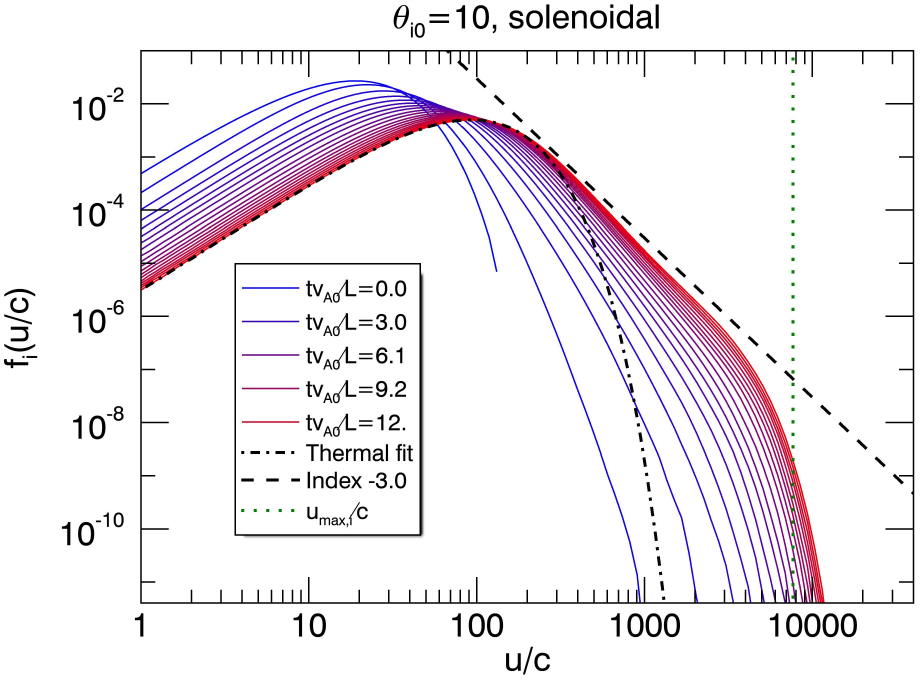}
 \includegraphics[width=0.95\columnwidth]{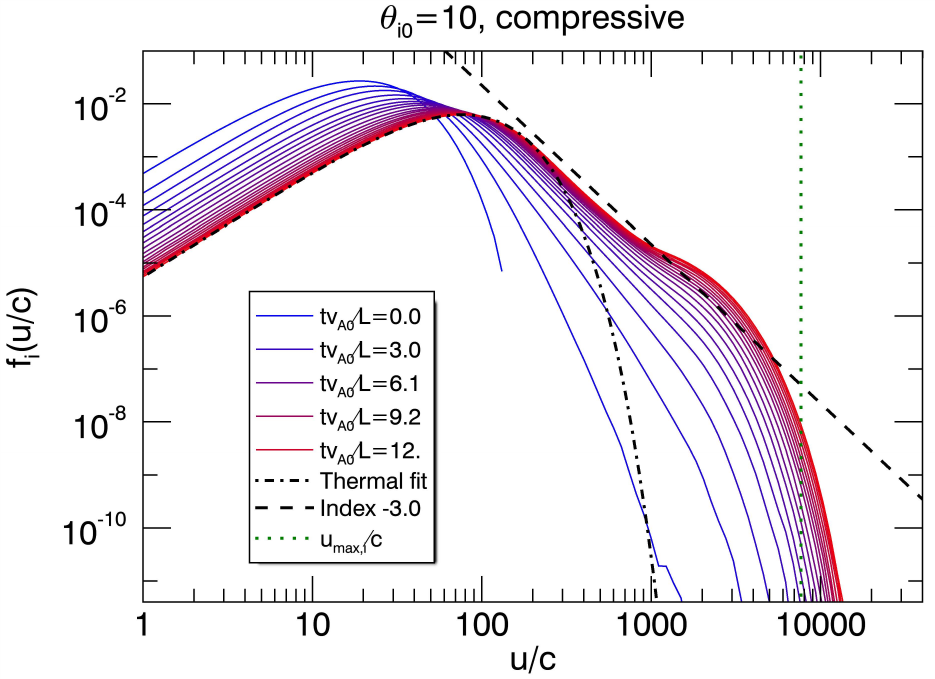}
      \centering
   \caption{\label{fig:distsi_thi10} Ion momentum distributions at various times for the $\theta_{i0} = 10$ simulations with solenoidal driving (top panel) and compressive driving (bottom panel). Also shown are a Maxwell-J\"{u}ttner fit to the peak (dash-dotted), a power law with index $-3$ (dashed), and the system-size limited momentum $u_{{\rm max},i}$. Note that electron distributions are very similar in these simulations (but shifted by a factor of $m_i/m_e$).}
 \end{figure}
 
  In the following, we denote the spatial components of the particle four-velocity by $\boldsymbol{u}/c = \boldsymbol{p}/m_s c$, where $\boldsymbol{p}$ is the momentum; we use $u/c$ as a dimensionless measure of the momentum. For relativistic particles, the kinetic energy is $E \approx m_s c u$, while for non-relativistic particles, it is $E \approx m_s u^2/2$. The four-velocity distributions described below are obtained by integrating the momentum distribution over angles, $f(p,t) = p^2 \int d\theta_p d\phi_p \sin{\theta_p} f(\boldsymbol{p},t)$, where the momentum vector in spherical coordinates is $\boldsymbol{p} = p (\cos{\phi_p} \sin{\theta_p}, \sin{\phi_p} \sin{\theta_p},\cos{\theta_p})$.
 
 We first describe the particle momentum distributions for the fiducial relativistic simulations. In Fig.~\ref{fig:distsi_thi10}, we show the evolution of the ion distributions $f_i(u/c)$ for the $\theta_{i0}=10$ simulations with solenoidal and compressive driving; electron distributions are not shown because they are nearly identical to the ions in this case (except for being shifted to Lorentz factors $m_i/m_e$ times larger), as expected from the $\theta_{i0} \gg 1$ relativistic mass symmetry. Consistent with previous works, we find efficient nonthermal particle acceleration in these simulations, such that the particle energy distributions acquire a power-law tail with indices close to $-3.0$ for the given $\sigma$ \citep[see, e.g.,][]{zhdankin_etal_2018b}. The power-law tail forms more rapidly for the compressive case than the solenoidal case, and as a consequence, there is a strong pile-up of particles at momenta that are comparable to the system-size limited momentum $p_{{\rm max}} = L e B_0/2 c$ at late times. Once a particle reaches $p_{\rm max}$, its gyroradius is comparable to the driving scale, so it can no longer absorb energy efficiently from turbulent fluctuations. We note that in comparison to the electromagnetically driven simulations in \cite{zhdankin_etal_2018b}, our simulations have an exponential cut-off that starts at energies a factor of $2$ or so smaller relative to $p_{\rm max}$; this may be due to the fact that magnetic fields cannot be sustained at the very largest scales in the domain unless they are produced directly by an electromagnetic driving mechanism.
 
 \begin{figure}
 \includegraphics[width=0.95\columnwidth]{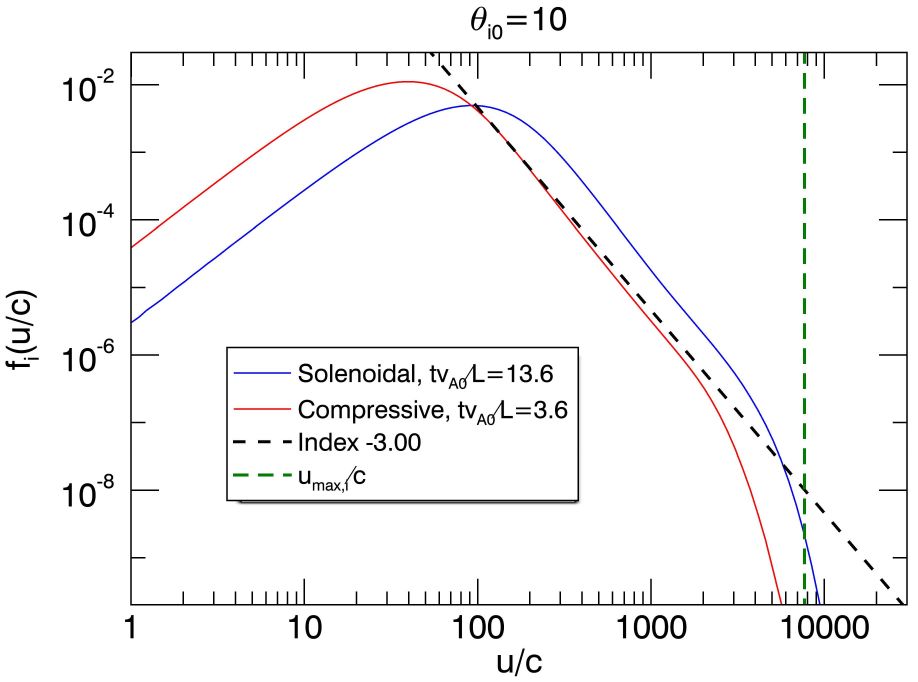}
      \centering
   \caption{\label{fig:dists_equiv} Ion momentum distributions for $\theta_{i0}=10$ simulations with solenoidal driving (blue) at $t v_{A0}/L = 13.6$ and compressive driving (red) at $t v_{A0}/L=3.6$. These times were chosen to be representative of the fully developed distribution before the pile-up forms at $u_{{\rm max},s} = p_{{\rm max}}/m_s$.}
 \end{figure}
 
 The indices of the momenta distributions can be compared when choosing times shortly before the formation of the high-energy pileup, as previously motivated in \cite{zhdankin_etal_2018b}. We find that this roughly corresponds to $t v_{A0}/L \sim 13.6$ for the solenoidal case and $t v_{A0}/L \sim 3.6$ for the compressive case. We show a side-by-side comparison of the ion distributions at these times in Fig.~\ref{fig:dists_equiv}. As seen from this comparison, the peak of the momentum distribution has shifted to larger values in the solenoidal case compared to the compressive case, consistent with the fact that there is a longer span of time for heating. In contrast, the shape of the power-law tail looks nearly identical in both cases, with the aforementioned index near $-3.0$. Thus, nonthermal acceleration occurs $\sim 4$ times faster in the compressively driven case than in the solenoidal case, despite a similar rate of external energy injection.
 
Currently, there is no established analytical theory for predicting the power-law indices arising from stochastic acceleration in a closed domain, where particles lack an explicit escape mechanism; see \cite{lemoine_malkov_2020} for a detailed discussion on this topic and suggested resolutions. \cite{comisso_sironi_2019} showed (in their PIC turbulence simulations) that the power-law index is insensitive to the mechanism by which particles are injected from the thermal population. Thus, one possible explanation for the similiarity between the nonthermal distributions in Fig.~\ref{fig:dists_equiv} is that particles undergo faster injection in the compressive case, but experience the same diffusive acceleration process. However, this does not explain why particles reach $p_{\rm max}$ faster in the compressive case. Instead, it seems that diffusive acceleration on a faster timescale is necessary to explain the enhanced acceleration in the compressive case. Quasilinear theories of diffusive particle acceleration that account for resonance broadening predict that the ratio of the fast-mode acceleration timescale to Alfv\'{e}nic acceleration timescale is proportional to $v_A/c$, but with a coefficient that depends on the system parameters \citep{demidem_etal_2020}; the observed difference in timescales is thus not unreasonable from a theoretical perspective. The process that regulates the power-law index may depend, e.g., on the advective contribution to particle acceleration, which is poorly understood theoretically \citep[see ][]{wong_etal_2020}.

  \begin{figure}
\includegraphics[width=0.95\columnwidth]{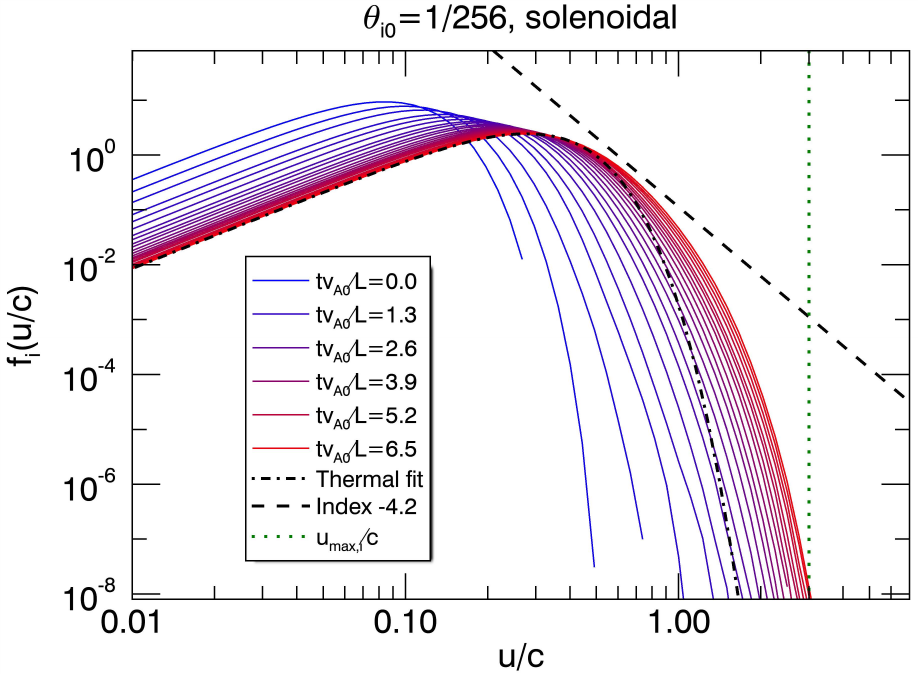}
\includegraphics[width=0.95\columnwidth]{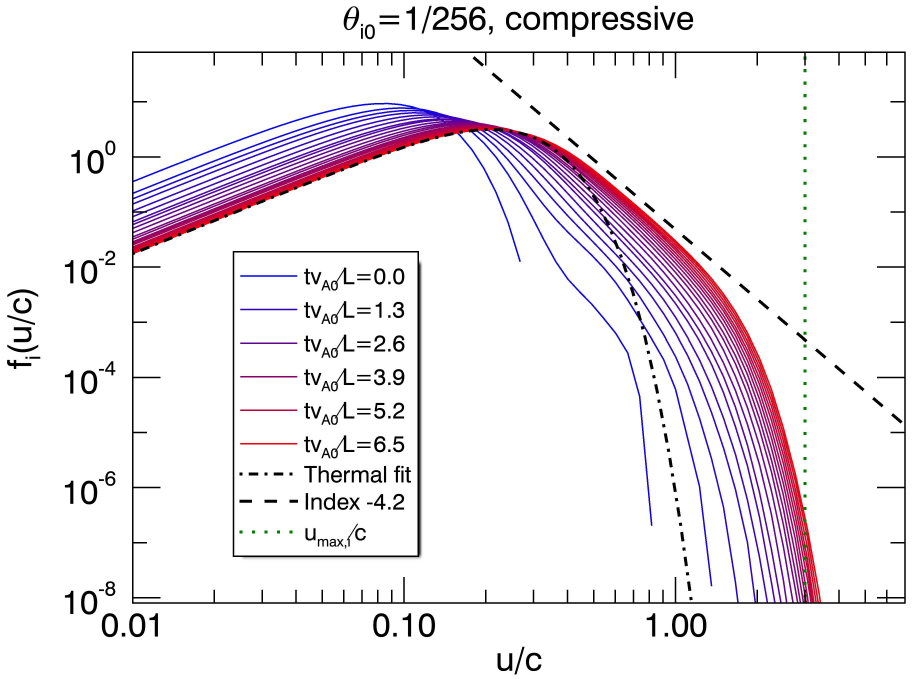}
      \centering
   \caption{\label{fig:distsi} Similar to Fig.~\ref{fig:distsi_thi10} except for ions in the $\theta_{i0} = 1/256$ simulations with solenoidal driving (top panel) and compressive driving (bottom panel).}
 \end{figure}
 
  \begin{figure}
 \includegraphics[width=0.95\columnwidth]{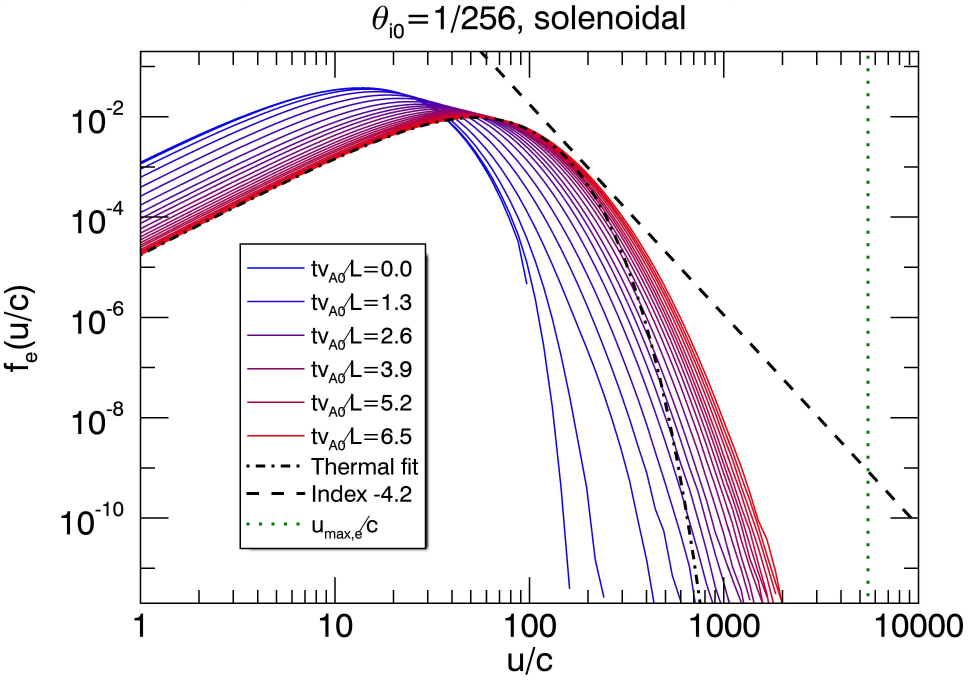}
\includegraphics[width=0.95\columnwidth]{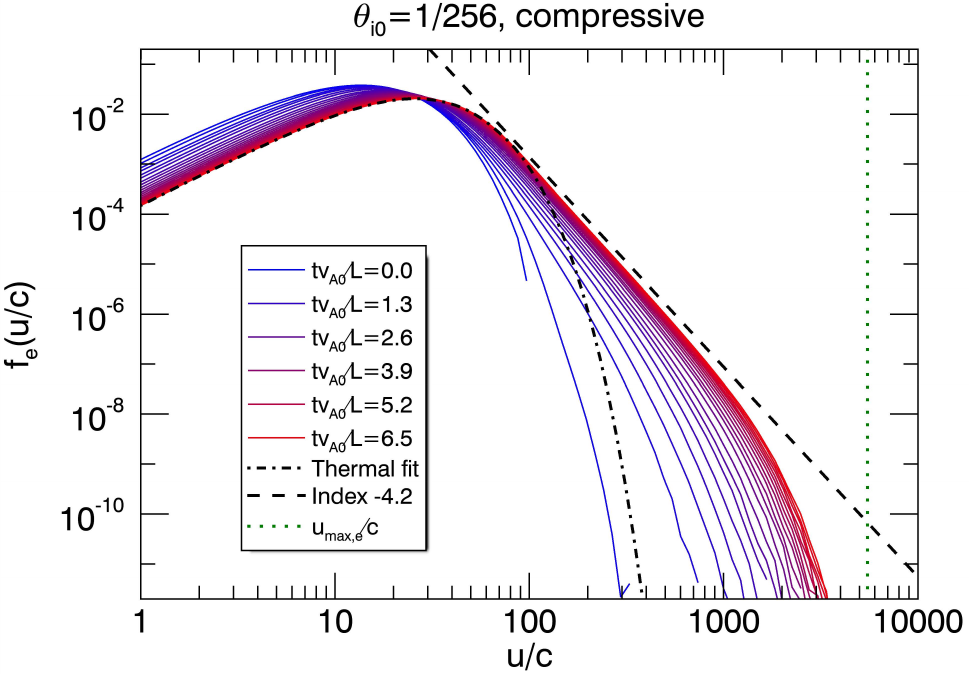}
      \centering
   \caption{\label{fig:distse} Similar to Fig.~\ref{fig:distsi} except for electrons in the $\theta_{i0} = 1/256$ simulations with solenoidal driving (top panel) and compressive driving (bottom panel).}
 \end{figure}

We next turn to the semirelativistic cases with $\theta_{i0} = 1/256$, for which we show the ion distributions in Fig.~\ref{fig:distsi} and the electron distributions in Fig.~\ref{fig:distse}. The primary result is that the solenoidally driven case does not exhibit any notable nonthermal particle acceleration, while the compressively driven case features a substantial nonthermal tail with a power-law index of roughly $-4.2$ (thus, being somewhat softer than the relativistic cases). The ions and electrons have qualitatively similar distributions, with the main difference being that for the compressive case, the electrons have a broader power law than ions. The longer extent of the electron power law can be attributed to the larger separation between the thermal momentum and $p_{\rm max}$ for electrons than ions.

The lack of significant nonthermal particle acceleration in the semirelativistic solenoidally driven simulation is consistent with the absence of particle energization by perpendicular electric fields, as previously shown in Fig.~\ref{fig:edotjspecperp}. It is plausible that slow or fast modes are necessary for efficient particle acceleration at low $\theta_{i}$, and more generally, in non-relativistic plasmas. In the relativistic case, on the other hand, the two driving mechanisms may give similar particle acceleration efficiency since fast modes and Alfv\'{e}n modes are strongly coupled when $\sigma \gtrsim 1$, so that the turbulent cascade properties are similar regardless of the driving \citep{takamoto_lazarian_2017}.
 
   \begin{figure}
\includegraphics[width=0.95\columnwidth]{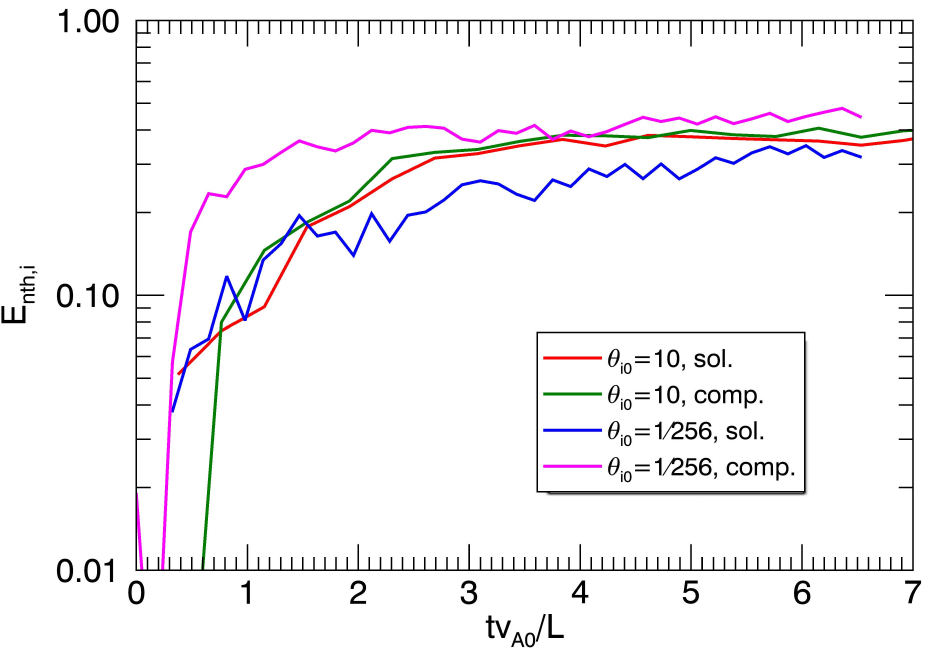}
\includegraphics[width=0.95\columnwidth]{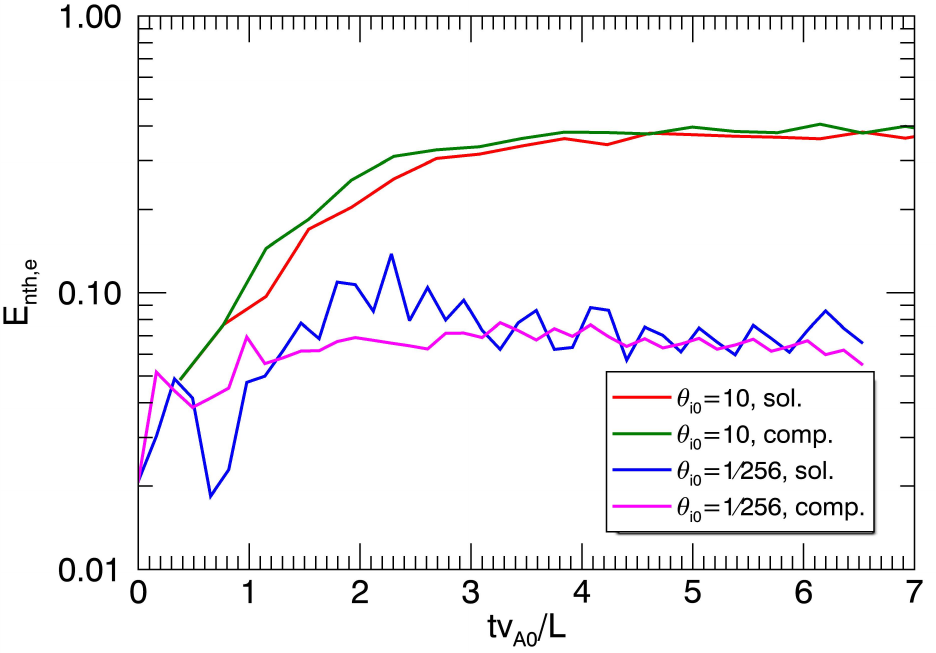}
      \centering
   \caption{\label{fig:enonthermal} Top panel: Evolution of the fraction of overall ion kinetic energy in the nonthermal population for the fiducial simulations, as indicated in the legend. Bottom panel: similar for electrons.}
 \end{figure}
 
To better characterize the significance of the nonthermal particle populations, we next measure the fraction of overall energy and particles in the nonthermal part of the distributions, following the procedure previously used in~\cite{zhdankin_etal_2019}. In short, we fit to the measured distribution with a Maxwell-J\"{u}ttner distribution that has the same peak value, and define this as the thermal part of the distribution, along with any excess of the measured distribution over the fitted distribution at energies below the peak location. We then define the nonthermal part to be the difference between the measured distribution and the thermal part. We denote $E_{{\rm nth},s}$ as the fraction of kinetic energy contained in the nonthermal part of $f_s$. Likewise, we denote $N_{{\rm nth},s}$ as the fraction of particles in the nonthermal part of $f_s$.
 
In Fig.~\ref{fig:enonthermal}, we show the nonthermal energy fraction $E_{{\rm nth},s}$ as a function of time in the four fiducial simulations; ions (top panel) and electrons (bottom panel) are shown separately. We find that at late times, $\sim 40\%$ of the ion energy is in the nonthermal population for all cases except for the solenoidal $\theta_{i0} = 1/256$ case, which has moderately less nonthermal ion energy ($\lesssim 30\%$). For the electrons, the fraction of energy in the nonthermal population is $\sim 40\%$ for the $\theta_{i0} = 10$ simulations (same as for ions, as expected from relativistic mass symmetry) and $\sim 6\%$ for the $\theta_{i0}=1/256$ simulations. We note that these nonthermal fractions appear to be converged with system size for the $\theta_{i0} = 10$ simulations, but may not yet be converged for the $\theta_{i0} = 1/256$ simulations (not shown), indicating that larger simulations will be necessary in the future.

    \begin{figure}
\includegraphics[width=0.95\columnwidth]{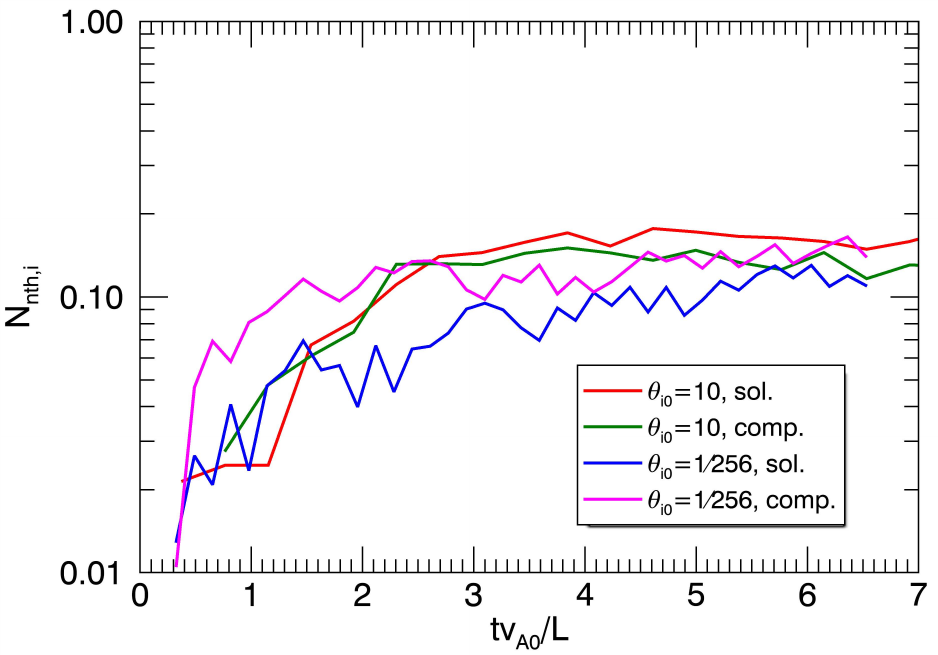}
\includegraphics[width=0.95\columnwidth]{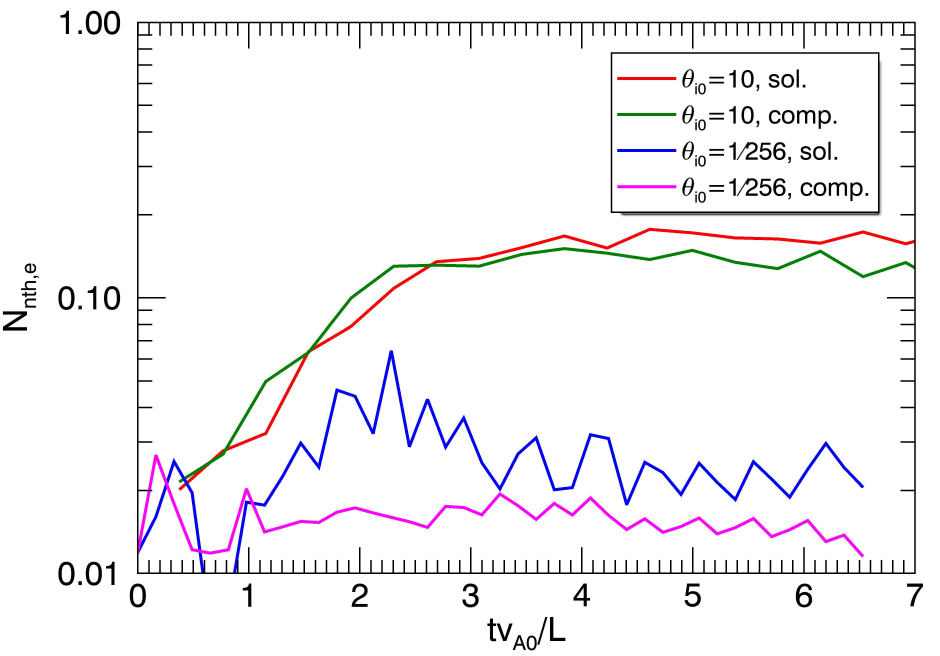}
      \centering
   \caption{\label{fig:nnonthermal} Top panel: Evolution of the fraction of ions in the nonthermal population for the fiducial simulations, as indicated in the legend. Bottom panel: similar for electrons.}
 \end{figure}
 
We show similar results for the nonthermal number fraction $N_{{\rm nth},s}$ for the four fiducial cases in Fig.~\ref{fig:nnonthermal}. Like for $E_{{\rm nth},s}$, all cases show a similar fraction of ions in the nonthermal populations (ranging between $10-20\%$) while the fraction of nonthermal electrons is reduced for the semirelativistic cases ($\sim 15\%$ for $\theta_{i0} = 10$ and $\sim 1-3 \%$ for $\theta_{i0} = 1/256$).
 
 Since the power-law tails in the measured distributions have an index steeper than $-2$, $E_{{\rm nth},s}$ and $N_{{\rm nth},s}$ are strongly weighted toward the low-energy end of the nonthermal population. We thus anticipate them to be fair representations of how many particles are injected into the nonthermal tail, rather than how hard the tail is. In this vein, these results suggest that the ions are injected into the nonthermal population with a similar efficiency for all cases, while the electron injection is diminished for low $\theta_i$. This is reasonable because for the semirelativistic case, typical electron gyroradii are much smaller than the spectral break from the MHD inertial range, which occurs at the characteristic ion gyroradius. Thus, for low $\theta_i$, it becomes challenging to inject electrons to energies where they may interact with inertial-range fluctuations. Regardless, once electrons do reach these scales, they can be efficiently accelerated to high energies.
 
  \subsection{Anisotropy of particle distributions}
 
    \begin{figure}
\includegraphics[width=0.95\columnwidth]{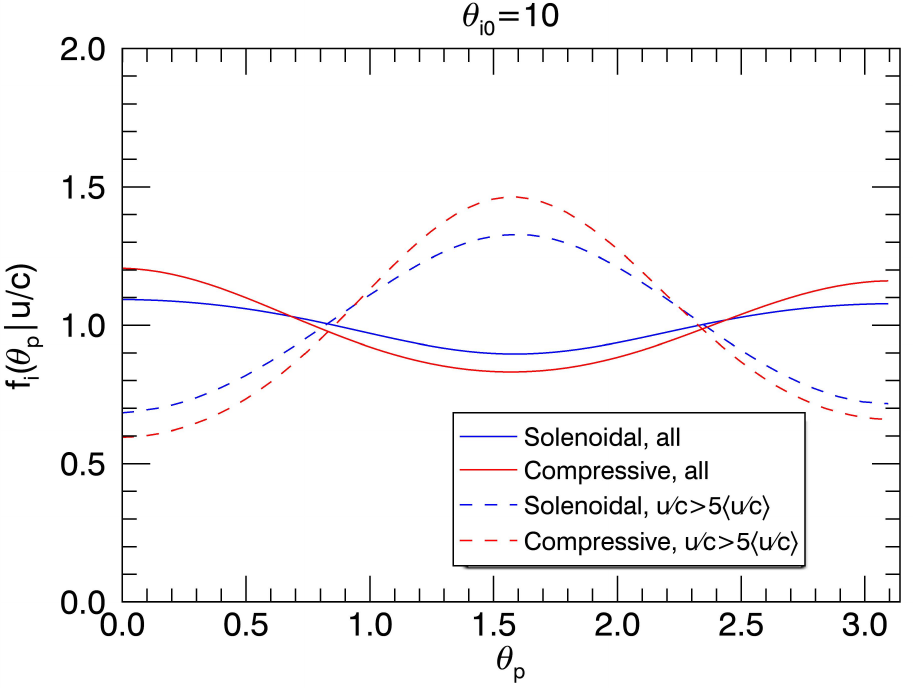}
\includegraphics[width=0.95\columnwidth]{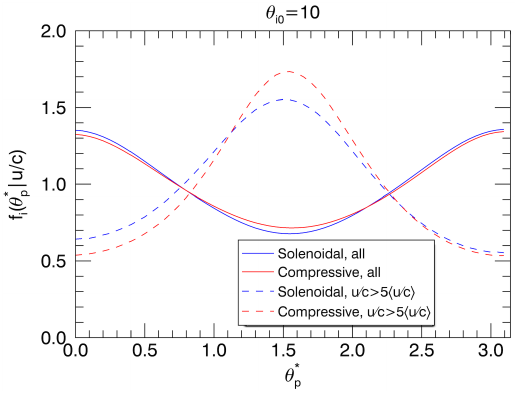}
      \centering
   \caption{\label{fig:ang_thetai10} Top panel: Distribution of pitch angles $\theta_p$ (relative to $\boldsymbol{B}_0$) for all particles (solid lines) and for high-energy particles with $u/c > 5 \langle u/c \rangle$ (dashed lines) for solenoidal (blue) and compressive (red) driving in the relativistic simulations ($\theta_{i0} = 10$). Bottom panel: Similar but for pitch angles $\theta^*_p$ relative to the local magnetic field $\boldsymbol{B}$, from the smaller ($384^3$-cell) simulations.}
 \end{figure}
 
 We conclude our presentation of results by commenting on the anisotropy of the particle momentum distributions. To do this, we first consider the distribution of particle pitch angles with respect to the global mean field, $\theta_p = \cos^{-1}(\hat{\boldsymbol{p}} \cdot \hat{\boldsymbol{z}})$. In the top panel of Fig.~\ref{fig:ang_thetai10}, we show the distribution of pitch angles $f_i(\theta_p)$ for all ions in the fiducial $\theta_{i0} = 10$ cases (solid lines), averaged from $t v_{A0}/L = 4.8$ to $t v_{A0}/L = 6.7$. We find that the distributions are moderately anisotropic in that particles have a slight tendency to move along $\pm \boldsymbol{B}_0$, rather than perpendicular to it. However, when we focus only on high-energy particles (with large $u/c$), the anisotropy changes to favor large pitch angles ($\theta_p \sim \pi/2$). This is demonstrated by the dashed lines of Fig.~\ref{fig:ang_thetai10}, which show the distributions conditioned on $u/c > 5 \langle u/c \rangle \approx 335$, where the brackets indicate an average over the same timescale. The anisotropy is qualitatively similar for both solenoidal and compressive driving, but is somewhat more pronounced in the compressive case.
 
The anisotropy is enhanced when the pitch angle is instead measured relative to the local magnetic field $\boldsymbol{B}$ rather than $\boldsymbol{B}_0$, which we denote by $\theta^*_p = \cos^{-1}(\hat{\boldsymbol{p}} \cdot \hat{\boldsymbol{B}})$. We show the distribution of $\theta^*_p$ from the smaller ($384^3$-cell) $\theta_{i0}=10$ simulations in the bottom panel of Fig.~\ref{fig:ang_thetai10}; this local quantity was not measured for the fiducial simulations but is expected to be similar. Interestingly, the distribution of $\theta^*_p$ is nearly identical for both types of driving when all particles are considered, with a significant preference for propagation along $\boldsymbol{B}$. At high energies, particles mainly propagate perpendicular to $\boldsymbol{B}$ and the anisotropy is slightly stronger with compressive driving.
 
 Ref.~\cite{comisso_sironi_2019} found a similar energy-dependent anisotropy in PIC simulations of decaying turbulence at high $\sigma$, which was attributed to a two-stage acceleration process involving particle injection by magnetic reconnection (through parallel electric fields) and diffusive acceleration by turbulence (through perpendicular electric fields); see \cite{comisso_etal_2020} for further discussion and implications. Our new results indicate that this process is insensitive to the driving mechanism.
 
 The semirelativistic ($\theta_{i0} = 1/256$) cases (not shown) exhibit a distribution of pitch angles that is much closer to isotropy than the relativistic case, with the shape being more strongly affected by turbulence variability rather than systematic structure.
 
 \section{Conclusions} \label{sec:conclusions}

In this paper, we investigated PIC simulations of kinetic turbulence with solenoidal and compressive external driving mechanisms. Motivated by high-energy astrophysical systems, we focused on relativistic ($\theta_e \gg 1$, $\theta_i \gg 1$, $\sigma \sim 1$) and semirelativistic ($\theta_e \gg 1$, $\theta_i \ll 1$, $\sigma \ll 1$) plasmas. We described similarities and differences in the turbulence properties, electron-ion energy partition, and nonthermal particle acceleration with the two methods of driving. The main results of our study are the following:
\begin{enumerate}
\item As expected, compressive driving triggers much stronger density fluctuations than solenoidal driving. These density fluctuations are mainly concentrated in clumps at large scales, and approximately obey an isothermal equation of state.
\item Regardless of the presence or absence of large-scale density fluctuations, the spectrum of turbulence looks very similar with both drives. In the MHD inertial range, the magnetic energy spectrum and structure functions are consistent with standard Goldreich-Sridhar phenomenology (as demonstrated by the relativistic simulations), suggesting that any fast-mode cascade is sub-dominant. In the kinetic range below the ion gyroscale (captured by the semirelativistic cases), there is insufficient scale separation to achieve a power-law magnetic energy spectrum, but the shape of the spectrum is similar for both drives.
\item In the semirelativistic regime (where mass symmetry between electrons and ions is broken), ions are heated preferentially over electrons for both driving mechanisms. However, the solenoidal driving causes a higher electron heating fraction (fit empirically by $\Delta E_{{\rm int},e}/\Delta E_{{\rm int},i} \sim \rho_e^{1/3}/\rho_i^{1/3}$) than the compressive case (fit by $\Delta E_{{\rm int},e}/\Delta E_{{\rm int},i} \sim \rho_e^{2/3}/\rho_i^{2/3}$). The extra ion energization in the compressive case comes from perpendicular electric fields at large scales (wavenumbers $k \rho_i \lesssim 1$).
\item In the relativistic regime, efficient nonthermal particle acceleration occurs for both solenoidal and compressive driving. Although both drives lead to similar power-law particle energy distributions (with index near $-3.0$), the nonthermal populations form much faster for compressive driving (with an implied acceleration timescale $\sim 4$ times shorter). A similar number of particles are injected into the nonthermal population for both drives.
\item In the semirelativistic regime, significant nonthermal particle acceleration occurs only for the compressively driven turbulence. The power law for this case is softer than the relativistic case, with an index near $-4.2$ for both electrons and ions. The solenoidal simulation exhibits a persistent quasi-thermal distribution. Although a significant amount of ions are injected to the nonthermal population, relatively few electrons are.
\end{enumerate}

These conclusions are broadly consistent with theoretical expectations discussed throughout the literature. Compressive fluctuations in a collisionless plasma are predicted to be strongly Landau damped, even at large scales \citep{barnes_1966}, which would explain our non-observation of an isotropic fast-mode cascade usually seen in MHD simulations \citep{cho_lazarian_2002, cho_lazarian_2003}. As shown by recent works operating in the non-relativistic gyrokinetic limit, this damping causes primarily ion heating in $\beta \lesssim 1$ plasmas \citep{schekochihin_etal_2019, kawazura_etal_2020}. Our work supports this theoretical picture in the relativistic and semirelativistic regimes, without relying on the gyrokinetic approximation.

Importantly, our work also indicates that a large fraction of the damped compressive energy goes into nonthermal particle acceleration. Several previous theoretical works in the MHD framework predicted that stochastic particle acceleration will be more efficient for a fast-mode cascade than an Alfv\'{e}nic cascade \citep[][]{schlickeiser_miller_1998, yan_lazarian_2002, yan_lazarian_2004, lazarian_etal_2012, demidem_etal_2020}. However, these studies did not account for collisionless damping of fast modes at large scales. Our work indicates that nonthermal particle acceleration occurs despite Landau damping, which may naively be expected to result in thermal heating.

Extrapolating to lower temperatures, our work suggests that a compressive component may be essential for obtaining turbulent particle acceleration in the fully non-relativistic regime ($\theta_i \ll 1$, $\theta_e \ll 1$, and $\sigma \ll 1$). Further numerical work on this regime is warranted.

All simulations had fixed initial plasma beta of $\beta_0 = 1$; more work is necessary to understand the effect of varying this parameter. We point out that qualitative differences may be expected the high-beta regime due to the effect of plasma microinstabilities such as the firehose and mirror instabilities, which arise from pressure anisotropy \citep{kunz_etal_2014}. In particular, it was recently suggested that these microinstabilities may impede Landau damping of large-scale compressive fluctuations \citep{kunz_etal_2020}, which would have consequences on the results described here.

Another parameter that may be varied in future studies is the turbulent Mach number $M = \delta {\mathcal V}_{\rm rms}/c_s$, which is related to the amplitude of driven fluctuations, fixed to be $\delta B_{\rm rms}/B_0 \sim 1$ in our study. Shocks were absent from our simulations because we focused on $M \lesssim 1$. However, for $M \gg 1$ (supersonic turbulence, which corresponds to super-Alfv\'{e}nic turbulence when $\beta \approx 1$), shocks are expected to become a major channel of energy dissipation \citep[see, e.g.,][]{stone_etal_1998, lemaster_stone_2009}. PIC simulations previously demonstrated that collisionless magnetized shocks may cause preferential ion heating \citep{tran_sironi_2020} and efficient nonthermal particle acceleration \citep{spitkovsky_2008, sironi_spitkovsky_2010}.

The results presented in this paper have implications for high-energy astrophysical systems. Examples of systems in the relativistic regime ($\theta_i \gg 1$, $\sigma \gtrsim 1$) include jets from active galactic nuclei, black-hole X-ray binaries, and gamma-ray bursts. Our results indicate that particle acceleration should be ubiquitous in such systems, regardless of how turbulence is driven. In situations where the turbulence is transient (and of modest $\sigma$), the faster acceleration timescale associated with compressively driven turbulence may make it a more viable explanation for observed nonthermal spectra. Recent PIC simulations indicated that the kink instability \citep{alves_etal_2018, alves_etal_2019, davelaar_etal_2020} and Kelvin-Helmholtz instability \citep{sironi_etal_2021} may trigger particle acceleration in jets. Theoretical work also suggested that internal shocks may drive turbulence in the gamma-ray burst scenario \citep[e.g.,][]{zhang_yan_2010}. In all of these situations, a significant compressive component may arise due to global inhomogeneities in the system.

The semirelativistic regime ($\theta_e \gg 1$, $\theta_i \ll 1$, $\sigma \lesssim 1$) is applicable to the inner regions of radiatively inefficient accretion flows onto supermassive black holes, such as the Event Horizon Telescope (EHT) targets of M87* and Sgr A* \citep{eht_2019a, eht_2019b}. Understanding nonthermal particle distributions and the electron-to-ion heating ratio is important for interpreting radiation spectra as well as its polarization \citep{eht_2021, eht_2021b}. General relativistic MHD (GRMHD) simulations were widely applied to connect the EHT observations with plasma physics theory \citep[e.g.,][]{ricarte_dexter_2015, porth_etal_2019, ripperda_etal_2020}, but are incapable of providing a self-consistent representation of the kinetic plasma properties. Thus, for producing maps of the observable emission, a thermal electron population with prescribed temperature is typically assumed. In particular, the EHT models applied a heating prescription parameterized by $T_i/T_e = R_{\rm high} \beta^2/(1+\beta^2) + 1/(1+\beta^2)$, where $\beta$ is the local plasma beta and $R_{\rm high}$ is a free parameter determining the cap on the temperature ratio \citep{moscibrodzka_etal_2016, eht_2019b}. Our empirical formula for the electron-to-ion heating ratio may be considered as an alternative scenario, given the electron and ion temperatures (although one must be cautious about the uncertainty in the $\beta$ dependence; note that \cite{zhdankin_etal_2019} did not observe a strong dependence of the heating ratio on $\beta$ in PIC simulations).

Recent GRMHD simulations of accretion flows incorporated the dynamical evolution of nonthermal electron populations \citep{ball_etal_2016, chael_etal_2017} as well as two-temperature plasmas \citep{ryan_etal_2017} using idealized prescriptions. PIC simulations may be applied to improve these prescriptions. In the longer term, kinetic shearing box simulations may provide a better opportunity to understand particle acceleration and heating in accreting systems \citep[e.g.,][]{riquelme_etal_2012, hoshino_2013, hoshino_2015, kunz_etal_2016, inchingolo_etal_2018}. We note, however, that global effects such as spiral shocks may drive a significant compressive component of turbulence, beyond the effects captured in a shearing box \citep[e.g.,][]{ju_etal_2016}. Based on the results in our paper, we emphasize that understanding compressive fluctuations driven at global scales will be an important ingredient in understanding nonthermal particle acceleration in accretion flows. 

\acknowledgements

The author thanks Matthew Kunz, Dmitri Uzdensky, and Martin Lemoine for helpful discussions, and the anonymous referee for suggestions that improved the paper. The author acknowledges support for this work from NASA through the NASA Hubble Fellowship grant \#HST-HF2-51426.001-A awarded by the Space Telescope Science Institute, which is operated by the Association of Universities for Research in Astronomy, Inc., for NASA, under contract NAS5-26555. This work used the Extreme Science and Engineering Discovery Environment (XSEDE), which is supported by National Science Foundation grant number ACI-1548562. This work used the XSEDE supercomputer Stampede2 at the Texas Advanced Computer Center (TACC) through allocation TG-PHY160032 \citep{xsede}.


\begin{thebibliography}{}
\makeatletter
\relax
\def\mn@urlcharsother{\let\do\@makeother \do\$\do\&\do\#\do\^\do\_\do\%\do\~}
\def\mn@doi{\begingroup\mn@urlcharsother \@ifnextchar [ {\mn@doi@}
  {\mn@doi@[]}}
\def\mn@doi@[#1]#2{\def\@tempa{#1}\ifx\@tempa\@empty \href
  {http://dx.doi.org/#2} {doi:#2}\else \href {http://dx.doi.org/#2} {#1}\fi
  \endgroup}
\def\mn@eprint#1#2{\mn@eprint@#1:#2::\@nil}
\def\mn@eprint@arXiv#1{\href {http://arxiv.org/abs/#1} {{\tt arXiv:#1}}}
\def\mn@eprint@dblp#1{\href {http://dblp.uni-trier.de/rec/bibtex/#1.xml}
  {dblp:#1}}
\def\mn@eprint@#1:#2:#3:#4\@nil{\def\@tempa {#1}\def\@tempb {#2}\def\@tempc
  {#3}\ifx \@tempc \@empty \let \@tempc \@tempb \let \@tempb \@tempa \fi \ifx
  \@tempb \@empty \def\@tempb {arXiv}\fi \@ifundefined
  {mn@eprint@\@tempb}{\@tempb:\@tempc}{\expandafter \expandafter \csname
  mn@eprint@\@tempb\endcsname \expandafter{\@tempc}}}

\bibitem[\protect\citeauthoryear{Akiyama et~al.,}{Akiyama
  et~al.}{2019a}]{eht_2019a}
Akiyama K.,  et~al., 2019a, The Astrophysical Journal Letters, 875, L4

\bibitem[\protect\citeauthoryear{Akiyama et~al.,}{Akiyama
  et~al.}{2019b}]{eht_2019b}
Akiyama K.,  et~al., 2019b, The Astrophysical Journal Letters, 875, L5

\bibitem[\protect\citeauthoryear{Akiyama et~al.,}{Akiyama
  et~al.}{2021a}]{eht_2021b}
Akiyama K.,  et~al., 2021a, The Astrophysical Journal Letters, 910, L12

\bibitem[\protect\citeauthoryear{Akiyama et~al.,}{Akiyama
  et~al.}{2021b}]{eht_2021}
Akiyama K.,  et~al., 2021b, The Astrophysical Journal Letters, 910, L13

\bibitem[\protect\citeauthoryear{Alexandrova, Saur, Lacombe, Mangeney,
  Mitchell, Schwartz  \& Robert}{Alexandrova
  et~al.}{2009}]{alexandrova_etal_2009}
Alexandrova O.,  Saur J.,  Lacombe C.,  Mangeney A.,  Mitchell J.,  Schwartz
  S.~J.,   Robert P.,  2009, Physical Review Letters, 103, 165003

\bibitem[\protect\citeauthoryear{Alves, Zrake  \& Fiuza}{Alves
  et~al.}{2018}]{alves_etal_2018}
Alves E.~P.,  Zrake J.,   Fiuza F.,  2018, Physical review letters, 121, 245101

\bibitem[\protect\citeauthoryear{Alves, Zrake  \& Fiuza}{Alves
  et~al.}{2019}]{alves_etal_2019}
Alves E.~P.,  Zrake J.,   Fiuza F.,  2019, Physics of Plasmas, 26, 072105

\bibitem[\protect\citeauthoryear{Arzamasskiy, Kunz, Chandran  \&
  Quataert}{Arzamasskiy et~al.}{2019}]{arzamasskiy_etal_2019}
Arzamasskiy L.,  Kunz M.~W.,  Chandran B.~D.,   Quataert E.,  2019, The
  Astrophysical Journal, 879, 53

\bibitem[\protect\citeauthoryear{Ball, {\"O}zel, Psaltis  \& Chan}{Ball
  et~al.}{2016}]{ball_etal_2016}
Ball D.,  {\"O}zel F.,  Psaltis D.,   Chan C.-k.,  2016, The Astrophysical
  Journal, 826, 77

\bibitem[\protect\citeauthoryear{Ball, Sironi  \& {\"O}zel}{Ball
  et~al.}{2018}]{ball_etal_2018}
Ball D.,  Sironi L.,   {\"O}zel F.,  2018, The Astrophysical Journal, 862, 80

\bibitem[\protect\citeauthoryear{Barnes}{Barnes}{1966}]{barnes_1966}
Barnes A.,  1966, The Physics of Fluids, 9, 1483

\bibitem[\protect\citeauthoryear{Beloborodov \& M{\'e}sz{\'a}ros}{Beloborodov
  \& M{\'e}sz{\'a}ros}{2017}]{beloborodov_meszaros_2017}
Beloborodov A.,  M{\'e}sz{\'a}ros P.,  2017, Space Science Reviews, 207, 87

\bibitem[\protect\citeauthoryear{Blandford, Simeon  \& Yuan}{Blandford
  et~al.}{2014}]{blandford_etal_2014}
Blandford R.,  Simeon P.,   Yuan Y.,  2014, Nuclear Physics B-proceedings
  supplements, 256, 9

\bibitem[\protect\citeauthoryear{B{\"o}ttcher}{B{\"o}ttcher}{2007}]{bottcher_2007}
B{\"o}ttcher M.,  2007, in , The Multi-Messenger Approach to High-Energy
  Gamma-Ray Sources.
Springer, pp 95--104

\bibitem[\protect\citeauthoryear{Brunetti \& Lazarian}{Brunetti \&
  Lazarian}{2007}]{brunetti_lazarian_2007}
Brunetti G.,  Lazarian A.,  2007, Monthly Notices of the Royal Astronomical
  Society, 378, 245

\bibitem[\protect\citeauthoryear{Cerri, Franci, Califano, Landi  \&
  Hellinger}{Cerri et~al.}{2017}]{cerri_etal_2017b}
Cerri S.,  Franci L.,  Califano F.,  Landi S.,   Hellinger P.,  2017, Journal
  of Plasma Physics, 83

\bibitem[\protect\citeauthoryear{Cerri, Gro{\v{s}}elj  \& Franci}{Cerri
  et~al.}{2019}]{cerri_etal_2019}
Cerri S.~S.,  Gro{\v{s}}elj D.,   Franci L.,  2019, Frontiers in Astronomy and
  Space Sciences, 6, 64

\bibitem[\protect\citeauthoryear{Cerri, Arzamasskiy  \& Kunz}{Cerri
  et~al.}{2021}]{cerri_etal_2021}
Cerri S.~S.,  Arzamasskiy L.,   Kunz M.~W.,  2021, arXiv preprint
  arXiv:2102.09654

\bibitem[\protect\citeauthoryear{Cerutti, Werner, Uzdensky  \&
  Begelman}{Cerutti et~al.}{2013}]{cerutti_etal_2013}
Cerutti B.,  Werner G.~R.,  Uzdensky D.~A.,   Begelman M.~C.,  2013, The
  Astrophysical Journal, 770, 147

\bibitem[\protect\citeauthoryear{Chael, Narayan  \& Sadowski}{Chael
  et~al.}{2017}]{chael_etal_2017}
Chael A.~A.,  Narayan R.,   Sadowski A.,  2017, Monthly Notices of the Royal
  Astronomical Society, 470, 2367

\bibitem[\protect\citeauthoryear{Chandran}{Chandran}{2000}]{chandran_2000}
Chandran B.~D.,  2000, Physical Review Letters, 85, 4656

\bibitem[\protect\citeauthoryear{Chew, Goldberger  \& Low}{Chew
  et~al.}{1956}]{cgl_1956}
Chew G.,  Goldberger M.,   Low F.,  1956, Proceedings of the Royal Society of
  London. Series A. Mathematical and Physical Sciences, 236, 112

\bibitem[\protect\citeauthoryear{Cho}{Cho}{2005}]{cho_2005}
Cho J.,  2005, The Astrophysical Journal, 621, 324

\bibitem[\protect\citeauthoryear{Cho \& Lazarian}{Cho \&
  Lazarian}{2002}]{cho_lazarian_2002}
Cho J.,  Lazarian A.,  2002, Physical Review Letters, 88, 245001

\bibitem[\protect\citeauthoryear{Cho \& Lazarian}{Cho \&
  Lazarian}{2003}]{cho_lazarian_2003}
Cho J.,  Lazarian A.,  2003, Monthly Notices of the Royal Astronomical Society,
  345, 325

\bibitem[\protect\citeauthoryear{Cho \& Vishniac}{Cho \&
  Vishniac}{2000}]{cho_vishniac_2000}
Cho J.,  Vishniac E.~T.,  2000, The Astrophysical Journal, 539, 273

\bibitem[\protect\citeauthoryear{Chou \& Hau}{Chou \&
  Hau}{2004}]{chou_hau_2004}
Chou M.,  Hau L.-N.,  2004, The Astrophysical Journal, 611, 1200

\bibitem[\protect\citeauthoryear{Comisso \& Sironi}{Comisso \&
  Sironi}{2018}]{comisso_sironi_2018}
Comisso L.,  Sironi L.,  2018, Physical review letters, 121, 255101

\bibitem[\protect\citeauthoryear{Comisso \& Sironi}{Comisso \&
  Sironi}{2019}]{comisso_sironi_2019}
Comisso L.,  Sironi L.,  2019, The Astrophysical Journal, 886, 122

\bibitem[\protect\citeauthoryear{Comisso, Sobacchi  \& Sironi}{Comisso
  et~al.}{2020}]{comisso_etal_2020}
Comisso L.,  Sobacchi E.,   Sironi L.,  2020, The Astrophysical Journal
  Letters, 895, L40

\bibitem[\protect\citeauthoryear{Dahlin, Drake  \& Swisdak}{Dahlin
  et~al.}{2016}]{dahlin_etal_2016}
Dahlin J.,  Drake J.,   Swisdak M.,  2016, Physics of Plasmas, 23, 120704

\bibitem[\protect\citeauthoryear{Davelaar, Philippov, Bromberg  \&
  Singh}{Davelaar et~al.}{2020}]{davelaar_etal_2020}
Davelaar J.,  Philippov A.~A.,  Bromberg O.,   Singh C.~B.,  2020, The
  Astrophysical Journal Letters, 896, L31

\bibitem[\protect\citeauthoryear{Demidem, Lemoine  \& Casse}{Demidem
  et~al.}{2020}]{demidem_etal_2020}
Demidem C.,  Lemoine M.,   Casse F.,  2020, Physical Review D, 102, 023003

\bibitem[\protect\citeauthoryear{Dong, Wang, Huang, Comisso  \&
  Bhattacharjee}{Dong et~al.}{2018}]{dong_etal_2018}
Dong C.,  Wang L.,  Huang Y.-M.,  Comisso L.,   Bhattacharjee A.,  2018,
  Physical review letters, 121, 165101

\bibitem[\protect\citeauthoryear{Federrath}{Federrath}{2016}]{federrath_2016}
Federrath C.,  2016, Journal of Plasma Physics, 82

\bibitem[\protect\citeauthoryear{Federrath, Klessen  \& Schmidt}{Federrath
  et~al.}{2008}]{federrath_etal_2008}
Federrath C.,  Klessen R.~S.,   Schmidt W.,  2008, The Astrophysical Journal
  Letters, 688, L79

\bibitem[\protect\citeauthoryear{Federrath, Chabrier, Schober, Banerjee,
  Klessen  \& Schleicher}{Federrath et~al.}{2011}]{federrath_etal_2011}
Federrath C.,  Chabrier G.,  Schober J.,  Banerjee R.,  Klessen R.~S.,
  Schleicher D.~R.,  2011, Physical Review Letters, 107, 114504

\bibitem[\protect\citeauthoryear{Fermi}{Fermi}{1949}]{fermi_1949}
Fermi E.,  1949, Physical Review, 75, 1169

\bibitem[\protect\citeauthoryear{Gaensler \& Slane}{Gaensler \&
  Slane}{2006}]{gaensler_slane_2006}
Gaensler B.~M.,  Slane P.~O.,  2006, Annu. Rev. Astron. Astrophys., 44, 17

\bibitem[\protect\citeauthoryear{Gedalin}{Gedalin}{1991}]{gedalin_1991}
Gedalin M.,  1991, Physics of Fluids B: Plasma Physics, 3, 1871

\bibitem[\protect\citeauthoryear{Goldreich \& Sridhar}{Goldreich \&
  Sridhar}{1995}]{goldreich_sridhar_1995}
Goldreich P.,  Sridhar S.,  1995, The Astrophysical Journal, 438, 763

\bibitem[\protect\citeauthoryear{Gruzinov}{Gruzinov}{1998}]{gruzinov_1998}
Gruzinov A.~V.,  1998, The Astrophysical Journal, 501, 787

\bibitem[\protect\citeauthoryear{Hopkins}{Hopkins}{2013}]{hopkins_2013}
Hopkins P.~F.,  2013, Monthly Notices of the Royal Astronomical Society, 430,
  1880

\bibitem[\protect\citeauthoryear{Hoshino}{Hoshino}{2013}]{hoshino_2013}
Hoshino M.,  2013, The Astrophysical Journal, 773, 118

\bibitem[\protect\citeauthoryear{Hoshino}{Hoshino}{2015}]{hoshino_2015}
Hoshino M.,  2015, Physical Review Letters, 114, 061101

\bibitem[\protect\citeauthoryear{Howes}{Howes}{2010}]{howes_2010}
Howes G.~G.,  2010, Monthly Notices of the Royal Astronomical Society: Letters,
  409, L104

\bibitem[\protect\citeauthoryear{Inchingolo, Grismayer, Loureiro, Fonseca  \&
  Silva}{Inchingolo et~al.}{2018}]{inchingolo_etal_2018}
Inchingolo G.,  Grismayer T.,  Loureiro N.~F.,  Fonseca R.~A.,   Silva L.~O.,
  2018, The Astrophysical Journal, 859, 149

\bibitem[\protect\citeauthoryear{Isliker, Pisokas, Vlahos  \&
  Anastasiadis}{Isliker et~al.}{2017}]{isliker_etal_2017}
Isliker H.,  Pisokas T.,  Vlahos L.,   Anastasiadis A.,  2017, The
  Astrophysical Journal, 849, 35

\bibitem[\protect\citeauthoryear{Ju, Stone  \& Zhu}{Ju
  et~al.}{2016}]{ju_etal_2016}
Ju W.,  Stone J.~M.,   Zhu Z.,  2016, The Astrophysical Journal, 823, 81

\bibitem[\protect\citeauthoryear{Kawazura, Barnes  \& Schekochihin}{Kawazura
  et~al.}{2019}]{kawazura_etal_2019}
Kawazura Y.,  Barnes M.,   Schekochihin A.~A.,  2019, Proceedings of the
  National Academy of Sciences, 116, 771

\bibitem[\protect\citeauthoryear{Kawazura, Schekochihin, Barnes, TenBarge,
  Tong, Klein  \& Dorland}{Kawazura et~al.}{2020}]{kawazura_etal_2020}
Kawazura Y.,  Schekochihin A.,  Barnes M.,  TenBarge J.,  Tong Y.,  Klein K.,
  Dorland W.,  2020, Physical Review X, 10, 041050

\bibitem[\protect\citeauthoryear{Klein \& Howes}{Klein \&
  Howes}{2016}]{klein_howes_2016}
Klein K.~G.,  Howes G.~G.,  2016, The Astrophysical Journal Letters, 826, L30

\bibitem[\protect\citeauthoryear{Klein, Howes, TenBarge  \& Valentini}{Klein
  et~al.}{2020}]{klein_etal_2020}
Klein K.~G.,  Howes G.~G.,  TenBarge J.~M.,   Valentini F.,  2020, Journal of
  Plasma Physics, 86

\bibitem[\protect\citeauthoryear{Kulsrud \& Ferrari}{Kulsrud \&
  Ferrari}{1971}]{kulsrud_ferrari_1971}
Kulsrud R.~M.,  Ferrari A.,  1971, Astrophysics and Space Science, 12, 302

\bibitem[\protect\citeauthoryear{Kunz, Schekochihin  \& Stone}{Kunz
  et~al.}{2014}]{kunz_etal_2014}
Kunz M.~W.,  Schekochihin A.~A.,   Stone J.~M.,  2014, Physical Review Letters,
  112, 205003

\bibitem[\protect\citeauthoryear{Kunz, Stone  \& Quataert}{Kunz
  et~al.}{2016}]{kunz_etal_2016}
Kunz M.~W.,  Stone J.~M.,   Quataert E.,  2016, Physical Review Letters, 117,
  235101

\bibitem[\protect\citeauthoryear{Kunz, Squire, Schekochihin  \& Quataert}{Kunz
  et~al.}{2020}]{kunz_etal_2020}
Kunz M.,  Squire J.,  Schekochihin A.,   Quataert E.,  2020, Journal of Plasma
  Physics, 86

\bibitem[\protect\citeauthoryear{Lazarian, Vlahos, Kowal, Yan, Beresnyak  \&
  Dal~Pino}{Lazarian et~al.}{2012}]{lazarian_etal_2012}
Lazarian A.,  Vlahos L.,  Kowal G.,  Yan H.,  Beresnyak A.,   Dal~Pino E.
  d.~G.,  2012, Space science reviews, 173, 557

\bibitem[\protect\citeauthoryear{Lemaster \& Stone}{Lemaster \&
  Stone}{2009}]{lemaster_stone_2009}
Lemaster M.~N.,  Stone J.~M.,  2009, The Astrophysical Journal, 691, 1092

\bibitem[\protect\citeauthoryear{Lemoine}{Lemoine}{2021}]{lemoine_2021}
Lemoine M.,  2021, arXiv preprint arXiv:2104.08199

\bibitem[\protect\citeauthoryear{Lemoine \& Malkov}{Lemoine \&
  Malkov}{2020}]{lemoine_malkov_2020}
Lemoine M.,  Malkov M.~A.,  2020, Monthly Notices of the Royal Astronomical
  Society, 499, 4972

\bibitem[\protect\citeauthoryear{Ley, Riquelme, Sironi, Verscharen  \&
  Sandoval}{Ley et~al.}{2019}]{ley_etal_2019}
Ley F.,  Riquelme M.,  Sironi L.,  Verscharen D.,   Sandoval A.,  2019, The
  Astrophysical Journal, 880, 100

\bibitem[\protect\citeauthoryear{Li, Howes, Klein, Liu  \& TenBarge}{Li
  et~al.}{2019}]{li_etal_2019}
Li T.~C.,  Howes G.~G.,  Klein K.~G.,  Liu Y.-H.,   TenBarge J.~M.,  2019,
  Journal of Plasma Physics, 85

\bibitem[\protect\citeauthoryear{Lithwick \& Goldreich}{Lithwick \&
  Goldreich}{2001}]{lithwick_goldreich_2001}
Lithwick Y.,  Goldreich P.,  2001, The Astrophysical Journal, 562, 279

\bibitem[\protect\citeauthoryear{Loureiro \& Boldyrev}{Loureiro \&
  Boldyrev}{2017a}]{loureiro_boldyrev_2017}
Loureiro N.~F.,  Boldyrev S.,  2017a, Physical review letters, 118, 245101

\bibitem[\protect\citeauthoryear{Loureiro \& Boldyrev}{Loureiro \&
  Boldyrev}{2017b}]{loureiro_boldyrev_2017b}
Loureiro N.~F.,  Boldyrev S.,  2017b, The Astrophysical Journal, 850, 182

\bibitem[\protect\citeauthoryear{Lynn, Quataert, Chandran  \& Parrish}{Lynn
  et~al.}{2014}]{lynn_etal_2014}
Lynn J.~W.,  Quataert E.,  Chandran B.~D.,   Parrish I.~J.,  2014, The
  Astrophysical Journal, 791, 71

\bibitem[\protect\citeauthoryear{Makwana \& Yan}{Makwana \&
  Yan}{2020}]{makwana_yan_2020}
Makwana K.,  Yan H.,  2020, Physical Review X, 10, 031021

\bibitem[\protect\citeauthoryear{Mallet, Schekochihin  \& Chandran}{Mallet
  et~al.}{2017a}]{mallet_etal_2017}
Mallet A.,  Schekochihin A.~A.,   Chandran B.~D.,  2017a, Journal of Plasma
  Physics, 83

\bibitem[\protect\citeauthoryear{Mallet, Schekochihin  \& Chandran}{Mallet
  et~al.}{2017b}]{mallet_etal_2017b}
Mallet A.,  Schekochihin A.,   Chandran B.,  2017b, Monthly Notices of the
  Royal Astronomical Society, 468, 4862

\bibitem[\protect\citeauthoryear{Mo{\'s}cibrodzka, Falcke  \&
  Shiokawa}{Mo{\'s}cibrodzka et~al.}{2016}]{moscibrodzka_etal_2016}
Mo{\'s}cibrodzka M.,  Falcke H.,   Shiokawa H.,  2016, Astronomy \&
  Astrophysics, 586, A38

\bibitem[\protect\citeauthoryear{N{\"a}ttil{\"a} \&
  Beloborodov}{N{\"a}ttil{\"a} \& Beloborodov}{2020}]{nattila_etal_2021}
N{\"a}ttil{\"a} J.,  Beloborodov A.~M.,  2020, arXiv preprint arXiv:2012.03043

\bibitem[\protect\citeauthoryear{Passot \& V{\'a}zquez-Semadeni}{Passot \&
  V{\'a}zquez-Semadeni}{1998}]{passot_etal_1998}
Passot T.,  V{\'a}zquez-Semadeni E.,  1998, Physical Review E, 58, 4501

\bibitem[\protect\citeauthoryear{Porth et~al.,}{Porth
  et~al.}{2019}]{porth_etal_2019}
Porth O.,  et~al., 2019, The Astrophysical Journal Supplement Series, 243, 26

\bibitem[\protect\citeauthoryear{Quataert}{Quataert}{1998}]{quataert_1998}
Quataert E.,  1998, The Astrophysical Journal, 500, 978

\bibitem[\protect\citeauthoryear{Quataert \& Gruzinov}{Quataert \&
  Gruzinov}{1999}]{quataert_gruzinov_1999}
Quataert E.,  Gruzinov A.,  1999, The Astrophysical Journal, 520, 248

\bibitem[\protect\citeauthoryear{Ricarte \& Dexter}{Ricarte \&
  Dexter}{2015}]{ricarte_dexter_2015}
Ricarte A.,  Dexter J.,  2015, Monthly Notices of the Royal Astronomical
  Society, 446, 1973

\bibitem[\protect\citeauthoryear{Ripperda, Bacchini  \& Philippov}{Ripperda
  et~al.}{2020}]{ripperda_etal_2020}
Ripperda B.,  Bacchini F.,   Philippov A.~A.,  2020, The Astrophysical Journal,
  900, 100

\bibitem[\protect\citeauthoryear{Riquelme, Quataert, Sharma  \&
  Spitkovsky}{Riquelme et~al.}{2012}]{riquelme_etal_2012}
Riquelme M.~A.,  Quataert E.,  Sharma P.,   Spitkovsky A.,  2012, The
  Astrophysical Journal, 755, 50

\bibitem[\protect\citeauthoryear{Rowan, Sironi  \& Narayan}{Rowan
  et~al.}{2017}]{rowan_sironi_narayan_2017}
Rowan M.~E.,  Sironi L.,   Narayan R.,  2017, The Astrophysical Journal, 850,
  29

\bibitem[\protect\citeauthoryear{Ryan, Ressler, Dolence, Tchekhovskoy, Gammie
  \& Quataert}{Ryan et~al.}{2017}]{ryan_etal_2017}
Ryan B.~R.,  Ressler S.~M.,  Dolence J.~C.,  Tchekhovskoy A.,  Gammie C.,
  Quataert E.,  2017, The Astrophysical Journal Letters, 844, L24

\bibitem[\protect\citeauthoryear{Sahraoui, Goldstein, Robert  \&
  Khotyaintsev}{Sahraoui et~al.}{2009}]{sahraoui_etal_2009}
Sahraoui F.,  Goldstein M.,  Robert P.,   Khotyaintsev Y.~V.,  2009, Physical
  Review Letters, 102, 231102

\bibitem[\protect\citeauthoryear{Schekochihin, Cowley, Dorland, Hammett, Howes,
  Quataert  \& Tatsuno}{Schekochihin et~al.}{2009}]{schekochihin_etal_2009}
Schekochihin A.,  Cowley S.,  Dorland W.,  Hammett G.,  Howes G.,  Quataert E.,
    Tatsuno T.,  2009, The Astrophysical Journal Supplement Series, 182, 310

\bibitem[\protect\citeauthoryear{Schekochihin, Kawazura  \&
  Barnes}{Schekochihin et~al.}{2019}]{schekochihin_etal_2019}
Schekochihin A.,  Kawazura Y.,   Barnes M.,  2019, Journal of Plasma Physics,
  85

\bibitem[\protect\citeauthoryear{Schlickeiser}{Schlickeiser}{1989}]{schlickeiser_1989}
Schlickeiser R.,  1989, The Astrophysical Journal, 336, 243

\bibitem[\protect\citeauthoryear{Schlickeiser \& Miller}{Schlickeiser \&
  Miller}{1998}]{schlickeiser_miller_1998}
Schlickeiser R.,  Miller J.~A.,  1998, The Astrophysical Journal, 492, 352

\bibitem[\protect\citeauthoryear{Shay, Drake  \& Swisdak}{Shay
  et~al.}{2007}]{shay_etal_2007}
Shay M.,  Drake J.,   Swisdak M.,  2007, Physical review letters, 99, 155002

\bibitem[\protect\citeauthoryear{Sironi \& Spitkovsky}{Sironi \&
  Spitkovsky}{2010}]{sironi_spitkovsky_2010}
Sironi L.,  Spitkovsky A.,  2010, The Astrophysical Journal, 726, 75

\bibitem[\protect\citeauthoryear{Sironi, Rowan  \& Narayan}{Sironi
  et~al.}{2021}]{sironi_etal_2021}
Sironi L.,  Rowan M.~E.,   Narayan R.,  2021, The Astrophysical Journal
  Letters, 907, L44

\bibitem[\protect\citeauthoryear{Spitkovsky}{Spitkovsky}{2008}]{spitkovsky_2008}
Spitkovsky A.,  2008, The Astrophysical Journal Letters, 682, L5

\bibitem[\protect\citeauthoryear{Stone, Ostriker  \& Gammie}{Stone
  et~al.}{1998}]{stone_etal_1998}
Stone J.~M.,  Ostriker E.~C.,   Gammie C.~F.,  1998, The Astrophysical Journal
  Letters, 508, L99

\bibitem[\protect\citeauthoryear{Takamoto \& Lazarian}{Takamoto \&
  Lazarian}{2016}]{takamoto_lazarian_2016}
Takamoto M.,  Lazarian A.,  2016, The Astrophysical Journal Letters, 831, L11

\bibitem[\protect\citeauthoryear{Takamoto \& Lazarian}{Takamoto \&
  Lazarian}{2017}]{takamoto_lazarian_2017}
Takamoto M.,  Lazarian A.,  2017, Monthly Notices of the Royal Astronomical
  Society, 472, 4542

\bibitem[\protect\citeauthoryear{TenBarge, Howes, Dorland  \& Hammett}{TenBarge
  et~al.}{2014}]{tenbarge_etal_2014}
TenBarge J.,  Howes G.~G.,  Dorland W.,   Hammett G.~W.,  2014, Computer
  Physics Communications, 185, 578

\bibitem[\protect\citeauthoryear{Thompson \& Blaes}{Thompson \&
  Blaes}{1998}]{thompson_blaes_1998}
Thompson C.,  Blaes O.,  1998, Physical Review D, 57, 3219

\bibitem[\protect\citeauthoryear{Towns et~al.,}{Towns et~al.}{2014}]{xsede}
Towns J.,  et~al., 2014, \mn@doi [Computing in Science \& Engineering]
  {10.1109/MCSE.2014.80}, 16, 62

\bibitem[\protect\citeauthoryear{Tran \& Sironi}{Tran \&
  Sironi}{2020}]{tran_sironi_2020}
Tran A.,  Sironi L.,  2020, The Astrophysical Journal Letters, 900, L36

\bibitem[\protect\citeauthoryear{Tsytovich}{Tsytovich}{1966}]{tsytovich_1966}
Tsytovich V.~N.,  1966, Physics-Uspekhi, 9, 370

\bibitem[\protect\citeauthoryear{Vlahos, Isliker  \& Lepreti}{Vlahos
  et~al.}{2004}]{vlahos_etal_2004}
Vlahos L.,  Isliker H.,   Lepreti F.,  2004, The Astrophysical Journal, 608,
  540

\bibitem[\protect\citeauthoryear{Walker, Boldyrev  \& Loureiro}{Walker
  et~al.}{2018}]{walker_etal_2018}
Walker J.,  Boldyrev S.,   Loureiro N.~F.,  2018, Physical Review E, 98, 033209

\bibitem[\protect\citeauthoryear{Werner, Uzdensky, Begelman, Cerutti  \&
  Nalewajko}{Werner et~al.}{2018}]{werner_etal_2018}
Werner G.,  Uzdensky D.,  Begelman M.,  Cerutti B.,   Nalewajko K.,  2018,
  Monthly Notices of the Royal Astronomical Society, 473, 4840

\bibitem[\protect\citeauthoryear{Wong, Zhdankin, Uzdensky, Werner  \&
  Begelman}{Wong et~al.}{2020}]{wong_etal_2020}
Wong K.,  Zhdankin V.,  Uzdensky D.~A.,  Werner G.~R.,   Begelman M.~C.,  2020,
  The Astrophysical Journal Letters, 893, L7

\bibitem[\protect\citeauthoryear{Xu \& Zhang}{Xu \&
  Zhang}{2017}]{xu_zhang_2017}
Xu S.,  Zhang B.,  2017, The Astrophysical Journal Letters, 846, L28

\bibitem[\protect\citeauthoryear{Yan \& Lazarian}{Yan \&
  Lazarian}{2002}]{yan_lazarian_2002}
Yan H.,  Lazarian A.,  2002, Physical review letters, 89, 281102

\bibitem[\protect\citeauthoryear{Yan \& Lazarian}{Yan \&
  Lazarian}{2004}]{yan_lazarian_2004}
Yan H.,  Lazarian A.,  2004, The Astrophysical Journal, 614, 757

\bibitem[\protect\citeauthoryear{Yang, Shi, Wan, Matthaeus  \& Chen}{Yang
  et~al.}{2016}]{yang_etal_2016}
Yang Y.,  Shi Y.,  Wan M.,  Matthaeus W.~H.,   Chen S.,  2016, Physical Review
  E, 93, 061102

\bibitem[\protect\citeauthoryear{Yang, Matthaeus, Shi, Wan  \& Chen}{Yang
  et~al.}{2017}]{yang_etal_2017}
Yang Y.,  Matthaeus W.~H.,  Shi Y.,  Wan M.,   Chen S.,  2017, Physics of
  Fluids, 29, 035105

\bibitem[\protect\citeauthoryear{Yuan \& Narayan}{Yuan \&
  Narayan}{2014}]{yuan_narayan_2014}
Yuan F.,  Narayan R.,  2014, Annual Review of Astronomy and Astrophysics, 52,
  529

\bibitem[\protect\citeauthoryear{Zhang \& Yan}{Zhang \&
  Yan}{2010}]{zhang_yan_2010}
Zhang B.,  Yan H.,  2010, The Astrophysical Journal, 726, 90

\bibitem[\protect\citeauthoryear{Zhdankin, Boldyrev, Perez  \& Tobias}{Zhdankin
  et~al.}{2014}]{zhdankin_etal_2014}
Zhdankin V.,  Boldyrev S.,  Perez J.~C.,   Tobias S.~M.,  2014, The
  Astrophysical Journal, 795, 127

\bibitem[\protect\citeauthoryear{Zhdankin, Werner, Uzdensky  \&
  Begelman}{Zhdankin et~al.}{2017}]{zhdankin_etal_2017}
Zhdankin V.,  Werner G.~R.,  Uzdensky D.~A.,   Begelman M.~C.,  2017, Physical
  Review Letters, 118, 055103

\bibitem[\protect\citeauthoryear{Zhdankin, Uzdensky, Werner  \&
  Begelman}{Zhdankin et~al.}{2018a}]{zhdankin_etal_2018a}
Zhdankin V.,  Uzdensky D.~A.,  Werner G.~R.,   Begelman M.~C.,  2018a, Monthly
  Notices of the Royal Astronomical Society, 474, 2514

\bibitem[\protect\citeauthoryear{{Zhdankin}, {Uzdensky}, {Werner}  \&
  {Begelman}}{{Zhdankin} et~al.}{2018b}]{zhdankin_etal_2018b}
{Zhdankin} V.,  {Uzdensky} D.~A.,  {Werner} G.~R.,   {Begelman} M.~C.,  2018b,
  The Astrophysical Journal Letters, 867, L18

\bibitem[\protect\citeauthoryear{Zhdankin, Uzdensky, Werner  \&
  Begelman}{Zhdankin et~al.}{2019}]{zhdankin_etal_2019}
Zhdankin V.,  Uzdensky D.~A.,  Werner G.~R.,   Begelman M.~C.,  2019, Physical
  review letters, 122, 055101

\bibitem[\protect\citeauthoryear{Zhdankin, Uzdensky, Werner  \&
  Begelman}{Zhdankin et~al.}{2020}]{zhdankin_etal_2020}
Zhdankin V.,  Uzdensky D.~A.,  Werner G.~R.,   Begelman M.~C.,  2020, Monthly
  Notices of the Royal Astronomical Society, 493, 603

\makeatother
\end{thebibliography}
\end{document}